\documentclass{jfm}

\usepackage[usenames,dvipsnames]{color} 
\usepackage{graphicx}
\usepackage{epstopdf}
\usepackage{natbib}
\usepackage{mathrsfs}
\usepackage{amsmath}
\usepackage{amssymb}
\usepackage{psfrag}

\usepackage{pifont}

\usepackage{multirow}
\usepackage{subfigure}

\usepackage{tikz}
\usetikzlibrary{shapes,arrows}

\tikzstyle{block} = [draw, fill=white, rectangle, minimum height=3em, minimum width=3em]
\tikzstyle{largeblock} = [draw, fill=white, rectangle, minimum height=3em, minimum width=3em]
\tikzstyle{smallblock} = [draw, fill=white, rectangle, minimum height=3em, minimum width=3em]
\tikzstyle{input} = [coordinate]
\tikzstyle{sum} = [draw, fill=white, circle, minimum size=2pt, inner sep=2pt, label={center:\tiny$+$}]
\tikzstyle{node} = [draw, fill=black, circle, minimum size=2pt, inner sep=0pt]
\tikzstyle{output} = [coordinate]
\tikzstyle{pinstyle} = [pin edge={to-,thin,black}]
\tikzstyle{line} = [draw, -latex']

\ifCUPmtlplainloaded \else
  \checkfont{eurm10}
  \iffontfound
    \IfFileExists{upmath.sty}
      {\typeout{^^JFound AMS Euler Roman fonts on the system,
                   using the 'upmath' package.^^J}%
       \usepackage{upmath}}
      {\typeout{^^JFound AMS Euler Roman fonts on the system, but you
                   dont seem to have the}%
       \typeout{'upmath' package installed. JFM.cls can take advantage
                 of these fonts,^^Jif you use 'upmath' package.^^J}%
      }
  \else
  \fi
\fi

\ifCUPmtlplainloaded \else
  \checkfont{msam10}
  \iffontfound
    \IfFileExists{amssymb.sty}
      {\typeout{^^JFound AMS Symbol fonts on the system, using the
                'amssymb' package.^^J}%
       \usepackage{amssymb}%
         \let\leq=\leqslant
         \let\geq=\geqslant
      }{}
  \fi
\fi

\ifCUPmtlplainloaded \else
  \IfFileExists{amsbsy.sty}
    {\typeout{^^JFound the 'amsbsy' package on the system, using it.^^J}%
     \usepackage{amsbsy}}
    {\providecommand\boldsymbol[1]{\mbox{\boldmath $##1$}}}
\fi

\newcommand\etal{\mbox{\textit{et al.}}}
\newcommand{\D}[2]{\frac{\partial #1}{\partial #2}}

\renewcommand{\div}{\nabla \cdot}

\renewcommand{\u}{{\mathbf{u}}}
\newcommand{\uz}{{\mathbf{u}}_{0}}
\newcommand{\uk}{{\mathbf{u}}_{\k}}
\newcommand{\f}{{\mathbf{f}}}
\newcommand{\fz}{\f_{0}}
\newcommand{\fk}{\f_{\k}}
\newcommand{\Lk}{{\mathcal{L}}_{\k}}
\renewcommand{\k}{\V{K}}
\newcommand{\x}{\V{x}}
\newcommand{\dlt}{\V{\delta}}
\newcommand{\e}{\mathrm{e}}
\newcommand{\phase}{\varphi}

\newcommand{\ud}{\mathrm{d}}

\newcommand{\mat}[2]{\left[\begin{array}{#1} #2\end{array}\right]}

\newcommand{\inprod}[2]{\left({#1},{#2}\right)}

\newcommand{\V}[1]{{\boldsymbol{#1}}}

\newcommand{\ave}[1]{\left< #1 \right>}

\newcommand{\ablabels}[2]%
	{
		\begin{tikzpicture}[overlay]
			\node [anchor=west] (a) at (-0.5\textwidth,#1) {(a)};
			\node [anchor=west] (b) at (-0.5\textwidth,#2) {(b)};
		\end{tikzpicture}
	}

\title[On coherent structure in wall turbulence]
  {On coherent structure in wall turbulence}

\author[A. S. Sharma and B. J. McKeon ]%
{A.\ns S. \ns S\ls H\ls A\ls R\ls M\ls A$^1$ \and B.\ns J.\ns M\ls c\ls K\ls E\ls O\ls N$^2$}

\affiliation{    $^1$Aerodynamics and Flight Mechanics, Faculty of Engineering and the Environment, University of Southampton, SO17 1BJ, UK\\[\affilskip]
    $^2$Graduate Aerospace Laboratories, California Institute of Technology, Pasadena, CA 91125, USA}

\pubyear{2013}
\volume{?}
\pagerange{?}
\date{?? and in revised form ??}

\begin{document}

\maketitle

\begin{abstract}
    A new theory of coherent structure in wall turbulence is presented.
    The theory is the first to predict packets of hairpin vortices and other structure in turbulence, and their dynamics, based on an analysis of the Navier-Stokes equations, under an assumption of a turbulent mean profile. The assumption of the turbulent mean acts as a restriction on the class of possible structures.
    It is shown that the coherent structure is a manifestation of essentially low-dimensional flow dynamics, arising from a critical layer mechanism.
	Using the decomposition presented in McKeon \& Sharma (\emph{J. Fluid Mech, {\bf{658}}, 2010}), complex coherent structure is recreated from minimal superpositions of response modes predicted by the analysis, which take the form of radially-varying travelling waves. The leading modes effectively constitute a low-dimensional description of the turbulent flow which is optimal in the sense of describing the resonant effects around the critical layer and which minimally predicts all types of structure.
    The approach is general for the full range of scales. By way of example, simple combinations of these modes are offered that predict hairpins and modulated hairpin packets.
    The example combinations are chosen to represent observed structure, consistent with the nonlinear triadic interaction for wavenumbers that is required for self-interaction of structures.
    The combination of the three leading response modes at streamwise wavenumbers $6,~1,~7$ and spanwise wavenumbers $\pm6,~\pm6,~\pm12$ respectively, with phase velocity $2/3$, is understood to represent a turbulence ``kernel'', which, it is proposed, constitutes a self-exciting process analogous to the near-wall cycle.
    Together, these interactions explain how the mode combinations may self-organise and self-sustain to produce experimentally observed structure.
    The phase interaction also leads to insight into skewness and correlation results known in the literature.
    It is also shown that the very large scale motions act to organise hairpin-like structures such that they co-locate with areas of low streamwise momentum, by a mechanism of locally altering the shear profile. 
These energetic streamwise structures arise naturally from the resolvent analysis, rather than by a summation of hairpin packets.	
    In addition, these packets are modulated through a ``beat" effect.
    The relationship between Taylor's hypothesis and coherence is discussed and both are shown to be the consequence of the localisation of the response modes around the critical layer.
    A pleasing link is made to the classical laminar inviscid theory, whereby the essential mechanism underlying the hairpin vortex is captured by two obliquely interacting Kelvin-Stuart (cat's eye) vortices.
Evidence for the theory is presented based on comparison to observations of structure in turbulent flow reported in the experimental and numerical simulation literature and to exact solutions reported in the transitional literature.
\end{abstract}

\section{Introduction}

The existence of coherent vortical structure in wall turbulence has been known for many decades, but the exact definition of coherent structure remains controversial, beyond the requirement that coherent motion has significant correlation in space and time \citep{Robinson91}. Attempts to describe additional features of the flow solely in terms of the velocity fields associated with the structures that are observed have had limited success. One notable exception is the attached eddy model proposed by \cite{Townsend1976} and subsequently developed by Perry and coworkers, e.g. \citep{Perrychong82, Perryhenchong, PerryMarusic95}. One weakness of the attached eddy model is that it does not incorporate the dynamics of the structures.
Over time, a split has developed in our understanding of how wall turbulence scales with increasing Reynolds number. One view derives from statistical measures of the velocity field and scaling arguments based on dimensional analysis in physical and spectral space. A second view focuses on the dynamics and spatial organisation of coherent structure.  
In the following, we refer to these views as ``statistical'' and ``structural'' respectively.
The analysis of \cite{McKeon2010} successfully modelled some selected structural features and Reynolds number scaling of velocity fluctuations in turbulent pipe flow. In this contribution, we use that framework to reconcile the distinct statistical and structural views, by demonstrating that the velocity response modes of the previous analysis are associated with strong evidence of coherent vortical structure. The superposition of just a few such modes captures the essence of even complex assemblies of individual structures such as hairpin vortices, and the relationship between large scale coherence and individual vortices.
In particular, we propose that hairpin vortices, hairpin packets and amplitude modulation arise naturally from the linear superposition of a small number of response modes. The characteristics of such superpositions quickly become complex as more response modes are added. However, the essential processes captured by simple packets are an innate feature of wall turbulence and we can extract them from the governing equations.
This permits very economical descriptions of classes of coherent structure with representative mode superpositions.

The structure of the paper is as follows.
Firstly, we continue this introduction with a brief review of the current understanding of coherent structures in wall turbulence relevant to the present study. We will then outline related modeling efforts and the connection with the dynamical systems viewpoint.
In section \ref{section:model} we present the theoretical development and analysis that provide the modes on which the work is based. Section \ref{section:skeleton} presents mode combinations representative of important characteristics of wall turbulence. Section \ref{section:structure} is a presentation and analysis of structure arising from the mode combinations, in particular, hairpins and hairpin packets. Their structural organisation and amplitude modulation across scales is also discussed. Next, section \ref{section:discussion} discusses the results in context of the current understanding of structure in wall turbulence and transitional flow. In the final section, conclusions are presented.

\subsection{Observations of coherent structure in wall turbulence}

The review of \cite{SmitsARFM11} identifies and describes four different classes of coherent structure, or organised motion, under current investigation in the literature, namely:
\begin{enumerate}
\item the streaks and quasi-streamwise vortices associated with the near-wall cycle;
\item hairpin or horseshoe vortices spanning from the wall into the overlap region in the mean velocity and perhaps beyond;
\item Large Scale Motions (LSMs), often associated with packets of hairpin vortices and stretching several outer lengthscales in the streamwise direction;
\item and Very Large Scale Motions (VLSMs) or superstructures, extending for approximately ten outer lengthscales in the streamwise direction and bearing an increasing proportion of streamwise energy and Reynolds shear stress for increasing Reynolds number.
\end{enumerate}
The main focus here will be on the origin and interactions of the second class of hairpin vortices, understood to have the distinctive, symmetrical hairpin shape only in the statistically representative or averaged sense, and subsequently make inferences about the likely relationships between assemblies of these structures and the last two long classes of structure. Readers with an interest in predictions regarding near-wall motions and other structure are referred to \cite{ArXiv10} and \cite{McKeonPoF13}.

The seminal, classical works of \cite{Theodorsen52} and \cite{HB81} utilised flow visualization to infer correctly many now-accepted features of coherent vortices. These include the geometrical similarity of the averaged structures, downstream-inclined horseshoes that reach up into the flow from the near-wall region and the dominance of hairpin heads with rotation in the sense of the mean shear and clustering of vortices into larger packets that could approach the outer lengthscale in cross-stream dimension. These conclusions were subsequently developed through the analysis of \cite{Robinson91} which resembled hairpins in the statistically averaged sense. \cite{Falco77} presented additional observations from smoke visualization and hot wire measurements, although these were interpreted somewhat differently by the author.

A concerted effort by the group of Adrian, as summarised in \cite{Adrian07}, has confirmed the existence of
hairpin vortices, and demonstrated the existence of hairpin packets throughout the
turbulent layer, culminating in LSMs. \cite{Kim99} also identified VLSMs and hypothesised that they are made up of aligned LSMs, in a picture that begins with very small
hairpins and ends with structures much longer than the thickness of the
turbulent layer.
The statistical characteristics of hairpin-like vortices in a turbulent boundary layer have been explored by \cite{Carlier05} (and subsequent work) and \cite{Gao11}, and linear stochastic estimation of likely vortical geometries and spatial relationships have been documented by \cite{Adrianinstfields00}, \cite{Natrajan07}, \cite{Sung11} and others, in both experiments and simulations of wall turbulence. \cite{Bharat03} confirmed the dynamical significance of hairpin vortices, quantified by the association of more than $25\%$ of the total Reynolds stress arising from Q2 events with hairpins occupying less than $4.5\%$ of space \citep{Bharat03}.

Despite a wealth of observational evidence for the existence of coherent hairpin vortices, there remains some controversy over their role in fully-developed turbulence, given the well-known occurrence of such structure in transitional flows and their apparent absence in the very high Reynolds number boundary layer of the atmospheric surface layer under near thermally neutral conditions \citep{Morris07}.
Correct identification of vortical structure in a complex, straining and rotational background velocity field has been the focus of much attention in recent years, as summarised in the works of \cite{Jeong95} and \cite{Chakraborty05}, who point out the strengths and weaknesses of different identification criteria in a range of flows most importantly associated with separating out coherent rotation in a field with a background vorticity (shear) likely of a similar magnitude to the (connected) vorticity associated with structural features.

A proposal for the relationship between the hairpin vortices, LSMs and VLSMs was given by \cite{Kim99} who described a possible organization in  which LSMs consisted of hairpin vortices aligned into packet structures, which themselves aligned to form very large scale structures.  Many authors have subsequently confirmed characteristic aspects of this description, in particular noting the common observations of wall-normal vortices ``straddling'' regions of constant low streamwise momentum, e.g. \cite{Bharat03}. The recent work of \cite{DennisVLSM11,Dennisvort11} developed these results in three dimensions, by projecting three-component time-resolved Particle Image Velocimetry (PIV) measurements performed in the cross-stream plane into the streamwise direction by use of Taylor's hypothesis.  They present elegant reconstructions of hairpin vortex packets and their collocation around extended regions of low streamwise momentum, constructed using averaging of the velocity field conditional on various swirling strength criteria.

\cite{Guala06} have noted that the VLSMs contain more than half of both the total kinetic energy of the streamwise velocity fluctuations and the Reynolds shear stress, at least in the moderate Reynolds number pipe flows of their studies. The subsequent studies of \cite{HutchinsPTRSA07} in the turbulent boundary layer have identified the further increasing energetic importance of the VLSMs as the Reynolds number increases, while \cite{Mathis09} provide a comprehensive report on the influence of the very large scales on the smaller, near-wall turbulence activity in the form of apparent amplitude modulation of the envelope of small scale streamwise turbulent fluctuations by the large scales.

\subsection{Models for generation, distribution and evolution of hairpin vortices}
Existing explanations of coherent structure in both wall turbulence and transition may be classified as either physically-argued models, those based on linear amplification of some kind, or those arguing from the association with exact, self-sustaining solutions to the Navier-Stokes equations.

As mentioned above, models for the origin, dynamics and evolution of coherent structure are generally sparse in the literature, with the notable exception of the near-wall, self-sustaining cycle elucidated by Waleffe and co-workers, e.g. \cite{Waleffe97}. Significant progress has also been made in terms of transient growth analyses, e.g. \cite{delAlamo06, Cossu09}.

In terms of the hairpin vortices proposed by \cite{Theodorsen52}, the attached eddy model originated by \cite{Townsend1976}, the dynamical model of \cite{SmithetalPTRSA91} and the hairpin vortex/packet paradigm developed by \cite{Adrian07} all address some aspects of their origin and development, but a complete picture appears to elude the community. We note also here the potential of the hierarchies of scales present in the mean momentum balance analysis of \cite{Klewicki07}, which have characteristics reminiscent of Townsend's attached eddies.

An important conceptual contribution to our understanding of the origin of structure has been made by workers looking to find self-sustaining solutions to the Navier-Stokes equations, either at asymptotically high Reynolds number, or ``edge states'' at the laminar-turbulent boundary in transitional flow. It seems likely that solutions existing in transitional flow continue in some way to turbulent Reynolds numbers. In this view, the phase-space flow around some of these solutions forms an attractor, concentrating the flow into regions that correspond to the coherent structures observed in turbulent flow. Thus the solutions shepherd any ensemble of flow trajectories to give the distinctive statistical and structural characteristics of turbulence. The work in this area has concentrated on finding these exact solutions and has associated these with archetypal structure. To efficiently and effectively enumerate the concentrations of flow trajectories in the region of state space near these solutions is to ``sketch'' turbulence.
There is some experimental evidence that this understanding is essentially correct in at least the case of transitional pipe flow  \citep{Mullin11}.
These solutions may be found by various methods, both numerical and analytical. It should be noted that the numerical approaches involve an iterative search and thus can be computationally expensive.
It is hoped that the present analysis will inform and facilitate the search for similar self-sustaining solutions at higher Reynolds numbers than has been possible up to now.

In a pipe, solutions have been found by \citet{Wedin04, Kerswell05, Duguet08, Duguet10, Pringle09}; these have tended to take the form of self-interacting rolls and streaks. A review of such work is given by \citet{Eckhardt07}.
\citet{Kerswell05} and others find that the travelling wave solutions are stationary states when viewed from the reference frame moving with the phase speed of the wave structure.
That such solutions are time quasi-periodic suggests that a frequency-domain treatment such as that applied in \cite{McKeon2010} is appropriate. Localised solutions may have a more complicated spectral signature.

The theoretical work of \citet{Benney69, Smith82} and others (see the review of \citet{Maslowe86}) found neutrally stable (self-exciting) nonlinear solutions for the inviscid equations. These asymptotic nonlinear solutions have analogues at lower Reynolds numbers, for example the vortex-wave interactions found numerically by \cite{Hall10}, and the near-wall cycle \citep{Waleffe97, Waleffe03}.

Most relevant to the current work, exact solutions appearing to consist of interacting hairpins and streaks were recently observed in various flows by \citet{Itano09, Generalis10, Gibson09, Deguchi10} and \cite{Cherubini11}.

\section{Modelling and analysis: Turbulence in the frequency-domain}
\label{section:model}
In the following, we develop a formulation of the Navier-Stokes equations designed to examine the highly selective response of turbulent pipe flow to (internal) forcing.
The first part of the analysis is a short recapitulation of the model presented by \citet{McKeon2010} with some improvements in notation. The second part makes some new observations about the interaction of the nonlinearity with the model and discusses consequences and some possible extensions of the development.
The conceptual picture underlying the analysis is one of a nonlinearity ``feeding back" to excite the linear dynamics, which in turn drive the nonlinearity.
As such, we concentrate on the linear amplification aspect of the turbulent process with the understanding that the velocity fields associated with individual response modes may be linearly superposed. The idea is that the linear dynamics in the nonlinear feedback loop are so selective in their response as to be the primary mechanism responsible for selecting structure in wall turbulence.
To provide closure, we \emph{assume} a turbulent mean velocity profile, thus restricting the search for structures to those that are supported by the assumed profile. By considering only velocity fields that are periodic in time and space, the search is also restricted to structures that do not decay and so, in turn, will contribute to the mean profile.
Thus, the chain of reasoning begins with the observed mean and infers flow structures that are consistent with it.
We find that the linear mechanism within the nonlinear feedback loop of the turbulent flow is highly selective (in a sense we quantify), with the consequence that analysis of a linear operator yields structures that are similar to those observed in wall turbulence.

\subsection{Formulation of the problem}

\begin{figure}
\centering
\begin{tikzpicture}[line join=round]
[\tikzset{>=latex}]\draw[arrows=->,thick](1.565,1.272)--(2.347,1.907);
\draw[thick](1.928,.993)--(1.924,.987)--(1.92,.982)--(1.916,.976)--(1.912,.971)--(1.908,.965)--(1.904,.96)--(1.9,.955)--(1.895,.949)--(1.891,.944)--(1.886,.94)--(1.881,.935)--(1.876,.93)--(1.871,.925)--(1.866,.921)--(1.861,.916)--(1.856,.912)--(1.851,.908)--(1.845,.904)--(1.84,.9)--(1.834,.896)--(1.829,.892)--(1.823,.888)--(1.817,.885)--(1.811,.882)--(1.805,.878)--(1.799,.875)--(1.793,.872)--(1.787,.869)--(1.781,.866)--(1.774,.864)--(1.768,.861)--(1.762,.859)--(1.755,.856)--(1.749,.854)--(1.742,.852)--(1.735,.85)--(1.729,.848)--(1.722,.847)--(1.715,.845)--(1.708,.844)--(1.701,.843)--(1.694,.841)--(1.687,.84)--(1.68,.84)--(1.673,.839)--(1.666,.838)--(1.659,.838)--(1.652,.837)--(1.645,.837)--(1.638,.837)--(1.63,.837)--(1.623,.837)--(1.616,.838)--(1.609,.838)--(1.601,.839)--(1.594,.839)--(1.587,.84)--(1.579,.841)--(1.572,.842)--(1.565,.843)--(1.557,.845)--(1.55,.846)--(1.543,.848)--(1.536,.85)--(1.528,.851)--(1.521,.853)--(1.514,.856)--(1.506,.858)--(1.499,.86)--(1.492,.863)--(1.485,.865)--(1.478,.868)--(1.47,.871)--(1.463,.874)--(1.456,.877)--(1.449,.88)--(1.442,.884)--(1.435,.887)--(1.428,.891)--(1.421,.895)--(1.415,.898)--(1.408,.902)--(1.401,.906)--(1.394,.911)--(1.388,.915)--(1.381,.919)--(1.374,.924)--(1.368,.928)--(1.362,.933)--(1.355,.938)--(1.349,.943)--(1.343,.948)--(1.336,.953)--(1.33,.958)--(1.324,.963)--(1.318,.969)--(1.312,.974)--(1.307,.98)--(1.301,.985)--(1.295,.991)--(1.29,.997)--(1.284,1.003)--(1.279,1.009)--(1.274,1.015)--(1.268,1.021)--(1.263,1.027)--(1.258,1.034)--(1.253,1.04)--(1.248,1.047)--(1.244,1.053)--(1.239,1.06)--(1.234,1.066)--(1.23,1.073)--(1.226,1.08)--(1.221,1.087)--(1.217,1.094)--(1.213,1.101)--(1.209,1.107)--(1.205,1.115)--(1.202,1.122)--(1.198,1.129)--(1.195,1.136)--(1.191,1.143)--(1.188,1.15)--(1.185,1.158)--(1.182,1.165)--(1.179,1.172)--(1.176,1.18)--(1.173,1.187)--(1.171,1.195)--(1.168,1.202)--(1.166,1.21)--(1.164,1.217)--(1.162,1.225)--(1.16,1.232)--(1.158,1.24)--(1.156,1.247)--(1.155,1.255)--(1.153,1.263)--(1.152,1.27)--(1.151,1.278)--(1.15,1.285)--(1.149,1.293)--(1.148,1.3)--(1.147,1.308)--(1.147,1.316)--(1.146,1.323)--(1.146,1.331)--(1.146,1.338)--(1.145,1.346)--(1.146,1.353)--(1.146,1.361)--(1.146,1.368)--(1.147,1.375)--(1.147,1.383)--(1.148,1.39)--(1.149,1.397)--(1.15,1.405)--(1.151,1.412)--(1.152,1.419)--(1.153,1.426)--(1.155,1.433)--(1.156,1.44)--(1.158,1.447)--(1.16,1.454)--(1.162,1.461)--(1.164,1.468)--(1.166,1.474)--(1.168,1.481)--(1.171,1.488)--(1.173,1.494)--(1.176,1.501)--(1.179,1.507)--(1.182,1.513)--(1.185,1.52)--(1.188,1.526)--(1.191,1.532)--(1.195,1.538)--(1.198,1.544)--(1.202,1.55)--(1.205,1.556)--(1.209,1.561)--(1.213,1.567)--(1.217,1.572)--(1.221,1.578)--(1.226,1.583)--(1.23,1.588)--(1.234,1.594)--(1.239,1.599)--(1.244,1.604)--(1.248,1.608)--(1.253,1.613)--(1.258,1.618)--(1.263,1.622)--(1.268,1.627)--(1.274,1.631)--(1.279,1.635)--(1.284,1.639)--(1.29,1.643)--(1.295,1.647)--(1.301,1.651)--(1.307,1.655)--(1.312,1.658)--(1.318,1.661)--(1.324,1.665)--(1.33,1.668)--(1.336,1.671)--(1.343,1.674)--(1.349,1.677)--(1.355,1.679)--(1.362,1.682)--(1.368,1.684)--(1.374,1.687)--(1.381,1.689)--(1.388,1.691);
\draw[arrows=<-,thick](3.242,.975)--(1.565,1.272);
\draw[dashed](.778,.633)--(1.565,1.272);
\draw[thick,->](1.355,1.679)--(1.456,1.704);
\draw[draw=black!50](.81,-.224)--(.782,-.221)--(.73,-.21)--(.677,-.195)--(.625,-.178)--(.574,-.157)--(.523,-.133)--(.474,-.106)--(.425,-.076)--(.378,-.043)--(.333,-.008)--(.289,.03)--(.248,.07)--(.208,.113)--(.171,.158)--(.136,.204)--(.104,.252)--(.074,.302)--(.048,.353)--(.024,.405)--(.003,.458)--(-.015,.512)--(-.03,.566)--(-.041,.621)--(-.05,.676)--(-.055,.73)--(-.056,.784)--(-.055,.812);
\filldraw[draw=black!50,fill=black!10,fill opacity=0.3,draw=none](1.612,.441)--(1.608,.397)--(1.603,.354)--(1.595,.311)--(1.584,.27)--(1.572,.229)--(1.557,.189)--(1.541,.151)--(1.522,.114)--(1.501,.079)--(1.478,.045)--(1.454,.012)--(1.427,-.018)--(1.399,-.047)--(1.369,-.074)--(1.337,-.099)--(1.304,-.122)--(1.269,-.143)--(1.233,-.162)--(1.196,-.179)--(1.157,-.194)--(1.118,-.206)--(1.078,-.216)--(1.036,-.223)--(.994,-.229)--(.952,-.232)--(.909,-.232)--(.866,-.23)--(.822,-.226)--(.778,-.219)--(.735,-.211)--(.691,-.199)--(.648,-.186)--(.605,-.17)--(.563,-.152)--(.521,-.132)--(.479,-.11)--(.439,-.086)--(.4,-.06)--(.361,-.032)--(.324,-.002)--(.288,.03)--(.253,.063)--(.22,.098)--(.188,.134)--(.158,.172)--(.13,.211)--(.103,.251)--(.079,.292)--(.056,.334)--(.035,.377)--(.016,.421)--(-.001,.465)--(-.015,.51)--(-.028,.555)--(-.038,.6)--(-.046,.645)--(-.051,.69)--(-.055,.735)--(-.056,.78)--(-.055,.825)--(-.051,.868)--(-.046,.912)--(-.038,.954)--(-.028,.996)--(-.015,1.036)--(-.001,1.076)--(.016,1.114)--(.035,1.151)--(.056,1.186)--(.079,1.22)--(.103,1.253)--(.13,1.283)--(.158,1.312)--(.188,1.339)--(.22,1.364)--(.253,1.388)--(.288,1.409)--(.324,1.428)--(.361,1.444)--(.4,1.459)--(.439,1.471)--(.479,1.481)--(.521,1.489)--(.563,1.494)--(.605,1.497)--(.648,1.497)--(.691,1.495)--(.735,1.491)--(.778,1.485)--(.822,1.476)--(.866,1.465)--(.909,1.451)--(.952,1.435)--(.994,1.417)--(1.036,1.397)--(1.078,1.375)--(1.118,1.351)--(1.157,1.325)--(1.196,1.297)--(1.233,1.267)--(1.269,1.235)--(1.304,1.202)--(1.337,1.167)--(1.369,1.131)--(1.399,1.093)--(1.427,1.054)--(1.454,1.014)--(1.478,.973)--(1.501,.931)--(1.522,.888)--(1.541,.844)--(1.557,.8)--(1.572,.756)--(1.584,.711)--(1.595,.665)--(1.603,.62)--(1.608,.575)--(1.612,.53)--(1.613,.485)--cycle;
\filldraw[draw=black!50,fill=black!10,fill opacity=0.3,draw=none](-1.399,-1.093)--(.158,.172)--(.188,.134)--(-1.369,-1.131)--cycle;
\filldraw[draw=black!50,fill=black!10,fill opacity=0.3,draw=none](-1.369,-1.131)--(.188,.134)--(.22,.098)--(-1.337,-1.167)--cycle;
\filldraw[draw=black!50,fill=black!10,fill opacity=0.3,draw=none](-1.427,-1.054)--(.13,.211)--(.158,.172)--(-1.399,-1.093)--cycle;
\filldraw[draw=black!50,fill=black!10,fill opacity=0.3,draw=none](-1.337,-1.167)--(.22,.098)--(.253,.063)--(-1.304,-1.202)--cycle;
\filldraw[draw=black!50,fill=black!10,fill opacity=0.3,draw=none](-1.454,-1.014)--(.103,.251)--(.13,.211)--(-1.427,-1.054)--cycle;
\filldraw[draw=black!50,fill=black!10,fill opacity=0.3,draw=none](-1.304,-1.202)--(.253,.063)--(.288,.03)--(-1.269,-1.235)--cycle;
\filldraw[draw=black!50,fill=black!10,fill opacity=0.3,draw=none](-1.478,-.973)--(.079,.292)--(.103,.251)--(-1.454,-1.014)--cycle;
\filldraw[draw=black!50,fill=black!10,fill opacity=0.3,draw=none](-1.269,-1.235)--(.288,.03)--(.324,-.002)--(-1.233,-1.267)--cycle;
\filldraw[draw=black!50,fill=black!10,fill opacity=0.3,draw=none](-1.501,-.931)--(.056,.334)--(.079,.292)--(-1.478,-.973)--cycle;
\filldraw[draw=black!50,fill=black!10,fill opacity=0.3,draw=none](-1.233,-1.267)--(.324,-.002)--(.361,-.032)--(-1.196,-1.297)--cycle;
\filldraw[draw=black!50,fill=black!10,fill opacity=0.3,draw=none](-1.522,-.888)--(.035,.377)--(.056,.334)--(-1.501,-.931)--cycle;
\filldraw[draw=black!50,fill=black!10,fill opacity=0.3,draw=none](-1.196,-1.297)--(.361,-.032)--(.4,-.06)--(-1.157,-1.325)--cycle;
\filldraw[draw=black!50,fill=black!10,fill opacity=0.3,draw=none](-1.541,-.844)--(.016,.421)--(.035,.377)--(-1.522,-.888)--cycle;
\filldraw[draw=black!50,fill=black!10,fill opacity=0.3,draw=none](-1.157,-1.325)--(.4,-.06)--(.439,-.086)--(-1.118,-1.351)--cycle;
\filldraw[draw=black!50,fill=black!10,fill opacity=0.3,draw=none](-1.557,-.8)--(-.001,.465)--(.016,.421)--(-1.541,-.844)--cycle;
\filldraw[draw=black!50,fill=black!10,fill opacity=0.3,draw=none](-1.118,-1.351)--(.439,-.086)--(.479,-.11)--(-1.078,-1.375)--cycle;
\filldraw[draw=black!50,fill=black!10,fill opacity=0.3,draw=none](-1.572,-.756)--(-.015,.51)--(-.001,.465)--(-1.557,-.8)--cycle;
\filldraw[draw=black!50,fill=black!10,fill opacity=0.3,draw=none](-1.078,-1.375)--(.479,-.11)--(.521,-.132)--(-1.036,-1.397)--cycle;
\filldraw[draw=black!50,fill=black!10,fill opacity=0.3,draw=none](-1.584,-.711)--(-.028,.555)--(-.015,.51)--(-1.572,-.756)--cycle;
\filldraw[draw=black!50,fill=black!10,fill opacity=0.3,draw=none](-1.036,-1.397)--(.521,-.132)--(.563,-.152)--(-.994,-1.417)--cycle;
\filldraw[draw=black!50,fill=black!10,fill opacity=0.3,draw=none](-1.595,-.665)--(-.038,.6)--(-.028,.555)--(-1.584,-.711)--cycle;
\filldraw[draw=black!50,fill=black!10,fill opacity=0.3,draw=none](-.994,-1.417)--(.563,-.152)--(.605,-.17)--(-.952,-1.435)--cycle;
\filldraw[draw=black!50,fill=black!10,fill opacity=0.3,draw=none](-1.603,-.62)--(-.046,.645)--(-.038,.6)--(-1.595,-.665)--cycle;
\filldraw[draw=black!50,fill=black!10,fill opacity=0.3,draw=none](-.952,-1.435)--(.605,-.17)--(.648,-.186)--(-.909,-1.451)--cycle;
\filldraw[draw=black!50,fill=black!10,fill opacity=0.3,draw=none](-1.608,-.575)--(-.051,.69)--(-.046,.645)--(-1.603,-.62)--cycle;
\filldraw[draw=black!50,fill=black!10,fill opacity=0.3,draw=none](-.909,-1.451)--(.648,-.186)--(.691,-.199)--(-.866,-1.465)--cycle;
\filldraw[draw=black!50,fill=black!10,fill opacity=0.3,draw=none](-1.612,-.53)--(-.055,.735)--(-.051,.69)--(-1.608,-.575)--cycle;
\filldraw[draw=black!50,fill=black!10,fill opacity=0.3,draw=none](-.866,-1.465)--(.691,-.199)--(.735,-.211)--(-.822,-1.476)--cycle;
\filldraw[draw=black!50,fill=black!10,fill opacity=0.3,draw=none](-1.613,-.485)--(-.056,.78)--(-.055,.735)--(-1.612,-.53)--cycle;
\filldraw[draw=black!50,fill=black!10,fill opacity=0.3,draw=none](-.822,-1.476)--(.735,-.211)--(.778,-.219)--(-.778,-1.485)--cycle;
\filldraw[draw=black!50,fill=black!10,fill opacity=0.3,draw=none](-1.612,-.441)--(-.055,.825)--(-.056,.78)--(-1.613,-.485)--cycle;
\draw[draw=black!50](-.055,.812)--(-.055,.838)--(-.053,.855);
\filldraw[draw=black!50,fill=black!10,fill opacity=0.3,draw=none](-.778,-1.485)--(.778,-.219)--(.822,-.226)--(-.735,-1.491)--cycle;
\draw[draw=black!50](.853,-.23)--(.835,-.228)--(.81,-.224);
\filldraw[draw=black!50,fill=black!10,fill opacity=0.3,draw=none](-1.608,-.397)--(-.051,.868)--(-.055,.825)--(-1.612,-.441)--cycle;
\draw[draw=black!50](-.053,.855)--(-.05,.89)--(-.041,.942)--(-.034,.975);
\filldraw[draw=black!50,fill=black!10,fill opacity=0.3,draw=none](-.735,-1.491)--(.822,-.226)--(.866,-.23)--(-.691,-1.495)--cycle;
\draw[draw=black!50](.973,-.231)--(.94,-.233)--(.888,-.232)--(.853,-.23);
\filldraw[draw=black!50,fill=black!10,fill opacity=0.3,draw=none](-1.603,-.354)--(-.046,.912)--(-.051,.868)--(-1.608,-.397)--cycle;
\filldraw[draw=black!50,fill=black!10,fill opacity=0.3,draw=none](-.691,-1.495)--(.866,-.23)--(.909,-.232)--(-.648,-1.497)--cycle;
\filldraw[draw=black!50,fill=black!10,fill opacity=0.3,draw=none](-1.595,-.311)--(-.038,.954)--(-.046,.912)--(-1.603,-.354)--cycle;
\filldraw[draw=black!50,fill=black!10,fill opacity=0.3,draw=none](-.648,-1.497)--(.909,-.232)--(.952,-.232)--(-.605,-1.497)--cycle;
\draw[draw=black!50](-.034,.975)--(-.032,.983);
\draw[draw=black!50](-.032,.983)--(-.03,.992)--(-.015,1.041)--(-.009,1.057);
\filldraw[draw=black!50,fill=black!10,fill opacity=0.3,draw=none](-1.584,-.27)--(-.028,.996)--(-.038,.954)--(-1.595,-.311)--cycle;
\draw[draw=black!50](.982,-.231)--(.973,-.231);
\draw[draw=black!50](1.058,-.221)--(1.042,-.224)--(.991,-.231)--(.982,-.231);
\filldraw[draw=black!50,fill=black!10,fill opacity=0.3,draw=none](-.605,-1.497)--(.952,-.232)--(.994,-.229)--(-.563,-1.494)--cycle;
\filldraw[draw=black!50,fill=black!10,fill opacity=0.3,draw=none](-1.572,-.229)--(-.015,1.036)--(-.028,.996)--(-1.584,-.27)--cycle;
\filldraw[draw=black!50,fill=black!10,fill opacity=0.3,draw=none](-.563,-1.494)--(.994,-.229)--(1.036,-.223)--(-.521,-1.489)--cycle;
\draw[draw=black!50](-.009,1.057)--(-.006,1.066);
\draw[draw=black!50](-.006,1.066)--(.003,1.089)--(.024,1.135);
\filldraw[draw=black!50,fill=black!10,fill opacity=0.3,draw=none](-1.557,-.189)--(-.001,1.076)--(-.015,1.036)--(-1.572,-.229)--cycle;
\draw[draw=black!50](1.067,-.22)--(1.058,-.221);
\draw[draw=black!50](1.139,-.202)--(1.091,-.215)--(1.067,-.22);
\filldraw[draw=black!50,fill=black!10,fill opacity=0.3,draw=none](-.521,-1.489)--(1.036,-.223)--(1.078,-.216)--(-.479,-1.481)--cycle;
\filldraw[draw=black!50,fill=black!10,fill opacity=0.3,draw=none](-1.541,-.151)--(.016,1.114)--(-.001,1.076)--(-1.557,-.189)--cycle;
\filldraw[draw=black!50,fill=black!10,fill opacity=0.3,draw=none](-.479,-1.481)--(1.078,-.216)--(1.118,-.206)--(-.439,-1.471)--cycle;
\draw[draw=black!50](.024,1.135)--(.027,1.141);
\draw[draw=black!50](.027,1.141)--(.048,1.178)--(.065,1.206);
\filldraw[draw=black!50,fill=black!10,fill opacity=0.3,draw=none](-1.522,-.114)--(.035,1.151)--(.016,1.114)--(-1.541,-.151)--cycle;
\draw[draw=black!50](1.146,-.2)--(1.139,-.202);
\draw[draw=black!50](1.216,-.173)--(1.186,-.186)--(1.146,-.2);
\filldraw[draw=black!50,fill=black!10,fill opacity=0.3,draw=none](-.439,-1.471)--(1.118,-.206)--(1.157,-.194)--(-.4,-1.459)--cycle;
\filldraw[draw=black!50,fill=black!10,fill opacity=0.3,draw=none](-1.501,-.079)--(.056,1.186)--(.035,1.151)--(-1.522,-.114)--cycle;
\filldraw[draw=black!50,fill=black!10,fill opacity=0.3,draw=none](-.4,-1.459)--(1.157,-.194)--(1.196,-.179)--(-.361,-1.444)--cycle;
\draw[draw=black!50](.065,1.206)--(.069,1.212);
\filldraw[draw=black!50,fill=black!10,fill opacity=0.3,draw=none](-1.478,-.045)--(.079,1.22)--(.056,1.186)--(-1.501,-.079)--cycle;
\draw[draw=black!50](.069,1.212)--(.074,1.22)--(.104,1.259)--(.115,1.271);
\draw[draw=black!50](1.223,-.17)--(1.216,-.173);
\filldraw[draw=black!50,fill=black!10,fill opacity=0.3,draw=none](-.361,-1.444)--(1.196,-.179)--(1.233,-.162)--(-.324,-1.428)--cycle;
\draw[draw=black!50](1.289,-.136)--(1.275,-.144)--(1.232,-.167)--(1.223,-.17);
\filldraw[draw=black!50,fill=black!10,fill opacity=0.3,draw=none](-1.454,-.012)--(.103,1.253)--(.079,1.22)--(-1.478,-.045)--cycle;
\filldraw[draw=black!50,fill=black!10,fill opacity=0.3,draw=none](-.324,-1.428)--(1.233,-.162)--(1.269,-.143)--(-.288,-1.409)--cycle;
\filldraw[draw=black!50,fill=black!10,fill opacity=0.3,draw=none](-1.427,.018)--(.13,1.283)--(.103,1.253)--(-1.454,-.012)--cycle;
\draw[draw=black!50](.115,1.271)--(.121,1.278);
\draw[draw=black!50](.121,1.278)--(.136,1.296)--(.171,1.33)--(.203,1.357);
\filldraw[draw=black!50,fill=black!10,fill opacity=0.3,draw=none](-.288,-1.409)--(1.269,-.143)--(1.304,-.122)--(-.253,-1.388)--cycle;
\draw[draw=black!50](1.297,-.131)--(1.289,-.136);
\draw[draw=black!50](1.33,-.109)--(1.317,-.119)--(1.297,-.131);
\filldraw[draw=black!50,fill=black!10,fill opacity=0.3,draw=none](-1.399,.047)--(.158,1.312)--(.13,1.283)--(-1.427,.018)--cycle;
\filldraw[draw=black!50,fill=black!10,fill opacity=0.3,draw=none](-.253,-1.388)--(1.304,-.122)--(1.337,-.099)--(-.22,-1.364)--cycle;
\draw[draw=black!50](1.388,-.063)--(1.356,-.09)--(1.33,-.109);
\filldraw[draw=black!50,fill=black!10,fill opacity=0.3,draw=none](-1.369,.074)--(.188,1.339)--(.158,1.312)--(-1.399,.047)--cycle;
\filldraw[draw=black!50,fill=black!10,fill opacity=0.3,draw=none](-.22,-1.364)--(1.337,-.099)--(1.369,-.074)--(-.188,-1.339)--cycle;
\filldraw[draw=black!50,fill=black!10,fill opacity=0.3,draw=none](-1.337,.099)--(.22,1.364)--(.188,1.339)--(-1.369,.074)--cycle;
\draw[draw=black!50](.203,1.357)--(.208,1.361)--(.248,1.39)--(.289,1.416)--(.333,1.438)--(.378,1.458)--(.425,1.474)--(.474,1.487)--(.523,1.496)--(.574,1.502)--(.625,1.505)--(.677,1.504)--(.73,1.5)--(.782,1.492)--(.835,1.481)--(.888,1.467)--(.94,1.449)--(.991,1.428)--(1.042,1.404)--(1.091,1.377)--(1.139,1.347)--(1.186,1.315)--(1.232,1.279)--(1.275,1.241)--(1.317,1.201)--(1.356,1.158)--(1.394,1.114)--(1.429,1.067)--(1.461,1.019)--(1.49,.969)--(1.517,.918)--(1.541,.866)--(1.562,.813)--(1.58,.759)--(1.595,.705)--(1.606,.651)--(1.614,.596)--(1.619,.542)--(1.621,.487);
\draw[dashed](-.778,-.633)--(.778,.633);
\filldraw[draw=black!50,fill=black!10,fill opacity=0.3,draw=none](-.188,-1.339)--(1.369,-.074)--(1.399,-.047)--(-.158,-1.312)--cycle;
\draw[draw=black!50](1.621,.487)--(1.619,.434)--(1.614,.381)--(1.606,.33)--(1.595,.279)--(1.58,.23)--(1.562,.183)--(1.541,.137)--(1.517,.093)--(1.49,.052)--(1.461,.012)--(1.429,-.024)--(1.394,-.059)--(1.388,-.063);
\filldraw[draw=black!50,fill=black!10,fill opacity=0.3,draw=none](-1.304,.122)--(.253,1.388)--(.22,1.364)--(-1.337,.099)--cycle;
\filldraw[draw=black!50,fill=black!10,fill opacity=0.3,draw=none](-.158,-1.312)--(1.399,-.047)--(1.427,-.018)--(-.13,-1.283)--cycle;
\filldraw[draw=black!50,fill=black!10,fill opacity=0.3,draw=none](-1.269,.143)--(.288,1.409)--(.253,1.388)--(-1.304,.122)--cycle;
\filldraw[draw=black!50,fill=black!10,fill opacity=0.3,draw=none](-.13,-1.283)--(1.427,-.018)--(1.454,.012)--(-.103,-1.253)--cycle;
\filldraw[draw=black!50,fill=black!10,fill opacity=0.3,draw=none](-1.233,.162)--(.324,1.428)--(.288,1.409)--(-1.269,.143)--cycle;
\filldraw[draw=black!50,fill=black!10,fill opacity=0.3,draw=none](-.103,-1.253)--(1.454,.012)--(1.478,.045)--(-.079,-1.22)--cycle;
\filldraw[draw=black!50,fill=black!10,fill opacity=0.3,draw=none](-1.196,.179)--(.361,1.444)--(.324,1.428)--(-1.233,.162)--cycle;
\filldraw[draw=black!50,fill=black!10,fill opacity=0.3,draw=none](-.079,-1.22)--(1.478,.045)--(1.501,.079)--(-.056,-1.186)--cycle;
\filldraw[draw=black!50,fill=black!10,fill opacity=0.3,draw=none](-1.157,.194)--(.4,1.459)--(.361,1.444)--(-1.196,.179)--cycle;
\filldraw[draw=black!50,fill=black!10,fill opacity=0.3,draw=none](-.056,-1.186)--(1.501,.079)--(1.522,.114)--(-.035,-1.151)--cycle;
\filldraw[draw=black!50,fill=black!10,fill opacity=0.3,draw=none](-1.118,.206)--(.439,1.471)--(.4,1.459)--(-1.157,.194)--cycle;
\filldraw[draw=black!50,fill=black!10,fill opacity=0.3,draw=none](-.035,-1.151)--(1.522,.114)--(1.541,.151)--(-.016,-1.114)--cycle;
\filldraw[draw=black!50,fill=black!10,fill opacity=0.3,draw=none](-1.078,.216)--(.479,1.481)--(.439,1.471)--(-1.118,.206)--cycle;
\filldraw[draw=black!50,fill=black!10,fill opacity=0.3,draw=none](-.016,-1.114)--(1.541,.151)--(1.557,.189)--(.001,-1.076)--cycle;
\filldraw[draw=black!50,fill=black!10,fill opacity=0.3,draw=none](-1.036,.223)--(.521,1.489)--(.479,1.481)--(-1.078,.216)--cycle;
\filldraw[draw=black!50,fill=black!10,fill opacity=0.3,draw=none](.001,-1.076)--(1.557,.189)--(1.572,.229)--(.015,-1.036)--cycle;
\filldraw[draw=black!50,fill=black!10,fill opacity=0.3,draw=none](-.994,.229)--(.563,1.494)--(.521,1.489)--(-1.036,.223)--cycle;
\filldraw[draw=black!50,fill=black!10,fill opacity=0.3,draw=none](.015,-1.036)--(1.572,.229)--(1.584,.27)--(.028,-.996)--cycle;
\filldraw[draw=black!50,fill=black!10,fill opacity=0.3,draw=none](-.952,.232)--(.605,1.497)--(.563,1.494)--(-.994,.229)--cycle;
\filldraw[draw=black!50,fill=black!10,fill opacity=0.3,draw=none](.028,-.996)--(1.584,.27)--(1.595,.311)--(.038,-.954)--cycle;
\filldraw[draw=black!50,fill=black!10,fill opacity=0.3,draw=none](-.909,.232)--(.648,1.497)--(.605,1.497)--(-.952,.232)--cycle;
\filldraw[draw=black!50,fill=black!10,fill opacity=0.3,draw=none](.038,-.954)--(1.595,.311)--(1.603,.354)--(.046,-.912)--cycle;
\filldraw[draw=black!50,fill=black!10,fill opacity=0.3,draw=none](-.866,.23)--(.691,1.495)--(.648,1.497)--(-.909,.232)--cycle;
\filldraw[draw=black!50,fill=black!10,fill opacity=0.3,draw=none](.046,-.912)--(1.603,.354)--(1.608,.397)--(.051,-.868)--cycle;
\filldraw[draw=black!50,fill=black!10,fill opacity=0.3,draw=none](-.822,.226)--(.735,1.491)--(.691,1.495)--(-.866,.23)--cycle;
\filldraw[draw=black!50,fill=black!10,fill opacity=0.3,draw=none](.051,-.868)--(1.608,.397)--(1.612,.441)--(.055,-.825)--cycle;
\filldraw[draw=black!50,fill=black!10,fill opacity=0.3,draw=none](-.778,.219)--(.778,1.485)--(.735,1.491)--(-.822,.226)--cycle;
\filldraw[draw=black!50,fill=black!10,fill opacity=0.3,draw=none](.055,-.825)--(1.612,.441)--(1.613,.485)--(.056,-.78)--cycle;
\filldraw[draw=black!50,fill=black!10,fill opacity=0.3,draw=none](-.735,.211)--(.822,1.476)--(.778,1.485)--(-.778,.219)--cycle;
\filldraw[draw=black!50,fill=black!10,fill opacity=0.3,draw=none](.056,-.78)--(1.613,.485)--(1.612,.53)--(.055,-.735)--cycle;
\filldraw[draw=black!50,fill=black!10,fill opacity=0.3,draw=none](-.691,.199)--(.866,1.465)--(.822,1.476)--(-.735,.211)--cycle;
\filldraw[draw=black!50,fill=black!10,fill opacity=0.3,draw=none](.055,-.735)--(1.612,.53)--(1.608,.575)--(.051,-.69)--cycle;
\filldraw[draw=black!50,fill=black!10,fill opacity=0.3,draw=none](-.648,.186)--(.909,1.451)--(.866,1.465)--(-.691,.199)--cycle;
\filldraw[draw=black!50,fill=black!10,fill opacity=0.3,draw=none](.051,-.69)--(1.608,.575)--(1.603,.62)--(.046,-.645)--cycle;
\filldraw[draw=black!50,fill=black!10,fill opacity=0.3,draw=none](-.605,.17)--(.952,1.435)--(.909,1.451)--(-.648,.186)--cycle;
\filldraw[draw=black!50,fill=black!10,fill opacity=0.3,draw=none](.046,-.645)--(1.603,.62)--(1.595,.665)--(.038,-.6)--cycle;
\filldraw[draw=black!50,fill=black!10,fill opacity=0.3,draw=none](-.563,.152)--(.994,1.417)--(.952,1.435)--(-.605,.17)--cycle;
\filldraw[draw=black!50,fill=black!10,fill opacity=0.3,draw=none](.038,-.6)--(1.595,.665)--(1.584,.711)--(.028,-.555)--cycle;
\filldraw[draw=black!50,fill=black!10,fill opacity=0.3,draw=none](-.521,.132)--(1.036,1.397)--(.994,1.417)--(-.563,.152)--cycle;
\filldraw[draw=black!50,fill=black!10,fill opacity=0.3,draw=none](.028,-.555)--(1.584,.711)--(1.572,.756)--(.015,-.51)--cycle;
\filldraw[draw=black!50,fill=black!10,fill opacity=0.3,draw=none](-.479,.11)--(1.078,1.375)--(1.036,1.397)--(-.521,.132)--cycle;
\filldraw[draw=black!50,fill=black!10,fill opacity=0.3,draw=none](.015,-.51)--(1.572,.756)--(1.557,.8)--(.001,-.465)--cycle;
\filldraw[draw=black!50,fill=black!10,fill opacity=0.3,draw=none](-.439,.086)--(1.118,1.351)--(1.078,1.375)--(-.479,.11)--cycle;
\filldraw[draw=black!50,fill=black!10,fill opacity=0.3,draw=none](.001,-.465)--(1.557,.8)--(1.541,.844)--(-.016,-.421)--cycle;
\filldraw[draw=black!50,fill=black!10,fill opacity=0.3,draw=none](-.4,.06)--(1.157,1.325)--(1.118,1.351)--(-.439,.086)--cycle;
\filldraw[draw=black!50,fill=black!10,fill opacity=0.3,draw=none](-.016,-.421)--(1.541,.844)--(1.522,.888)--(-.035,-.377)--cycle;
\filldraw[draw=black!50,fill=black!10,fill opacity=0.3,draw=none](-.361,.032)--(1.196,1.297)--(1.157,1.325)--(-.4,.06)--cycle;
\filldraw[draw=black!50,fill=black!10,fill opacity=0.3,draw=none](-.035,-.377)--(1.522,.888)--(1.501,.931)--(-.056,-.334)--cycle;
\filldraw[draw=black!50,fill=black!10,fill opacity=0.3,draw=none](-.324,.002)--(1.233,1.267)--(1.196,1.297)--(-.361,.032)--cycle;
\filldraw[draw=black!50,fill=black!10,fill opacity=0.3,draw=none](-.056,-.334)--(1.501,.931)--(1.478,.973)--(-.079,-.292)--cycle;
\filldraw[draw=black!50,fill=black!10,fill opacity=0.3,draw=none](-.288,-.03)--(1.269,1.235)--(1.233,1.267)--(-.324,.002)--cycle;
\filldraw[draw=black!50,fill=black!10,fill opacity=0.3,draw=none](-.079,-.292)--(1.478,.973)--(1.454,1.014)--(-.103,-.251)--cycle;
\filldraw[draw=black!50,fill=black!10,fill opacity=0.3,draw=none](-.253,-.063)--(1.304,1.202)--(1.269,1.235)--(-.288,-.03)--cycle;
\filldraw[draw=black!50,fill=black!10,fill opacity=0.3,draw=none](-.103,-.251)--(1.454,1.014)--(1.427,1.054)--(-.13,-.211)--cycle;
\filldraw[draw=black!50,fill=black!10,fill opacity=0.3,draw=none](-.22,-.098)--(1.337,1.167)--(1.304,1.202)--(-.253,-.063)--cycle;
\filldraw[draw=black!50,fill=black!10,fill opacity=0.3,draw=none](-.13,-.211)--(1.427,1.054)--(1.399,1.093)--(-.158,-.172)--cycle;
\filldraw[draw=black!50,fill=black!10,fill opacity=0.3,draw=none](-.188,-.134)--(1.369,1.131)--(1.337,1.167)--(-.22,-.098)--cycle;
\filldraw[draw=black!50,fill=black!10,fill opacity=0.3,draw=none](-.158,-.172)--(1.399,1.093)--(1.369,1.131)--(-.188,-.134)--cycle;
\filldraw[draw=black!50,fill=black!10,fill opacity=0.3,draw=none](.056,-.78)--(.055,-.735)--(.051,-.69)--(.046,-.645)--(.038,-.6)--(.028,-.555)--(.015,-.51)--(.001,-.465)--(-.016,-.421)--(-.035,-.377)--(-.056,-.334)--(-.079,-.292)--(-.103,-.251)--(-.13,-.211)--(-.158,-.172)--(-.188,-.134)--(-.22,-.098)--(-.253,-.063)--(-.288,-.03)--(-.324,.002)--(-.361,.032)--(-.4,.06)--(-.439,.086)--(-.479,.11)--(-.521,.132)--(-.563,.152)--(-.605,.17)--(-.648,.186)--(-.691,.199)--(-.735,.211)--(-.778,.219)--(-.822,.226)--(-.866,.23)--(-.909,.232)--(-.952,.232)--(-.994,.229)--(-1.036,.223)--(-1.078,.216)--(-1.118,.206)--(-1.157,.194)--(-1.196,.179)--(-1.233,.162)--(-1.269,.143)--(-1.304,.122)--(-1.337,.099)--(-1.369,.074)--(-1.399,.047)--(-1.427,.018)--(-1.454,-.012)--(-1.478,-.045)--(-1.501,-.079)--(-1.522,-.114)--(-1.541,-.151)--(-1.557,-.189)--(-1.572,-.229)--(-1.584,-.27)--(-1.595,-.311)--(-1.603,-.354)--(-1.608,-.397)--(-1.612,-.441)--(-1.613,-.485)--(-1.612,-.53)--(-1.608,-.575)--(-1.603,-.62)--(-1.595,-.665)--(-1.584,-.711)--(-1.572,-.756)--(-1.557,-.8)--(-1.541,-.844)--(-1.522,-.888)--(-1.501,-.931)--(-1.478,-.973)--(-1.454,-1.014)--(-1.427,-1.054)--(-1.399,-1.093)--(-1.369,-1.131)--(-1.337,-1.167)--(-1.304,-1.202)--(-1.269,-1.235)--(-1.233,-1.267)--(-1.196,-1.297)--(-1.157,-1.325)--(-1.118,-1.351)--(-1.078,-1.375)--(-1.036,-1.397)--(-.994,-1.417)--(-.952,-1.435)--(-.909,-1.451)--(-.866,-1.465)--(-.822,-1.476)--(-.778,-1.485)--(-.735,-1.491)--(-.691,-1.495)--(-.648,-1.497)--(-.605,-1.497)--(-.563,-1.494)--(-.521,-1.489)--(-.479,-1.481)--(-.439,-1.471)--(-.4,-1.459)--(-.361,-1.444)--(-.324,-1.428)--(-.288,-1.409)--(-.253,-1.388)--(-.22,-1.364)--(-.188,-1.339)--(-.158,-1.312)--(-.13,-1.283)--(-.103,-1.253)--(-.079,-1.22)--(-.056,-1.186)--(-.035,-1.151)--(-.016,-1.114)--(.001,-1.076)--(.015,-1.036)--(.028,-.996)--(.038,-.954)--(.046,-.912)--(.051,-.868)--(.055,-.825)--cycle;
\draw[draw=black!50](-1.356,-1.158)--(-1.394,-1.114)--(-1.422,-1.076);
\draw[draw=black!50](-1.223,-1.286)--(-1.232,-1.279)--(-1.275,-1.241)--(-1.317,-1.201)--(-1.356,-1.158);
\draw[draw=black!50](-1.422,-1.076)--(-1.429,-1.067)--(-1.461,-1.019)--(-1.49,-.969)--(-1.517,-.918);
\draw[draw=black!50](-.888,-1.467)--(-.94,-1.449)--(-.991,-1.428)--(-1.042,-1.404)--(-1.091,-1.377)--(-1.139,-1.347)--(-1.186,-1.315)--(-1.223,-1.286);
\draw[draw=black!50](-1.517,-.918)--(-1.541,-.866)--(-1.562,-.813)--(-1.571,-.787);
\draw[draw=black!50](-1.571,-.787)--(-1.58,-.759)--(-1.595,-.705)--(-1.597,-.695);
\draw[draw=black!50](-1.597,-.695)--(-1.606,-.651)--(-1.613,-.604);
\draw[draw=black!50](-1.613,-.604)--(-1.614,-.596)--(-1.619,-.542)--(-1.62,-.515);
\draw[draw=black!50](-.72,-1.5)--(-.73,-1.5)--(-.782,-1.492)--(-.835,-1.481)--(-.888,-1.467);
\draw[draw=black!50](-1.62,-.515)--(-1.621,-.487)--(-1.619,-.434)--(-1.618,-.424);
\draw[draw=black!50](-1.618,-.424)--(-1.614,-.381)--(-1.607,-.337);
\draw[draw=black!50](-.633,-1.505)--(-.677,-1.504)--(-.72,-1.5);
\draw[draw=black!50](-1.607,-.337)--(-1.606,-.33)--(-1.595,-.279)--(-1.588,-.255);
\draw[draw=black!50](-.549,-1.499)--(-.574,-1.502)--(-.625,-1.505)--(-.633,-1.505);
\draw[draw=black!50](-1.588,-.255)--(-1.58,-.23)--(-1.562,-.183)--(-1.558,-.174);
\draw[draw=black!50](-.465,-1.484)--(-.474,-1.487)--(-.523,-1.496)--(-.549,-1.499);
\draw[draw=black!50](-1.558,-.174)--(-1.541,-.137)--(-1.521,-.099);
\draw[draw=black!50](-.385,-1.46)--(-.425,-1.474)--(-.465,-1.484);
\draw[draw=black!50](-1.521,-.099)--(-1.517,-.093)--(-1.49,-.052)--(-1.476,-.033);
\draw[draw=black!50](-.312,-1.427)--(-.333,-1.438)--(-.378,-1.458)--(-.385,-1.46);
\draw[draw=black!50](-1.476,-.033)--(-1.461,-.012)--(-1.429,.024)--(-1.422,.031);
\draw[draw=black!50](-.241,-1.385)--(-.248,-1.39)--(-.289,-1.416)--(-.312,-1.427);
\draw[draw=black!50](-1.422,.031)--(-1.394,.059)--(-1.362,.086);
\draw[draw=black!50](-.176,-1.334)--(-.208,-1.361)--(-.241,-1.385);
\draw[dashed](-1.565,-1.272)--(-.778,-.633);
\draw[draw=black!50](-1.362,.086)--(-1.356,.09)--(-1.317,.119)--(-1.297,.131);
\draw[draw=black!50](-.121,-1.278)--(-.136,-1.296)--(-.171,-1.33)--(-.176,-1.334);
\draw[draw=black!50](-1.297,.131)--(-1.275,.144)--(-1.232,.167)--(-1.223,.17);
\draw[draw=black!50](-.069,-1.212)--(-.074,-1.22)--(-.104,-1.259)--(-.121,-1.278);
\draw[draw=black!50](-1.223,.17)--(-1.216,.173);
\draw[draw=black!50](-1.216,.173)--(-1.186,.186)--(-1.146,.2);
\draw[draw=black!50](-.065,-1.206)--(-.069,-1.212);
\draw[draw=black!50](-.027,-1.141)--(-.048,-1.178)--(-.065,-1.206);
\draw[arrows=<-,thick](-.447,-.695)--(.056,-.784);
\draw[draw=black!50](-1.146,.2)--(-1.139,.202);
\draw[draw=black!50](-1.139,.202)--(-1.091,.215)--(-1.067,.22);
\draw[draw=black!50](-.024,-1.135)--(-.027,-1.141);
\draw[draw=black!50](.006,-1.066)--(-.003,-1.089)--(-.024,-1.135);
\draw[draw=black!50](-1.067,.22)--(-1.058,.221);
\draw[draw=black!50](-1.058,.221)--(-1.042,.224)--(-.991,.231)--(-.982,.231);
\draw[draw=black!50](.009,-1.057)--(.006,-1.066);
\draw[draw=black!50](.032,-.983)--(.03,-.992)--(.015,-1.041)--(.009,-1.057);
\draw[draw=black!50](-.982,.231)--(-.973,.231);
\draw[draw=black!50](-.973,.231)--(-.94,.233)--(-.895,.233);
\draw[draw=black!50](.048,-.898)--(.041,-.942)--(.032,-.983);
\draw[draw=black!50](-.895,.233)--(-.888,.232)--(-.835,.228)--(-.81,.224);
\draw[draw=black!50](.055,-.812)--(.055,-.838)--(.05,-.89)--(.048,-.898);
\draw[draw=black!50](-.81,.224)--(-.782,.221)--(-.73,.21)--(-.677,.195)--(-.625,.178)--(-.574,.157)--(-.523,.133)--(-.474,.106)--(-.425,.076)--(-.378,.043)--(-.333,.008)--(-.289,-.03)--(-.248,-.07)--(-.208,-.113)--(-.171,-.158)--(-.136,-.204)--(-.104,-.252)--(-.074,-.302)--(-.048,-.353)--(-.024,-.405)--(-.003,-.458)--(.015,-.512)--(.03,-.566)--(.041,-.621)--(.05,-.676)--(.055,-.73)--(.056,-.784);
\draw[draw=black!50](.056,-.784)--(.055,-.812);
\path (2.347,1.907) node[above] {$x,~u$}
                               (3.242,.975) node[right] {$r,~v$};\path (-.447,-.695) node[below] {$y,~v'$};\node at (1.02,1.925) {$\theta,~w$};\end{tikzpicture}

\caption{A schematic of pipe geometry and nomenclature.}
\label{fig:pipe}
\end{figure}
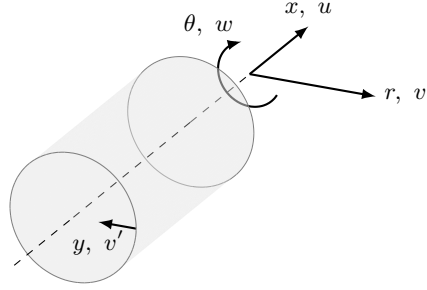

The non-dimensional Navier-Stokes equations for incompressible pipe flow with constant viscosity are given by
\begin{equation}
\partial_t \tilde{\u} = -\nabla p -\tilde{\u}\cdot\nabla\tilde{\u}  + \frac{1}{Re}\nabla^2 \tilde{\u}, \quad \div \tilde{\u} = 0
\label{eq:pipe}
\end{equation}
For convenience of comparison across different studies, these equations of motion are non-dimensionalised with respect to the pipe diameter and the bulk, volume-averaged velocity, $U_\mathrm{bulk}$, so that the Reynolds number in Equation~\ref{eq:pipe} is
\[Re=\frac{U_\textrm{bulk}D}{\nu},\]
with $D=2R$ the pipe diameter, $p$ the pressure field and $\nu$ the kinematic viscosity. We retain the boundary layer terminology by fixing $y=1-r$, and $u$, $v(=-v')$ and $w$ as corresponding to the streamwise, wall-normal and azimuthal velocities such that $\tilde{\mathbf{u}}=(v',w,u)$, as shown in figure~\ref{fig:pipe}.

Next, the problem is Fourier transformed in the streamwise, azimuthal and temporal directions. The frequency-domain approach permits computational searches at high Reynolds number without neglecting viscosity and avoids the question of whether initial conditions are ``forgotten''. We use a different non-dimensionalisation when considering the critical layer effects because we find comparison to the centreline velocity (rather than bulk velocity) to be easier to work with in this context. Accordingly we non-dimensionalise variables with the pipe radius and turbulent mean centerline velocity, such that $k=k'R$, $n=n'R$ and $\omega = \omega' R/U_{CL}$, where the prime denotes the dimensional variables. Thence the wavespeed is $c=\omega/k$ and we define the wavenumber set $\k = (k,n,c)$ for convenience so that $\k$ with all positive elements refers to a downstream travelling wave.
The velocity field $\tilde{\u}$ is expressed as a sum of harmonic, radially varying traveling waves.
\begin{equation}
    \tilde{\u}(r,x,\theta,t) = \sum_n \int_{-\infty}^\infty \int_{-\infty}^\infty \tilde{\u}_\k(r) e^{i\k \cdot \x} \ud k \ud \omega
\end{equation}
where, for convenience, we define the notation $\k\cdot\x \equiv kx+n\theta-\omega t$ (which is not a dot product).
As such, $\tilde{\mathbf{u}}_\mathbf{k}$ for a particular $\k$ maps a point in $r$ to the appropriate complex Fourier coefficient of the velocity field at that point.

As yet, only the wall-normal direction remains untransformed; the problem is to find a suitable basis for these functions of $r$.
We introduce a projection onto a divergence free basis $\left\{\xi_m(r)\right\}$ in the radial direction, Fourier modes in the homogeneous spatial directions and in time, so that only harmonic forcing and response is considered.
With coefficients $\{\chi_{m\k}\}$,
\begin{align}
    u(r,x,\theta,t) &= \sum_{m',n}\int_{-\infty}^{\infty}\int_{-\infty}^{\infty} \chi_{m'\k} \xi_{m'}(r)\e^{i\k\cdot \x} \ud k \ud \omega  \notag\\
    &= \sum_{m,n}\int_{-\infty}^{\infty}\int_{-\infty}^{\infty} \chi_{m\k} \xi_{m\k}(r)\ud k \ud \omega.
\end{align}
In order to eliminate pressure, the basis functions are required to have the special properties
\begin{align}
\inprod{\xi_a}{\xi_b}_r   &= \delta_{ab},\\
\div \xi(r)_{m\k} &= 0.
\end{align}

The mean equation falls out naturally as the steady Fourier component of the Navier-Stokes equations (the component for which for which $\k\cdot\x=0$  always and everywhere).
This component of the velocity field is essentially the velocity field averaged over all homogeneous directions. Knowing this component implies correct Reynolds stresses in the rest of the model via its appearance in the resolvent.
Accordingly, an assumed mean is used to close the equations. In principle, another Fourier component could have been chosen instead.
It is convenient to denote this mean velocity field by $\uz$. The field with the mean subtracted is denoted $\u =\tilde{\u}-\uz$.

The term $\f=-\u\cdot\nabla\u$ describes the nonlinear, triadic interaction represented as an internal forcing to an otherwise linear system. It is similarly decomposed into constituent $\fk$ and its $(k,n,\omega)=(0,0,0)$ component similarly denoted $\fz$. This finally yields equations for the fluctuations that are linear in $\uk$, and a base flow equation,
\begin{align}
-i\omega\uk &= \Lk \uk + \fk, \quad \forall (k,n,\omega)\neq(0,0,0),
	\label{eq:linu}\\
0 &= \fz -\uz\cdot\nabla\uz + \frac{1}{Re}\nabla^2\uz.
	\label{eq:u0}
\end{align}
The response of the flow at a particular wavenumber combination subjected to the harmonic forcing $\fk$ is given by rearrangement of equation~\ref{eq:linu},
\begin{align}
\uk &= (-i\omega - \Lk)^{-1} \fk\\
 &= \mat{ccc}{
    i(k \uz - \omega) - Re^{-1}D & -{2in}{r^{-2} Re^{-1}}              &   0   \\
    {2in}{r^{-2} Re^{-1}}    & i(k \uz-\omega) - Re^{-1}D          &   0   \\
    -\partial_r \uz          & 0               &i(k \uz - \omega) - Re^{-1}(D+ r^{-2})
}^{-1} \fk
\label{eq:model}
\end{align}
with $D=\partial_r^2 + r^{-1}\partial_r - r^{-2}(n^2+1) - k^2$, and the states being the radial, azimuthal and axial velocities expressed in a divergence-free basis.
The operator $(-i\omega - \Lk)^{-1}$ is called the resolvent and is the focus of our analysis.
For large enough Reynolds number, we expect high gain in regions of high shear where the $\partial_r \u_0$ is large and at the critical layer where $c$ is equal to the local mean velocity.
This localisation is the physical basis for there being a highly selective response to forcing and we posit that it \emph{will be somehow present in any simplifying theory of coherent structure}.

\subsection{Most amplified modes}

\cite{McKeon2010} use the Schmidt decomposition of the resolvent (the singular value decomposition in the discrete case) at particular $\k$ corresponding to those observed in real wall turbulence to rank the response to forcing in an energetic sense, i.e.
\begin{equation}
    (-i\omega I - \Lk)^{-1} = \sum_{l=1}^{\infty} \psi_{l\k}(r)\sigma_{l\k}\phi^{*}_{l\k}(r)
\label{eq:decomposeG}
\end{equation}
with the orthogonality condition
\begin{equation}
    \inprod{\phi_{l\k}(r)}{\phi_{m\k}(r)}_y =\delta_{lm}, \quad
    \inprod{\psi_{l\k}(r)}{\psi_{m\k}(r)}_y =\delta_{lm}
\end{equation} 
and ordering \(\sigma_{l\k}\geq \sigma_{l+1,\k} \geq 0.\)

The inner product on $r$ (equivalently $y$) only is required,
\begin{equation}
    \inprod{a_{\k}}{b_{\k}}_r = \int^1_0 a^*_{\k}(r) b_{\k}(r)r dr.
\end{equation}
This defines the forcing ($\phi$) and response ($\psi$) modes associated with the response of the flow. Note that these are normalised with respect to the energy integrated over the radius of the pipe. The velocity of a mode at any given wall-normal location will therefore be dependent  on the allocation between velocity components and the wall-normal location.
The $\phi_{l\k}$ and $\psi_{l\k}$ form the right and left Schmidt bases for the forcing and velocity fields at a given $\k$ and the real $\sigma_{l\k}$ are the associated singular values. Assuming $\Lk$ is always stable, this decomposition exists for real $\omega$. The decomposition is unique up to a pre-multiplying unitary complex factor on both bases corresponding to a phase shift and up to the ordering in $l$ of $\sigma_{l\k}$s, hence in what follows the relative phases are fixed with respect to the first coefficient.

This basis pair can then be used to decompose some arbitrary forcing and the resulting velocity at any particular Fourier
component
\begin{eqnarray}
    \fk(r) = \sum_{l=1}^{\infty} \phi_{l\k}(r)\chi_{l\k}
	\label{decompf}
	\label{eq:fk}\\
    \uk(r) = \sum_{l=1}^{\infty} \sigma_{l\k}\psi_{l\k}(r)\chi_{l\k}.
	\label{eq:uk}
\label{decompv}
\end{eqnarray}
The energy of the same Fourier component of the resulting disturbance velocity is
\begin{equation}
    E_{\k} = \inprod{\uk}{\uk} = \sum_{l=1}^{\infty} \sigma_{l\k}^{2}\chi_{l\k}^{2}.
	\label{eq:Ek}
\end{equation}
The rank-$m$ approximation of the resolvent at a given $\k$ by $\sum_{l=1}^m \psi_{l\k} \sigma_{l\k} \phi_{l\k}^*$ is optimal in the sense of the energy defined in \eqref{eq:Ek} for equal $\chi_{l\k}$.
The forcing shape that gives the largest energy gain at a particular $\k$ is given with $\chi_{l\k}=0$ for ${l \neq 1}$.
This approach describes the dependence of maximum energy amplification on the radial form of the forcing in the wavenumber and frequency domain. The norm of the resolvent at $\k$ induced by the inner product over the radius is the leading singular value, $\sigma_{1\k}$. This means that the normalised harmonic forcing that gives the largest disturbance energy in the $\inprod{\cdot}{\cdot}_y$ sense is $\fk=\phi_{1\k}$, with a gain of $\sigma_{1\k}$. The next largest arises from $\fk=\phi_{2\k}$ and so on, at a particular wavenumber pair and frequency. The corresponding flow response modes are given by the related $\uk = \psi_{1\k}, \psi_{2\k}$, etc.

For the purposes of this work, we consider only real $\omega$, $k$ and $n$. This corresponds to statistical homogeneity in those directions. Consideration of transient or spatially developing flows would require complex frequency or wavenumber respectively.
We remark that consideration of only real $\omega$ gives a great simplification of the problem. Indeed a further simplification is possible: as in the streamwise direction, where projection into a streamwise periodic unit limits the lower range of $k$, the idea of a periodic decomposition in time implies a fundamental frequency setting the lower limit in $\omega$. The relationship between the implied ``turbulence period'' and the unit length is given by the timescale associated with the largest structures captured in the box. Similarly, we need not consider $\omega$ at frequencies above those where viscosity dominates. The periodicity also results in discrete $\omega$. The equivalent of this in the wall-normal direction is a natural truncation where the amplitude of a response mode, namely the product of the singular value and the magnitude of the forcing, is below a given threshold $\varepsilon$, i.e. $\sigma_i \inprod{-\psi_a \cdot \nabla \psi_b}{\phi_i} < \varepsilon$.

As in \citet{McKeon2010}, the computational analysis of $\Lk$ was performed using a modified version of the spectral code of \cite{Meseguer03}.
The turbulent mean profiles for the higher Reynolds number were determined directly from experimental data from the Princeton/ONR \emph{Superpipe} and reported by \cite{mckeonmean04}. Unless stated otherwise, all results shown here will be for $Re=75\times10^3$ ($R^+=D u_\tau/(2 \nu)=1800$), where $u_\tau=\sqrt{\tau_w/\rho}$ is the friction velocity, $\tau_w$ is the mean wall shear stress, $\rho$ is the density and $\nu$ is the kinematic viscosity. The lower Reynolds number considered in Section \ref{sec:VLSM} uses the mean profile from \cite{toonder} at $Re=10^4$ ($R^+=315$).

\subsection{Summary of analysis}
Here, the steps in the analysis are summarised:
\begin{enumerate}
\item Fourier decomposition in symmetric directions (axial, azimuthal, time),
\item decomposition of nonlinear forcing and linear resolvent operator,
\item assumption of the zero-frequency, zero-wavenumber Fourier mode to close the model, which turns out to be the turbulent mean profile,
\item decomposition of resolvent operator by singular value decomposition to select maximal gain directions consistent with the mean,
\item restriction to real frequencies and wavenumbers, neglecting transients, consistent with statistically stationary, homogeneous turbulence,
\item choice of example wavenumbers, frequencies and mode numbers to give modes representative of observed structures of interest,
\item selection of a number of leading singular values (in the case of this paper, one) for each chosen response mode corresponding to the desired model fidelity,
\item choice of amplitudes and phases for constructed mode combinations.
\end{enumerate}
Of these, only the first four steps are critical to the model formulation. The remaining steps are chosen for reasons of simplification or exposition.
At the moment the choice of amplitudes and phases for mode combinations is the result of informed intuition. In principle, this choice is governed by the nonlinear interaction between modes, and work is ongoing to formalise this choice.

\section{Characterising the skeleton of wall turbulence}
\label{section:skeleton}

In order to describe the structure that has been observed in turbulence, and to aid comparison with the literature, we examine wavenumbers and convection velocities that are representative of important spectral results in the literature. Most key features of the chosen combinations persist away from the precise values of parameters. That is, the structural features to be demonstrated are quite robust to the wavenumber, frequency and amplitude shown. Response modes at these wavenumbers form a ``skeleton'' of wall turbulence that can describe such structure in a compact way.
Close to the wall, i.e. in the logarithmic region of the mean velocity and below, we assume at least some semblance of universality, such that the results from all the canonical wall flows can be interpreted to inform the synthesis of dominant mode combinations. In the following, the choice of various combinations of streamwise and spanwise wavenumbers, relative streamwise convection velocities and mode amplitudes is outlined and justified.

\subsection{Selection of representative wavenumber combinations}

At sufficiently high Reynolds number, wall flows are known to exhibit at least two energetically important combinations of lengthscales, namely those associated with the near-wall cycle centered with energetic contribution concentrated around $y^+ = 15$, $(\lambda_x^+, \lambda_z^+)\approx (1000,100)$, and the so-called Very Large Scale Motions (VLSMs, \citet{Kim99}). The importance of the latter has been emphasised only in the past decade \citep{SmitsARFM11}. The recent comparison between canonical flows performed by \cite{Monty09} describes the slight differences in the VLSM lengthscales between the canonical flows, but for the purposes of this manuscript, we adopt a scale of $\lambda_x/\delta =2\pi$, such that $k=1$. The spanwise lengthscale associated with the VLSMs has been investigated by \cite{Hutchins07}, \cite{BaileyPOD09} and \cite{Monty07} and found to be approximately equal to the outer lengthscale, $n \approx 6$. The emergence of these two energetic spanwise wavelengths are also consistent with the linear amplification analyses of, e.g. \cite{delAlamo06} and \cite{Cossu09} in channel and boundary layer flow respectively, which reveal that the globally most amplified perturbations are associated with two modes with inner and outer scaling.

Thus we expect there to be two combinations of streamwise and spanwise lengthscales which can be considered to constitute the spectral ``skeleton'' of turbulence in the near-wall and logarithmic region: namely $(\lambda_x^+, \lambda_z^+)\approx (1000,100)$ and $(k,n)=(1,6)$. We focus here on the latter combination, centered in or around the logarithmic region, but note that there are likely similar mechanisms to those discussed below associated with the near-wall motions. We will discuss the appropriate convection velocities in the following section.

The recent work of \cite{Dennisvort11, DennisVLSM11} used conditional averaging of time-resolved, cross-stream PIV to reconstruct hairpin packets with spanwise dimension $z/\delta \sim 0.5$. That study used a much greater wall-normal extent than those identified in \cite{Adrianinstfields00}, which was at lower Reynolds number and in pipe flow, and a streamwise extent of at least $6 \delta$. These packets were observed to span long low-momentum regions, assumed to be associated with the low-speed part of the VLSMs. A typical streamwise vortex spacing of approximately $\delta$ suggests consideration of a mode combination with $(k,n)=(6,\pm 6)$ (mode $K_A$) with the VLSM. This combination is termed here
the ``ideal packet'', which we represent with packet $K_B$.

We will also consider a shorter, narrower mode with $(k,n)=(7,\pm 12)$ such that there are consistent triadic interactions with $(k,n)=(6,\pm 6)$ and $(k,n)=(1,\pm 6)$. Packet $K_C$ represents what we will ultimately call the ``modulating packet'', in which the modes are aligned in the spanwise direction. The ``decorrelated modulating packet'', $K_D$, contains modes that are still triadically consistent but have a less simple azimuthal periodicity.
Of course, in real flows, these will be subject to significant spread or jitter around these representative values, perhaps best interpreted in the context of a continuous spectrum of energetic scales.

We note also that studies that rank scales based on their contribution to the energy integrated in the wall-normal direction, e.g. the proper orthogonal decomposition (POD) of \cite{Hellstroem11}, identify scales that are even longer in the streamwise direction than the VLSMs. Thus we also consider the extremely large-scale mode corresponding to \citet{Hellstroem11}, which, as will be shown below, corresponds to $(k,n)=(0.3, \pm 3)$. We term these ``globally energetic'' modes and represent them with packet $K_E$.

The modes under investigation in this paper, which should be considered as representative of the skeleton of wall turbulence, are summarised in table \ref{tab:modes}.

\begin{table}
    \centering
    \begin{tabular}{lccccc}
        & & $k$ & $n$ & $c$ & $A$\\ \hline
        \multirow{1}{*}{$K_A$: single response mode} & $\k_1$ & $6$ & $\pm 6$ & $2/3$ & 1.00i \\ \hline

        \multirow{2}{*}{$K_B$: ideal packet} & $\k_1$ & $6$ & $\pm 6$ & $2/3$ & 1.00i \\
        & $\k_2$ & $1$  & $\pm 6$   & $2/3$ & -4.50~ \\ \hline

        \multirow{3}{*}{$K_C$: modulating packet} & $\k_1$ & $6$ & $\pm 6$ & $2/3$ & 1.00i \\
        & $\k_2$ & $1$  & $\pm 6$   & $2/3$ & -4.50~ \\
        & $\k_3$ & $7$ & $\pm 12$ & $2/3$ & 0.83i \\ \hline

        \multirow{4}{*}{$K_D$: decorrelated modulating packet} & $\k_1^*$ & $6$ & $\pm 8$ & $2/3$ & 1.00i \\
        & $\k_2^*$ & $1$  & $\pm 6$   & $2/3$ & -4.50~ \\
        & $\k_3^*$ & $7$ & $\pm 14$ & $2/3$ & ~0.83i \\ \hline

        \multirow{5}{*}{$K_E$: globally energetic modes} & $\k_4$ & $0.3$ & $\pm 3$ & $2/3$ & 0.30 \\
        & $\k_5$ & $1.5$ & $\pm 4$ & $2/3$ & 1.00\\
        & $\k_6$ & $2.1$ & $\pm 5$ & $2/3$ & 3.00\\
        & $\k_2$ & $1$   & $\pm 6$ & $2/3$ & 2.00\\ \hline
    \end{tabular}
    \caption{Representative mode combinations for each wavenumber packet. $A$ is the complex amplitude of the response mode ($A_i=\sigma_i \chi_i$ as per equation \ref{decompv}). The phase angle of complex amplitude $A$ determines the relative phase of that mode.}
    \label{tab:modes}
\end{table}

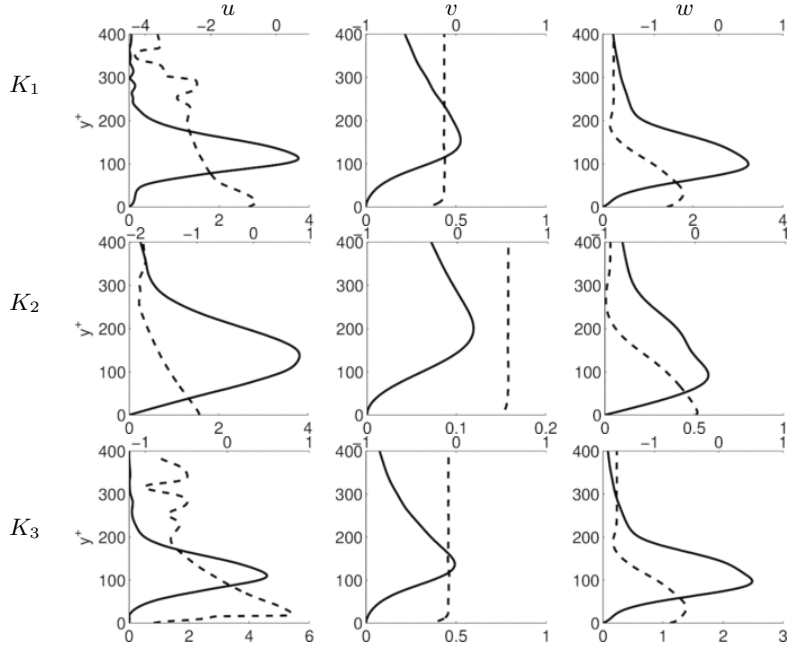
\begin{figure}
	\centering
	\begin{tikzpicture}[overlay]
		\node [] (u) at (-0.2\textwidth,-0.1) {$u$};
		\node [] (v) at (0.02\textwidth,-0.1) {$v$};
		\node [] (w) at (0.25\textwidth,-0.1) {$w$};
		\node [] (k1) at (-0.4\textwidth,-1.1) {$K_1$};
		\node [] (k2) at (-0.4\textwidth,-4) {$K_2$};
		\node [] (k3) at (-0.4\textwidth,-7) {$K_3$};
	\end{tikzpicture}\\\vspace{.2cm}
	\includegraphics[width=.7\textwidth]{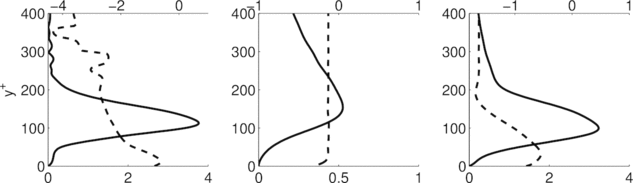}
	\includegraphics[width=.7\textwidth]{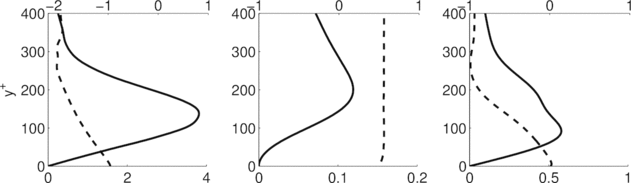}
	\includegraphics[width=.7\textwidth]{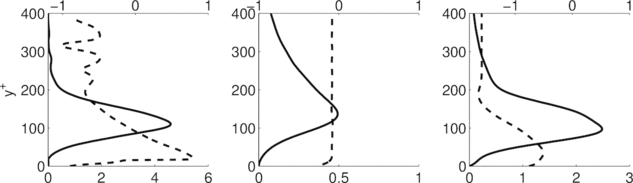}
	\caption{Amplitude and phase variation over wall-normal position of $\psi_{1\k}(r)$, for single response modes $\k_1$ (top row), $\k_2$ (middle row) and $\k_3$ (bottom row) each at unit magnitude. The amplitude is shown by the solid lines (lower abscissa) and the phase, in $\pi$ radians, by the dotted lines (upper abscissa). Components of streamwise, radial and azimuthal velocity are shown in the left, middle and right columns respectively. }
	\label{fig:ampphase_Ks}
\end{figure}

\subsection{Choice of convective velocity}
While a precise definition of ``coherent structure'' has proven elusive over the years, at minimum it must have the property of coherence over significant space and time. With this in mind, the streamwise convective velocity $c$ for each mode was selected with reference to two constraints:
\begin{enumerate}
\item the relative velocity of one mode relative to the other must be consistent with observations of the evolution and dispersion of real hairpin packets;
\item $\omega$ must be sufficiently low for each mode such that the footprint of the mode reaches down to the wall, since the majority of observations of the structure under consideration here are have been assumed to be attached to the wall in the sense of Townsend.
\end{enumerate}
The latter constraint is simply met by examination of the velocity and swirl fields associated with the selected modes once the frequency has been selected. The hairpin packet modes recorded in table~\ref{tab:modes} all meet these criteria.

With regard to the first constraint, we require limited dispersion of the assembled wavepacket, akin to enforcing similar streamwise modal convection velocities.  In terms of hairpin packets, \citet{Adrian00} reported less than $7\%$ dispersion of packets in their PIV study, equivalent to noting that individual hairpin structures convected with similar streamwise convection velocities. This is mandated in some sense for the packet structure to be observable in a turbulent flow: if the difference is too large between convective velocities, then any packet structure observed will be highly dispersive, with such a short lifetime that it will be unlikely to be detected in a real flow.

In the extreme, the validity of Taylor's hypothesis of frozen turbulence over times and distances which are small relative to the largest scales in the flow (equivalent to a zeroth order functional expansion) implies zero dispersion at a given wall-normal location and therefore wave packet with all components traveling at the same convective velocity. In order to investigate only the most persistent structure (therefore most likely to be observed in a real flow), we opt to follow Taylor's hypothesis in our selection of convection velocity $c$ and identify structure associated with idealised, non-dispersive wavepackets.

\begin{figure}
	\centering
	\includegraphics[width=0.6\textwidth]{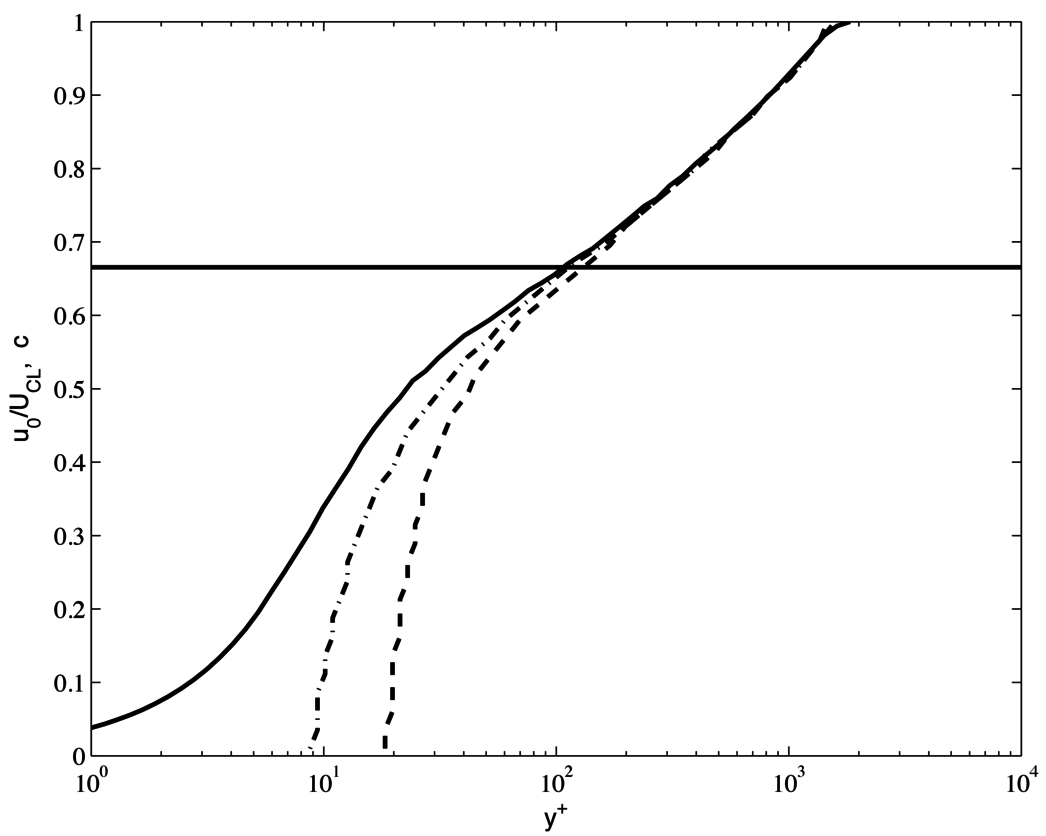}
    \caption{Variation of the radial location of peak streamwise velocity fluctuation for the response modes associated with the wavenumbers of $K_B$ (table \ref{tab:modes}) with increasing convection velocity, $c$. The lines denote $- \cdot -$: $(k,n)=(6,\pm 6)$ and $--$: $(k,n)=(1,\pm 6)$. The mean profile at this Reynolds number, $R^+=1800$, is given by the solid curve. The horizontal line identifies $c=2/3$, identified as the location of the critical layer for the VLSM by \cite{McKeon2010}, corresponding to a wall-normal location of $y^+ \approx 140$ at this Reynolds number. }
    \label{fig:peaklocs1}
\end{figure}

The amplitude and phase of the main component modes ($\k_1,~\k_2,~\k_3$) making up table~\ref{tab:modes} are shown in figure \ref{fig:ampphase_Ks}. Figure~\ref{fig:peaklocs1} shows the variation of the wall-normal location of the peak in streamwise velocity intensity associated with the component modes of $K_B$ from table~\ref{tab:modes} as convective velocity, $c$, of each mode is increased. The appropriately-scaled mean velocity profile is also shown to highlight the transition of each mode from a configuration in which the energy can be considered to be ``attached'' to the wall, i.e. the footprint reaches down to the wall and the location of the peak energy varies little with increasing $c$, to a critical mode where the wall-normal locus of the peak tracks the wall-normal location where $c$ corresponds to the local mean velocity. The selection of modes with the same convective velocity is akin to picking modes with peak streamwise intensities that fall on the horizontal line in figure~\ref{fig:peaklocs1}. A convection velocity corresponding to the outer (VLSM) mode, $c=2/3$, discussed in \cite{McKeon2010}, was selected, and appears to correspond exactly to the critical layer for $\k_2$ and close to the critical layer for $\k_1$. Inspection of figure~\ref{fig:peaklocs1} reveals the wall-normal location of the critical layer associated with the modes is about $y^+=140$; it will be seen that this is a fairly good predictor of the location of the hairpin head.

Our use of something equivalent to the frozen turbulence hypothesis is roughly equivalent to the projection downstream of cross-stream, time-resolved PIV using Taylor's hypothesis, as for example in the work of \cite{Dennisvort11} and \cite{Hellstroem11}.  In the former study, averaging conditioned on swirl criteria was employed, effectively winnowing away activity without time-correlation and leaving structure that is coherent in time.  The resulting structures can be viewed as reflecting the effect of filtering the full flow field based on the identification of structure and characterizing the dominant structural scale distribution in $k$, $n$ and $y$. Use of a different filtering event would result in a different identified field. By contrast, the current approach works in the opposite direction, effectively filtering on $k$ and $n$ and determining the $y$ distribution of velocity associated with the response mode, permitting subsequent identification of any associated coherent structure.

\subsection{Mode amplitudes}

Observations of the full, three-dimensional $(k,n,\omega)$ spectrum are relatively rare in the literature, but the measurements of the streamwise velocity by \cite{LeHew11} give some guidance on the appropriate selection of relative mode amplitudes (which will be important when summations of modes are considered). These data suggest that the amplitudes of the streamwise velocities of the turbulence skeleton should be of the same order of magnitude, with the highest amplitude associated with the outer mode (VLSM). The selected amplitudes are given in table~\ref{tab:modes}.

\subsection{Tools for identification of coherent vortical structure}

The challenges associated with identifying coherent vortical structure in a turbulent field have been well-studied, e.g. \cite{Jeong95}, \cite{Chakraborty05}.
The swirl, $\lambda$, is given by the imaginary part of the complex conjugate eigenvalue pair associated with the velocity gradient tensor. In what follows, we will use the swirl to identify regions of rotation as opposed to the combined influence of shear and rotation identified by vorticity; note that the use of any of the other common identifiers of coherent structure, e.g. \cite{Chakraborty05}, give similar results in this simplified flow model.

\section{Structure from response modes}
\label{section:structure}

We begin our discussion of the coherent structure associated with the response modes by investigating single modes, before demonstrating that structures such as hairpin packets arise naturally from the linear superposition of such modes.

\subsection{Structure associated with individual wall modes}

\begin{figure}
\centering
\begin{tikzpicture}
	\node [] (a) at (-0.3\textwidth,-0.1) {(a)};
	\node [] (b) at (0\textwidth,-0.1) {(b)};
	\node [] (c) at (0.3\textwidth,-0.1) {(c)};
\end{tikzpicture}\\
\includegraphics[width=0.3\textwidth]{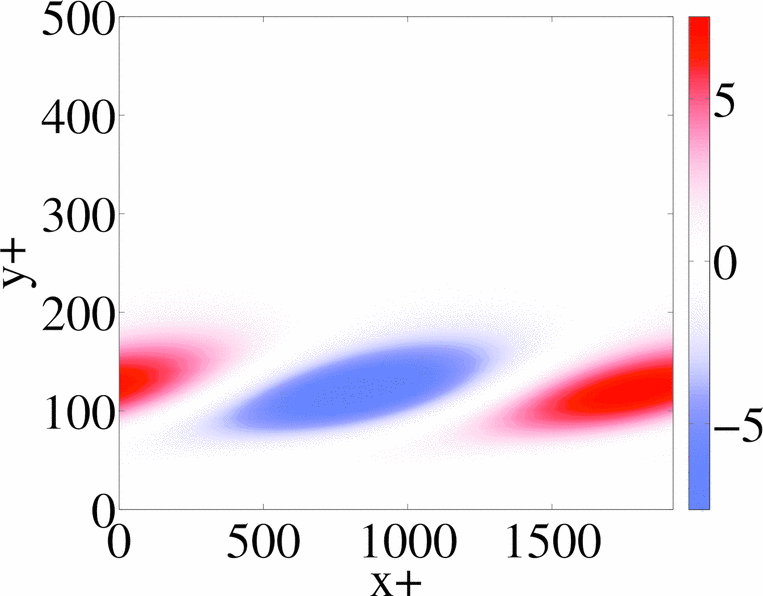}
\includegraphics[width=0.3\textwidth]{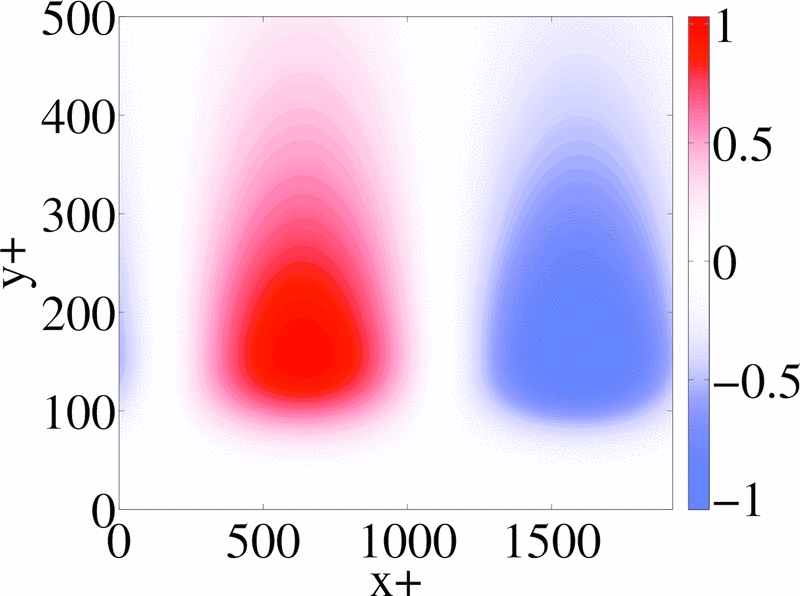}
\includegraphics[width=0.3\textwidth]{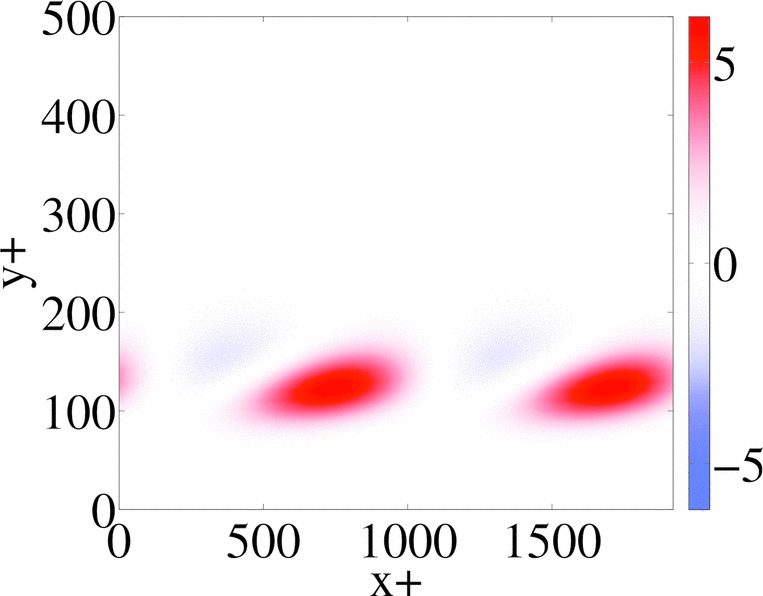}
\caption{Intensity maps of the velocity distributions associated with $K_A$ at $R^+=1800$, showing (a) $u$, (b) $v$ and (c) $-uv$. The azimuthal component (at this azimuthal location) is negligibly small.}
\label{fig:1stmodeshapes}
\end{figure}
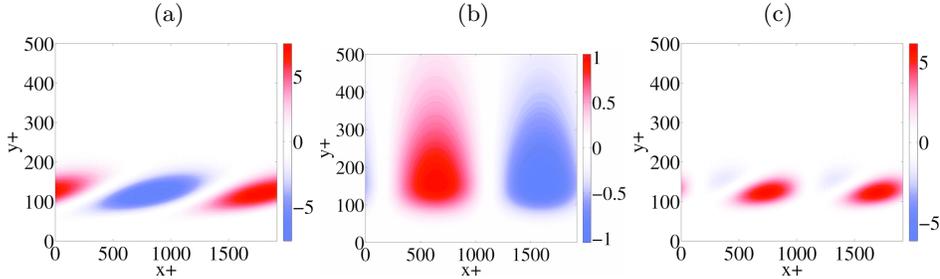

The streamwise and wall-normal velocity components, $u(y)$ and $v(y)$ and the Reynolds stress, $uv(y)$, of the response mode associated with $K_A$ are shown in figure~\ref{fig:1stmodeshapes}. Note the distinct features of the various velocity components, namely the inclination of isocontours of the wall parallel components in the downstream direction and the corresponding lack of phase variation in the wall normal direction in the wall normal velocity (and thus, lack of inclination). The aspect ratio of the response mode is approximately proportional to the ratio of the wall-normal distance to the peak intensity to half the wavelength, and this is also reflected in the inclination angle of the wall-parallel components to the wall. Thus different combinations of wavenumbers will necessarily lead to different inclination angles of coherent regions in the velocity field. We return to this point later in this section.

\begin{figure}
\centering
\subfigure[]{\includegraphics[width = .9\textwidth]{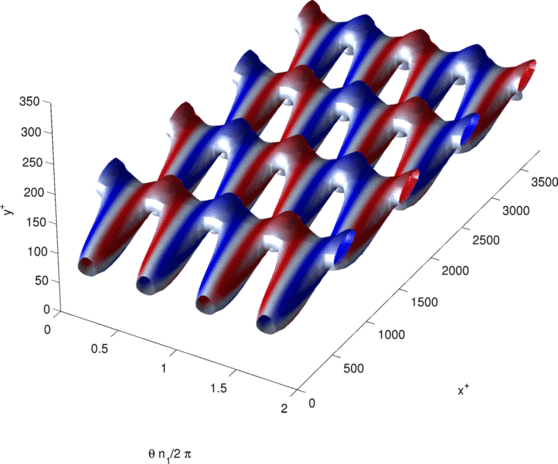}}\\
\subfigure[]{\includegraphics[width = .9\textwidth]{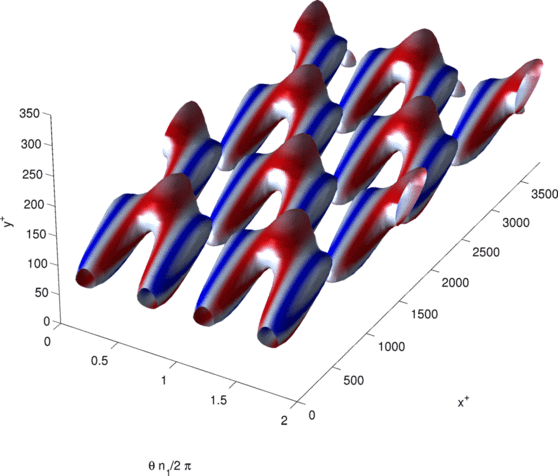}}\\
\caption{(a-b) Isosurfaces of constant swirling strength at $50\%$ of the absolute maximum over the volume for $K_A$ (three wavelengths are shown in the streamwise and two in the spanwise directions), coloured with the azimuthal vorticity. Red and blue denote rotation in and counter to the sense of the classical hairpin vortex respectively. (a) Equal numbers of prograde and retrograde vortices when just the velocity response mode is considered. (b) With the mean velocity profile added, i.e. effectively under a Galilean transformation with constant (zero) convection velocity subtracted throughout the field of view, the retrograde vortices are suppressed and the prograde ones are strengthened.  The ratio of centerline mean velocity to $|A_1|$ is $70:1$.
}
\label{fig:hairpin isosurfaces}
\end{figure}

Figure~\ref{fig:hairpin isosurfaces}(a) shows the distribution of isocontours of swirl associated with the $K_A$ response mode, coloured by the sense of the azimuthal vorticity. The periodic distribution of the velocity components give rise to a periodic array of coherent, downstream-inclined regions of swirl with alternating sense of azimuthal rotation in the head. There is obvious similarity between the geometry of these structures and the hairpin vortices observed in real flows via conditional averaging. Both prograde and retrograde hairpins (vortices with azimuthal rotation in and counter to the sense of the mean shear) respectively, have been identified in the literature, e.g. \cite{Wu06}, and similar phenomena have been reported by e.g. \cite{Falco77}, \cite{Falco91}, \cite{Carlier05}. The equal distribution of both types here is a natural consequence of not including the turbulent mean velocity in the calculation, since the vorticity field associated with each mode must necessarily integrate to zero over a wavelength.

A more recognizable distribution of vortices is obtained by superposing the mean velocity profile and the velocity response mode, as shown in figure~\ref{fig:hairpin isosurfaces}(b).  For the relative amplitudes of the response mode and mean profile and the swirl threshold selected for this figure, there is no observed signature of the retrograde vortices.  While the formulation permits linear superposition of the velocity modes, the nonlinear nature of the swirl operator is reflected in the suppression of retrograde vortices and strengthening of prograde ones. Note that the apparent strength of this effect can be altered by the choice of different amplitude and swirl threshold combinations. \cite{ArXiv10} and \cite{Dennisvort11} have pointed out also that the operations of temporal averaging and calculating swirl are not commutative; $\mathrm{swirl}(\mathrm{ave}(\u(t))) \neq \mathrm{ave}(\mathrm{swirl}(\u(t)))$.

Consider the azimuthal component of two-dimensional swirl. In Cartesian coordinates (to simplify the discussion) this is given by
\begin{equation}
\lambda_{az} = \frac{1}{2}\mathrm{Im} \sqrt{\left(\frac{\partial u}{\partial x} + \frac{\partial v}{\partial y}\right)^2 - 4\left(\frac{\partial u}{\partial x}\frac{\partial v}{\partial y}-\frac{\partial u}{\partial y} \frac{\partial v}{\partial x}\right)}.
\label{eqn:2Dswirl}
\end{equation}
This can be used to identify hairpin vortex heads in the streamwise wall-normal plane.  It also clearly identifies the nonlinear dependence of the diagnostic on the velocity and local velocity gradients (see also \citet{Chernyshenko06}). Here we examine the importance of this effect in the presence of a complex velocity field.
Consider the swirl associated with hairpin heads as given by equation~\ref{eqn:2Dswirl}. For a sufficiently large contribution from the mean profile to the $\D{u}{y}$ term appearing in the second bracket on the right-hand side, the term under the square root becomes less negative, ultimately yielding only a real component and therefore no retrograde swirl when \(\D{v}{x} > 0\) (since $\D{U}{y}>0$ everywhere). The opposite effect is true when $\D{v}{x}<0$, and the apparent strength of the prograde structures is increased. Since $\D{v}{x}<0$ through a prograde hairpin head, the swirl field under the Galilean transformation (effectively a subtraction of a constant, zero velocity from an equivalent real field with mean shear) in figure~\ref{fig:hairpin isosurfaces}(b) is contaminated by the effect of the mean shear. Conversely, the Reynolds decomposition of figure~\ref{fig:hairpin isosurfaces}(a) (the local mean velocity is absent) shows an even distribution of pro- and retro-grade vortices. We note that \cite{Adrianinstfields00} have discussed identifying structure using both a Reynolds decomposition (subtraction of the local mean velocity from an equivalent real field) and a Galilean transformation (subtraction of a constant, zero velocity from an equivalent real field).

This effect is further-reaching, however, since the (linear) superposition of velocity response modes will lead to local velocity gradients that will alter the swirl function in the same way. We will exploit this effect in the reconstruction of assemblies of hairpin vortices into packets.

In \cite{McKeon2010}, we showed that the velocity response modes have characteristics consistent with real turbulent flows. The simple development above demonstrates that a single attached velocity response mode leads to structural features highly reminiscent of hairpin vortices, while the superposition of the mean velocity profile (or use of a Galilean transformation) effectively reduces the observations of retrograde vortices. These effects are robust with respect to the selection of relative amplitudes and plotting thresholds.

The inclination to the wall of the vortices associated with $\k_1$ is around $10^\circ$, considerably less than the average values of approximately $25-45^\circ$ observed in real flows, however we note that the aspect ratio of the structure is a simple consequence of the $\k$ combination selected; therefore we proceed with $\k_1$ as the base response mode and build up the complexity of the wavenumber packets to observe the structural features associated with the superposition of multiple response modes.

\subsection{Formation of idealised hairpin vortex packets}

We now consider structure arising from the wavenumber-frequency combination identified as $K_B$ in table~\ref{tab:modes}.  It will be shown that these modes, consisting of matched azimuthal wavenumbers, azimuthal phase and streamwise convection velocities generate ``ideal'', non-dispersive, azimuthally-aligned hairpin packets.

\begin{figure}
	\centering
	\includegraphics[width=1\textwidth]{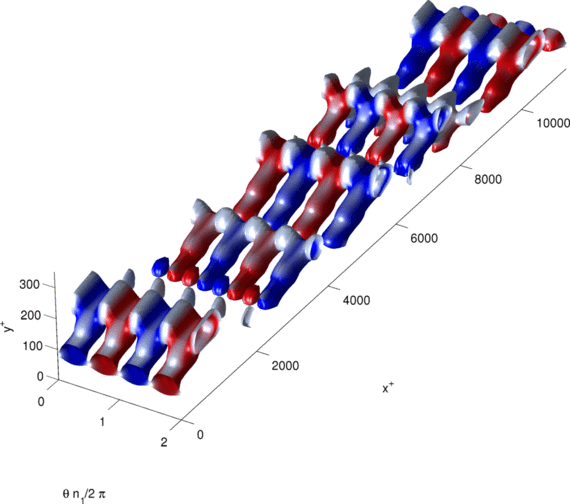}
    {\includegraphics[width=1\textwidth]{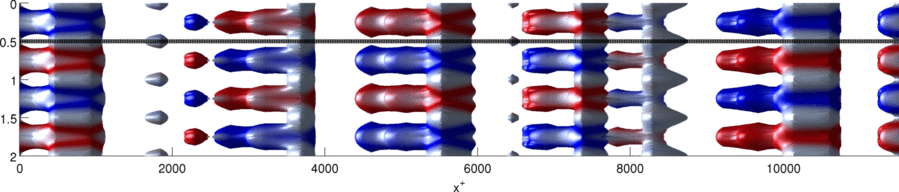}}
    {\includegraphics[width=1\textwidth]{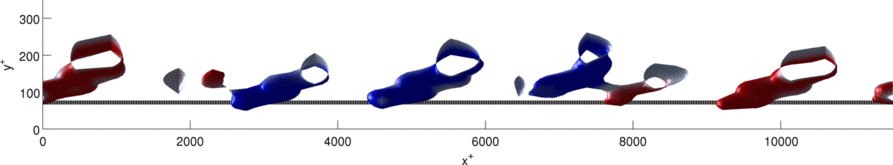}}
	\caption{Isosurfaces of constant swirling strength ($33\%$ of maximum over the volume) for the mode combination $K_B$, coloured by wall-normal vorticity.
		The lower two views show where the cuts for figures \ref{fig:uv_outer} and \ref{fig:uw_outer} are taken: the cut at constant $\theta$ for figure \ref{fig:uv_outer} and at constant $y^+$ for figure \ref{fig:uw_outer}.
   }
   \label{fig:swirl_outer}
\end{figure}

\begin{figure}
	\centering
    {\includegraphics[width=1\textwidth]{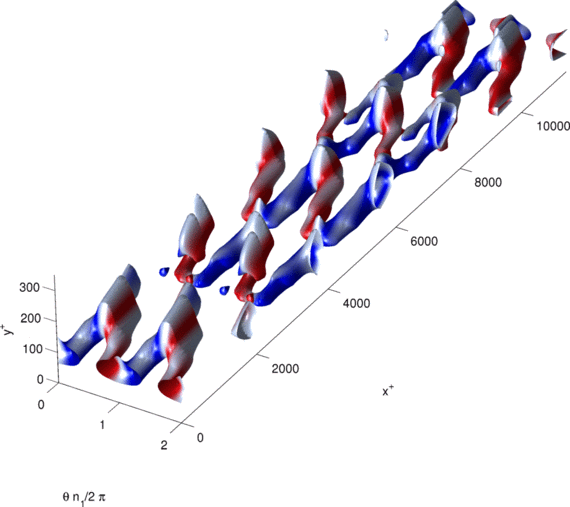}}
    \caption{Isosurfaces of constant swirling strength ($50\%$ of maximum over the volume) for the mode combination $K_B$ in the presence of mean shear, coloured by wall-normal vorticity. The ratio of the centerline mean velocity to $|A_1|$ is 1000:1, to $|A_2|$ is 1000:4.5. All hairpin heads identified using this threshold have prograde rotation.
    }
	\label{fig:swirlshear_outer}
\end{figure}

Isocontours of swirl for the two-mode combination $K_B$ are shown in figure~\ref{fig:swirl_outer}. The superposition of modes that in isolation generate an array of hairpin vortices with period determined by their $\k$ leads to a more complex pattern of coherent vorticity which is reminiscent of a hairpin packet inclined in the downstream direction. In the absence of mean shear, packets of both prograde and retrograde vortices are observed, with a lengthscale determined by the longer value of streamwise wavenumber in the combination. As expected, the effect of adding mean shear (and with it mean vorticity; figure~\ref{fig:swirlshear_outer}) is to suppress the retrograde vortices, almost eliminating the retrograde packets for this ratio of mode amplitude to mean shear. Due to the ratio of streamwise wavenumbers (6:1), these packets appear to contain six vortices, with the last vortex in a packet occurring close to the streamwise location of the first vortex in the following packet with the same sense of swirl.


\begin{figure}
\begin{minipage}[b]{\textwidth}
	\centering
	\includegraphics[angle=0,width=\textwidth]{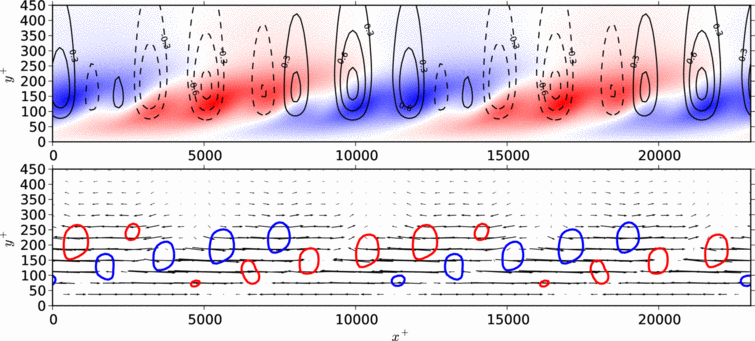}
	\ablabels{4.5}{2.1}
	\caption{(a) Intensity maps of streamwise velocity for the mode combination $K_B$ overlaid with isocontours of wall-normal velocity in the streamwise/wall-normal plane centered at the hairpin heads, $n_2\theta/2\pi = 0.5$. (b) Corresponding two-dimensional vector plot with isocontours of swirl at $50\%$ of the maximum over the slice, coloured by the sense of azimuthal vorticity. We identify a packet as, for example, the succession of prograde vortical motions (red) in rising formation from $x^+\simeq 0$ to $x^+ \simeq 10000$. The packets of pro- and retro-grade hairpins form due to the local shear provided by the long streamwise streaks of $K_2$. These act to organise the vorticity and modulate the shorter modes.}
		\label{fig:uv_outer}
\end{minipage}
\begin{minipage}[b]{\textwidth}
	\centering
	\includegraphics[angle=0,width=\textwidth]{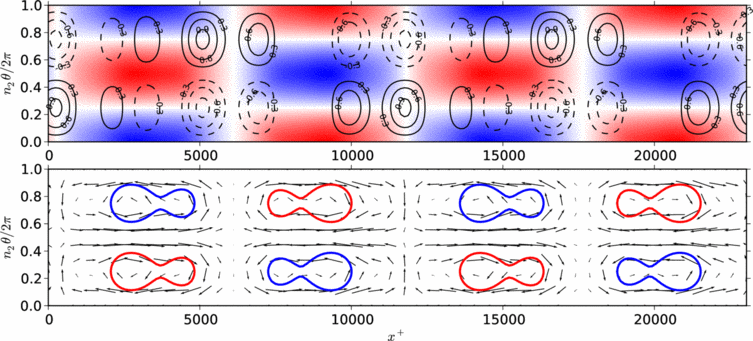}
	\ablabels{4.5}{2.1}
	\caption{(a) Intensity map of streamwise velocity for the mode combination $K_B$ overlaid with isocontours of azimuthal velocity in the streamwise/spanwise plane at $y^+=45$. (b) Corresponding two-dimensional vector plot with isocontours of swirl at $65\%$ of the maximum over the slice, coloured by the sense of wall-normal vorticity. Note that the view is effectively looking from `underneath' figure \ref{fig:swirl_outer}. }
		\label{fig:uw_outer}
\end{minipage}
\end{figure}

Figures~\ref{fig:uv_outer} and \ref{fig:uw_outer} show the velocity and swirl distributions in cuts on the plane of symmetry of the hairpin heads (at one quarter of the spanwise wavelength) and at $y^+ = 45$ (through the trailing legs of the hairpins), respectively, in the absence of the mean shear. The distributions of in-plane velocity shown in the top panels of figures~\ref{fig:uv_outer} and \ref{fig:uw_outer} for the streamwise/wall-normal and wall-parallel planes show that while there is some footprint of the shorter mode on the larger mode response in the streamwise component (filled contours), it is the dominant contributor to the wall-normal and azimuthal velocities (line contours) for this combination of response mode amplitudes.  The maximum swirl associated with the isolated shorter mode far exceeds that of the longer mode because the velocity gradients are larger, even though the peak velocity amplitudes are much smaller. These composite velocity lobes lead to swirling motion in both planes, consistent with the single mode results of figures~\ref{fig:hairpin isosurfaces}, but now clustered into ``packets'' of vortices with like-signed azimuthal vorticity (figure~\ref{fig:uv_outer}(b)).
The main effect of the superposition of the large and small streamwise wavelength modes is the cause an apparent variation of the wall-normal location of the heads of hairpin vortices associated with the lengthscales of the shorter mode because of the large-scale velocity gradients associated with the longer one. The average diameter of the cores of the hairpin heads appears to be approximately 50 viscous units, although this has some sensitivity to the swirl threshold selected.

\begin{figure}
    \centering
    \subfigure[]{
        \begin{minipage}[c]{.58\textwidth}
            \includegraphics[width=\textwidth]{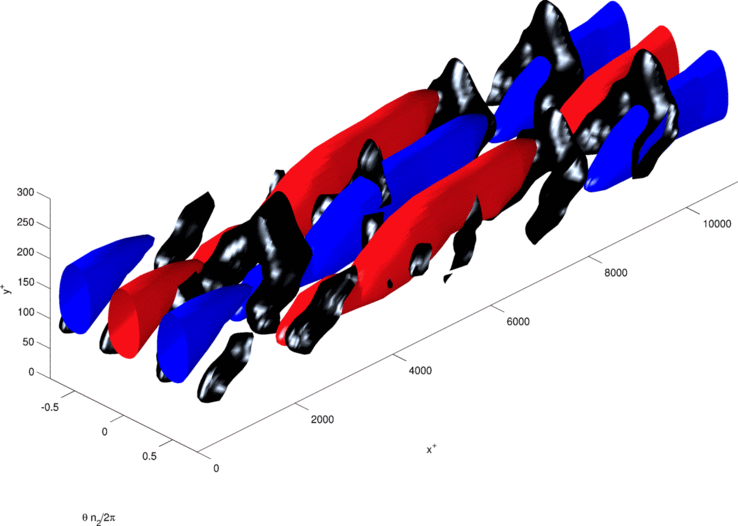}
        \end{minipage}
        \begin{minipage}[c]{0.4\textwidth}
            \includegraphics[width=\textwidth]{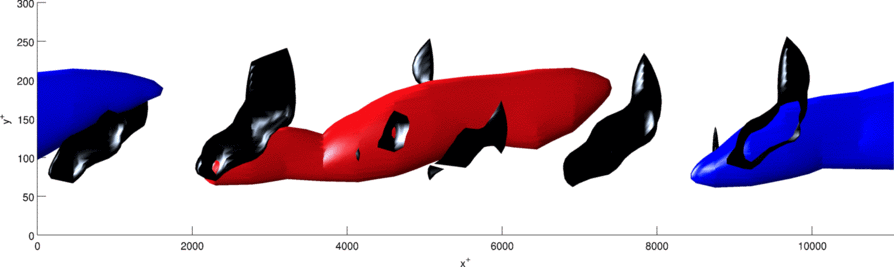}\\
            \includegraphics[width=\textwidth]{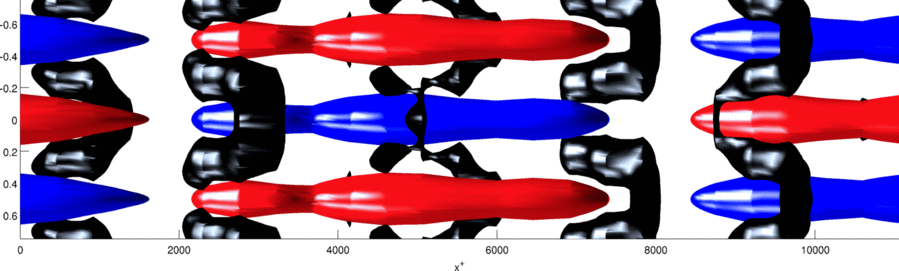}
        \end{minipage}
    }
    \subfigure[]{\includegraphics[width=0.9\textwidth]{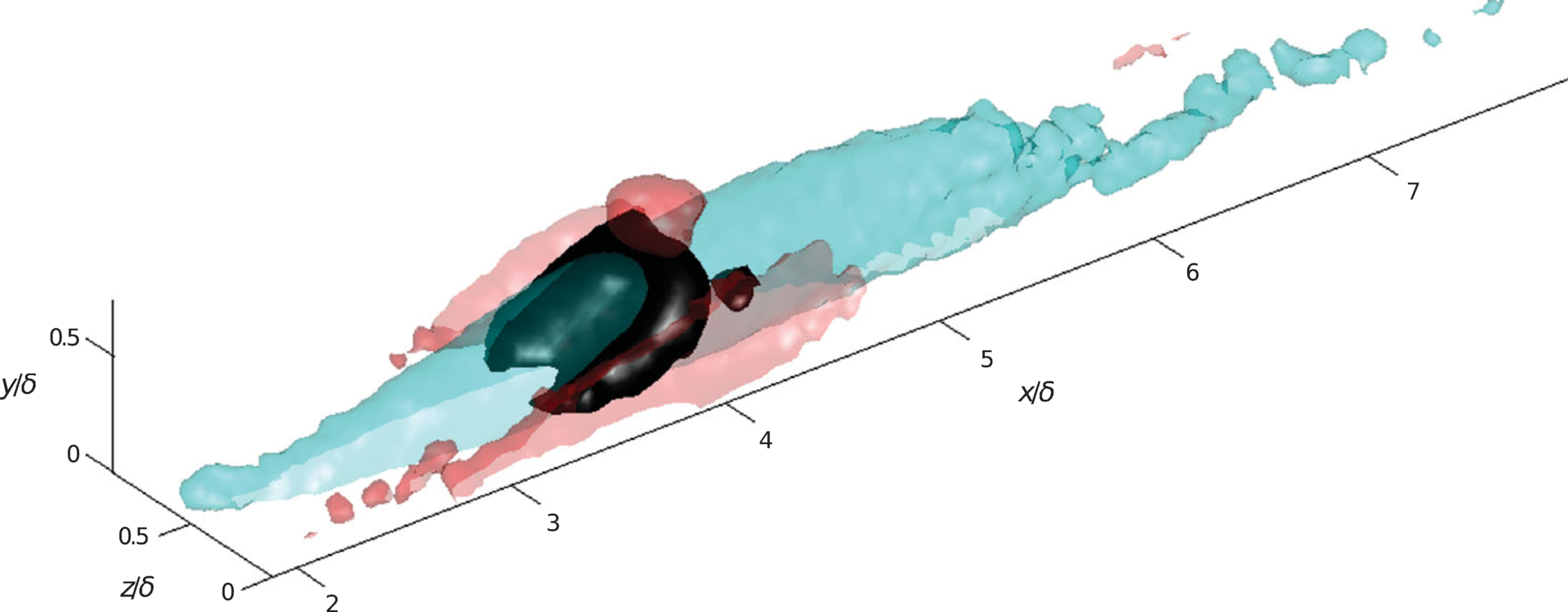}}
    \caption{The relative position of swirl (isosurfaces in black) and isosurfaces of streamwise velocity fluctuation (red and blue for high and low momentum, respectively): (a) model prediction of the swirl field (50\% of maximum) and streamwise velocity ($\pm$50\% of maximum) arising from the mode combination $K_B$, with mean shear of relative amplitude 500 and (b) reproduction of figure 18 from \citet{Dennisvort11} showing structure conditionally-averaged on occurrence of a spanwise swirl at $y/\delta=0.42$. While exact scale comparisons are not meaningful, the arrangement and mechanism appears to be similar.}
    \label{fig:hairpins}
\end{figure}

The lower panel of figure~\ref{fig:uw_outer} gives an indication of the location of the prograde swirling motions relative to the low momentum regions of the VLSM mode; figure~\ref{fig:hairpins}(a) shows a more visual representation. The low speed momentum region appears to exert an organizing effect on the structures, such that they straddle it, tying the hairpin packet to occur at the same streamwise location as the low speed region. The results from this two-mode model closely resemble the conditional structure identified by \cite{Dennisvort11} and reproduced here in figure~\ref{fig:hairpins}(b), although note that the condition in the latter data was set further from the wall than where we observe spanwise swirl for $K_B$. Recall that, in the absence of mean shear, alternating pro- and retro-grade vortices are associated with this $\k$ combination, so that the initial relative phase between the large and small streamwise scales is not the cause of this collocation in space. Rather, the large wall-normal extent of the streamwise velocity variation associated with the large scale response mode, and in particular the shear, $\D{u}{y}$, preferentially biases the swirling motion to be prograde and occur along the location of maximum shear corresponding to the large scale low momentum region. This observation explains the relative locations of experimental observations of structure and large scale low momentum regions, e.g. \cite{Meinhart95, Adrian00, Bharat03}, however it essentially reverses the causality proposed by \cite{HB81}, who comment on shear layers consisting of a forest of inclined hairpins. Note also that figures~\ref{fig:uv_outer}, \ref{fig:uw_outer} and \ref{fig:hairpins}(a) hint at spanwise scale growth with wall distance, which is at least approximately linear in the overlap region of the mean velocity, detailed by attached eddy theory, experimental data on the spanwise scale associated with conditionally-averaged eddies \citep{Tomkins05} and the spanwise velocity spectra of the streamwise velocity spectra of the streamwise velocity from the direct numerical simulation (DNS) of \cite{Schlatter09long}. Visual comparison between the wall-normal distance of a hairpin head in figure \ref{fig:uv_outer} and its footprint in the wall-parallel plane (figure \ref{fig:uw_outer}) suggests that the taller a hairpin, the larger the separation of its legs and therefore the larger its spanwise scale.

Of course, we have considered here ``ideal'' combinations of $\k$, in the sense that the spanwise wavenumbers and phases, and the streamwise convection velocities of the modes are matched, leading to azimuthal symmetry of the vortical structures and packets of structures that do not evolve in time.  These constraints are surely the exception rather than the rule in real flows.  However, examination of these idealised structures has permitted a demonstration of the swirling motion associated with individual near-wall response modes from the model and shed light on their organisation due to the interaction of response modes.  By including a wider range of $\k$s, it is straightforward to obtain more complex swirl fields, which, not surprisingly, are harder to interpret.

\subsection{Structure self-organisation, skewness and amplitude modulation of small scale fluctuations by the large scales}
\label{sec:beats}

In the preceding sections, it was demonstrated that a superposition of response modes could lead to clustering of vortices, while addition of the shear associated with the mean velocity profile weakened or suppressed retrograde vortices, leading to recognisable packets of prograde vortices. Here we demonstrate that the latter phenomenon can arise for subsets of modes with appropriate phase relationships in the absence of mean shear, effectively giving a mechanism by which structure can self-organise, and that in such a scenario the mode combination also reproduces other statistical results associated with wall turbulence, including the relationship between skewness of the streamwise velocity fluctuations and the apparent amplitude modulation of the small scale turbulent activity.

Consider the ``modulating packet'' of case $K_C$ in table~\ref{tab:modes}.  This consists, essentially, of the $K_B$ modes with the addition of a further small scale wavenumber combination $\k_3=(7,\pm 12,2/3)$, chosen such that beat frequency between the smaller scale modes occurs at the same $\k$ as the appropriate VLSM mode, $\k_2=(1,\pm6,2/3)$. Equivalently, this is a consistent set of modes in the sense of nonlinear triadic interaction. All modes travel at the same convective velocity and we have chosen a priori a phase lag of $\pi/2$ for the VLSM relative to the envelope of the small scale activity at the critical layer, based on observations of real flows \citep{Chung09,Hutchins11}.

The spatial interaction of the two smaller modes can be identified by considering the general case of two mode pairs with wavenumber vectors $\k_1$ and $\k_3$, ${\mathbf{u}}_{\k_1}\e^{i\k_1\cdot\x}$ and ${\mathbf{u}}_{\k_3}\e^{i\k_3\cdot\x},$ differing by a wavenumber $\dlt$, that is, $\k_1=\k-\dlt/2$ and $\k_3=\k+\dlt/2$. Let the streamwise velocity component, $u$ of the two waves have radially-dependent real amplitudes $A_1(y)$ and $A_3(y)$ and phases $\phase_1(y)$ and $\phase_3(y)$, i.e.
\[u(y)=u_{\k_1}(y)+u_{\k_3}(y)\]
with
\[u_{\k_1}(y) = A_1(y) \left(\e^{i[\k_1\cdot\x+\phase_1(y)]} + \e^{-i[\k_1\cdot\x+\phase_1(y)]}\right)\]
\[u_{\k_3}(y) = A_3(y) \left(\e^{i[\k_3\cdot\x+\phase_3(y)]} + \e^{-i[\k_3\cdot\x+\phase_3(y)]}\right)\]
It is straightforward to verify that
\begin{align}
    u(y)=& \frac{A_3(y)-A_1(y)}{A_3(y)}u_{\k_3}+\nonumber \\
&A_1(y)\left(\e^{i [\k\cdot\x+\phase_1(y)+\Delta \phase]}+\e^{-i[\k\cdot\x+\phase_1(y)+\Delta \phase]}\right)
        \left(\e^{i[\frac{\dlt}{2}\cdot\x+\Delta \phase]}+\e^{-i[\frac{\dlt}{2}\cdot\x+\Delta \phase]}\right)
		\label{eqn:beating}
\end{align}
with $\Delta \phase=(\psi_3-\psi_1)/2$. The second term on the right-hand side identifies a traditional amplitude modulation associated with the beating of $\k_1$ and $\k_3$, namely a signal with streamwise wavenumber $k = 0.5$ and phase $\Delta \phase$. For the simplest case with $A_1(y)=A_3(y)$, this is the only component of $u$. The traditional measure of amplitude modulation takes large values twice per period of the modulating signal; in terms of the large scale amplitude modulation terminology currently favored in the turbulence literature, this is equivalent to a signal with $k=k_2=1$ and phase $\Delta \phase$ that takes large values once per period.

The wall-normal variation of the amplitudes and phases for the three components of $K_C$ are shown in figure~\ref{fig:ampphase_Ks}. For most of the radius, $A_{\k_1}(y) \approx A_{\k_3}(y)$ and, for a wide range of wall-normal distance around the critical layer, the two shorter modes have a constant phase difference. Note there is a difficulty in resolving the phase far from the wall where the amplitude is negligible. In addition, the VLSM mode $\k_2$ has a difference of phase of $3\pi/2$ at the critical layer, giving the aforementioned phase lag of $\pi/2$ for the VLSM mode with respect to the beating envelope at the critical layer.

\begin{figure}
    \centering
    \subfigure[]{\includegraphics[width=.45\columnwidth,angle=0]{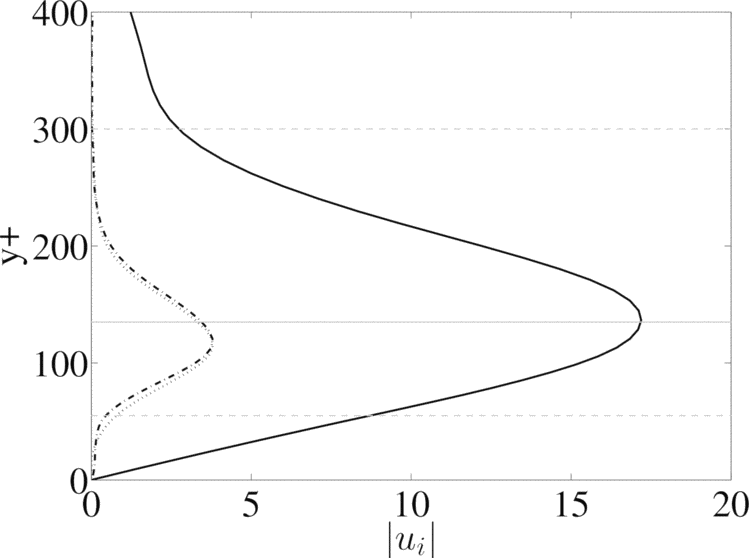}}
    \subfigure[]{\includegraphics[width=.45\columnwidth,angle=0]{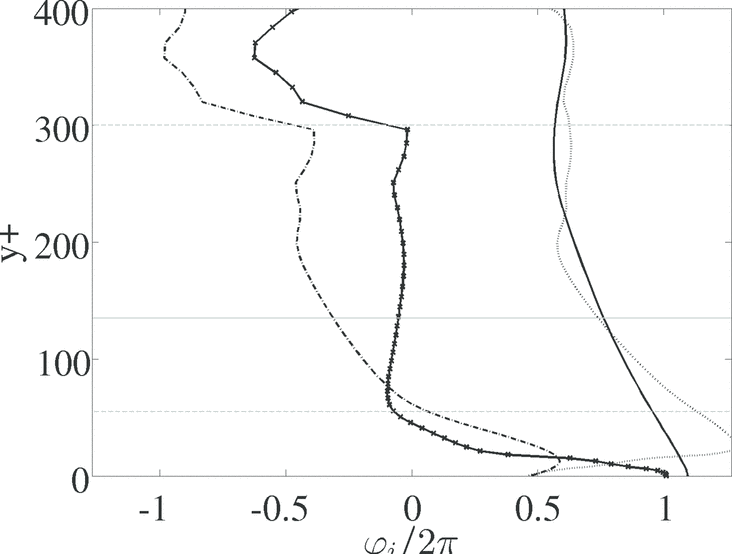}}
\caption{Wall-normal variation of (a) amplitude of the streamwise velocity component and (b) phase in multiples of $2\pi$ for the three modes making up $K_C$. The lines denote $-\cdot-$: $\k_1~(k=6)$, --- : $\k_2~(k=1)$, $\cdot \cdot \cdot$: $\k_3~(k=7)$, $-\times-$: $\phase_1-\phase_3$. The solid horizontal lines show the location of the critical layer and the dashed horizontal lines demarcate the region where $\phase_1 \approx \phase_3$.}
\label{fig:ampphase}
\end{figure}

\begin{figure}
\begin{minipage}[b]{\textwidth}
	\centering
	\includegraphics[angle=0,width=\textwidth]{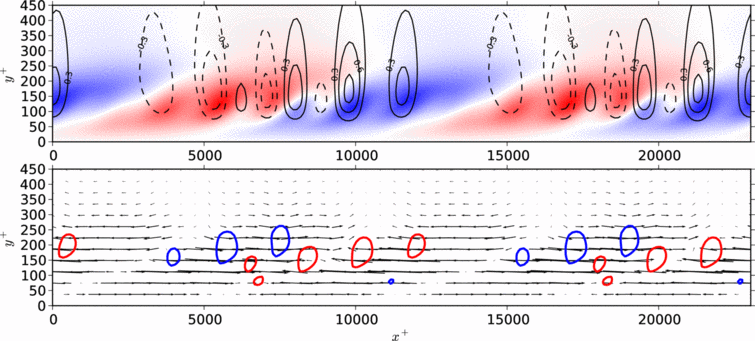}
	\ablabels{4.5}{2.1}
	\caption{(a) Intensity map of streamwise velocity for the mode combination $K_C$ overlaid with isocontours of wall-normal velocity in the streamwise/wall-normal plane centered at the dominant hairpin heads, $n_2\theta/2\pi = 0.5$. (b) Corresponding two-dimensional vector plot with isocontours of swirl at $50\%$ of the maximum over the slice, coloured by the sense of azimuthal vorticity. The packet of prograde hairpins is seen as the rising array of prograde vortices (red) from $x^+\simeq 0$ to $x^+\simeq 10000$. Note the weakened retrograde vortex motion at the leading and trailing ends of the packet. }
		\label{fig:uv_outer2}
\end{minipage}
\begin{minipage}[b]{\textwidth}
	\centering
	\includegraphics[angle=0,width=\textwidth]{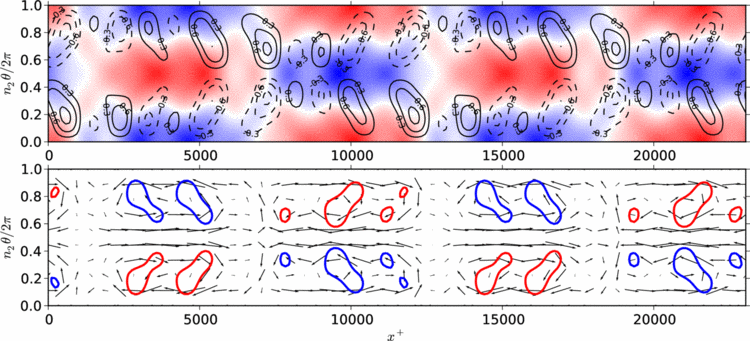}
	\ablabels{4.5}{2.1}
\caption{(a) Intensity map of streamwise velocity for the mode combination $K_C$ overlaid with isocontours of azimuthal velocity in the streamwise/spanwise plane at $y^+=72$. (b) Corresponding two-dimensional vector plot with isocontours of swirl at $65\%$ of the maximum over the slice, coloured by the sense of wall-normal vorticity. Comparison with the preceding figure shows that the narrower structures are also shorter in wall-normal extent.}
	\label{fig:uw_outer2}
\end{minipage}
\end{figure}

The structural organisation associated with $K_C$ is shown in figures~\ref{fig:uv_outer2} and \ref{fig:uw_outer2}, respectively. The location of the planes taken through the three-dimensional field is shown in figure \ref{fig:swirl_outer}. The upper panels show that the beating identified for the streamwise velocity in equation~\ref{eqn:beating} affects all velocity components of the small scale modes.  More interestingly, the lower panels show that the same threshold used for wavenumber combination $K_B$ in figures~\ref{fig:uv_outer} and \ref{fig:uw_outer} now identifies a complex swirl distribution centered on a packet with prograde swirl that appears to grow in the downstream direction. This can be understood by the action of the velocity gradients (preferentially $\D{u}{y}$ and $\D{u}{z}$) associated with the VLSM mode: the large amplitude beating of the two (locally) shorter modes is spatially collocated with large positive $\D{u}{y}$ from the VLSM, which has the same action on the swirl as the mean shear, as discussed in section~\ref{section:structure} above. Regions of retrograde swirl are observed upstream of and closer to the wall than the leading edge of the prograde packet, and upstream and further from the wall than the trailing edge. We note that this distribution is at least qualitatively similar to the preferred orientations of prograde vortices with respect to retrograde ones identified by \cite{Natrajan07} using cross-correlation techniques.

\begin{figure}
	\centering
    \begin{minipage}[c]{.7\textwidth}
    	\includegraphics[width=\textwidth]{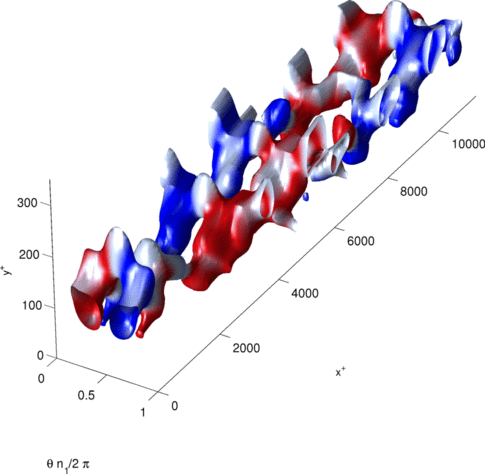}
    \end{minipage}
    \begin{minipage}[c]{\textwidth}
    	\includegraphics[width=\textwidth]{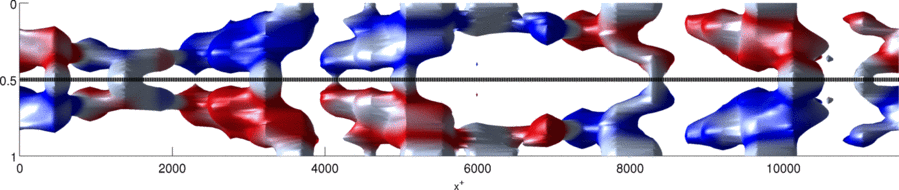}
    	\includegraphics[width=\textwidth]{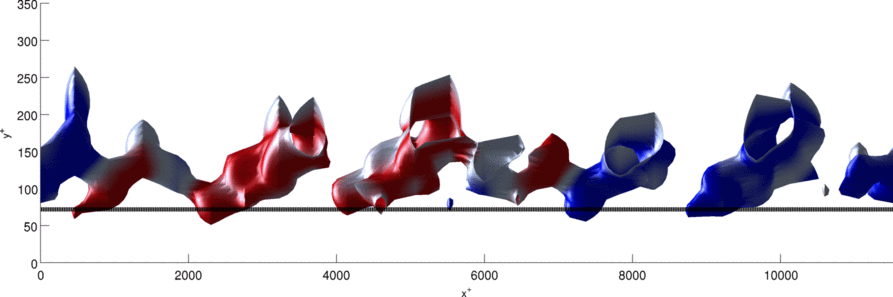}
    \end{minipage}
    \caption{Isosurfaces of constant swirling strength ($25\%$ of maximum over volume) for the modulating packet $K_C$, coloured by wall-normal vorticity.
    The local shear is provided by the $K_2$ mode, which acts to organise the hairpins around regions of low-speed momentum. The lower two views show where the cuts for the previous two figures are taken: the cut at constant $\theta$ for figure \ref{fig:uv_outer2} and at constant $y^+$ for figure \ref{fig:uw_outer2}. }
	\label{fig:swirl_outer2}
\end{figure}

In terms of the three-dimensional swirl field, shown in figure~\ref{fig:swirl_outer2}, it seems that the VLSM exerts an organisational effect that leads to discrete packets of prograde hairpins distributed with streamwise and spanwise period corresponding to the beating between the smaller modes. Note that an increase of mean shear would serve only to strengthen the prograde vortices and suppress the retrograde ones rather than change the organised distribution.

A phase difference of $\pi/2$ between the VLSM and beating envelope at the critical layer maximises the differentiation between prograde and retrograde vortices, however this choice was also informed by results in the recent literature concerning the skewness of the streamwise velocity fluctuation \citep{Mathis09,Mathis11} and amplitude modulation of the small scale fluctuations by the larger scales \citep{Bandy84,Mathis09}.  We now investigate the variation of these two diagnostic functions for $K_C$.

For this combination of three response modes, an expansion of the cube of the superposed velocity fields shows that the skewness is given by
\begin{equation}
    \hat{S}_u(y) = \frac{\ave{u^3}}{\ave{u^2}^{3/2}} =  \frac{2 A_1 A_2 A_3 \cos \left[ \phase_2(y)+\phase_1(y)-\phase_3(y) \right]}{(A_1^2 + A_2^2+A_3^2)^{3/2}},
\end{equation}
where angle brackets $\ave{\cdot}$ denote averaging in the sense of $\x$.
A proxy for the skewness (essentially an unscaled version) is therefore
\begin{equation}
{S} = \cos\left[\phase_2(y)+\phase_1(y)-\phase_3(y)\right].
\label{eqn:skewkc}
\end{equation}
Note that the contributions to the mean skewness from the full velocity field will always be governed by triadic interactions in which the three wavenumbers sum to zero, such that our modulating packet can be considered as one representative of a wide range of triads in the real flow.

Following \cite{Mathis11}, the Pearson correlation coefficient is defined in terms of the large scale signal $u_L$ and the envelope of the small-scale activity, $E_L(u_S)$ (where the large and small scales can be defined using a sharp Fourier filter or Hilbert transform)
\begin{equation}
R^*=\frac{\left<u_L E_L(u_S) \right>}{\left(\left< u_L^2 \right> \left< E_L(u_S)^2\right>\right)^{1/2}}.
\label{eqn:corrcoeff}
\end{equation}

The zero of this correlation coefficient in experiment can be interpreted either as zero net amplitude modulation \citep{Mathis09} or a $\pi/2$ difference in phase between the signals \citep{Bandy84,Chung09} at this wall-normal distance. Recent work by \cite{Jacobi12} has determined that the sense of this phase is to give the small scale envelope the lead in the spatial sense, which is consistent with the conditional averaging results of \cite{Hutchins11}.  The zero crossing has been observed to occur at the wall normal location corresponding to both the VLSM peak in streamwise fluctuation, explored in more detail in \cite{McKeon2010}, and the region of zero skewness of the streamwise velocity fluctuation \citep{Mathis09,Mathis11}. For packet $K_C$, the expression for the correlation coefficient reduces simply to
\begin{equation}
    R=\frac{\left(\ave{A^2_2(y)}\ave{A^2_{env}(y)}\right)^{1/2}\cos\left[\phase_2(y)-\Delta\phase(y)\right]}{\left(\ave{A^2_2(y)}\ave{A^2_{env}(y)}\right)^{1/2}}=\cos\left[\phase_2(y)+\phase_1(y)-\phase_3(y)\right].
\label{eqn:corrcoeffkc}
\end{equation}
Thus the expressions for $R$ and ${S}$ are identical for this three-mode combination. The qualitative agreement with experiment is very encouraging, and expands on the hereto puzzling \citep{Mathis09,SchlatterOrlu10,Mathis11} connection between skewness and amplitude modulation away from the near-wall region.

Recall that the component wavenumbers and frequencies comprising $K_C$ were selected from structural considerations, however they appear to capture the connection between two well-studied statistical measures surprisingly well. Both $R$ and ${S}$ can be considered to be effectively unscaled versions of what would be observed in the real flow containing many more than three modes. While the form of the skewness will always be dictated by wavenumber triads such as $K_C$, a fact that becomes quite apparent when studying modes decomposed into $\k$ space, this agreement suggests that the apparent amplitude modulation effect can be interpreted as a consequence of the phase relationship between the VLSM mode and pairs of wavenumber-frequency combinations that create a part of (consistent) triadic interactions with it. That the full correlation coefficient and skewness, i.e. the statistics formed in the presence of a wide range of modes, are both zero at the VLSM critical layer in the real flow suggests a similar phase relationship (or at least alternatives that sum to zero) for the other modes present in the real flow, with any significant differences confined to the near-wall and core regions. The theoretical determination of the phase relationships between triadic subsets is the subject of ongoing work.

\subsection{Spanwise decorrelation: Clustering of coherent structures and the largest scale motions}
\label{sec:VLSM}

\begin{figure}
\begin{minipage}[b]{\textwidth}
	\centering
	\includegraphics[angle=0,width=\textwidth]{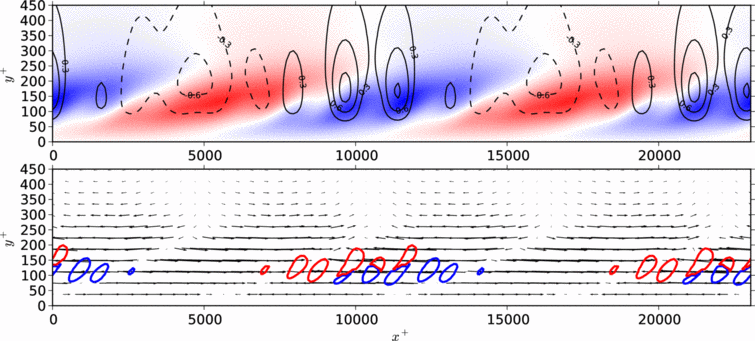}
	\ablabels{4.5}{2.1}
    \caption{(a) Intensity map of streamwise velocity associated with $K_D$ overlaid with isocontours of wall-normal velocity in the streamwise/wall-normal plane at the azimuthal location of maximum azimuthal swirl, i.e. at the hairpin heads. (b) Corresponding two-dimensional vector plot with isocontours of swirl at $50\%$ of the maximum over the slice, coloured by the sense of azimuthal vorticity.
    }\label{fig:uv_outerx}
\end{minipage}
\begin{minipage}[b]{\textwidth}
	\centering
	\includegraphics[angle=0,width=\textwidth]{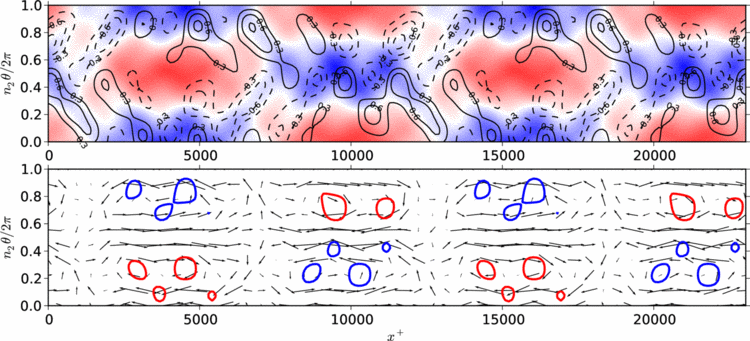}
	\ablabels{4.5}{2.1}
    \caption{(a) Intensity map of streamwise velocity associated with $K_D$ overlaid with isocontours of azimuthal velocity in the streamwise/spanwise plane at $y^+=72$. (b) Corresponding two-dimensional vector plot with isocontours of swirl at $65\%$ of the maximum over the slice, coloured by wall-normal vorticity.}
		\label{fig:uw_outer2x}
\end{minipage}
\end{figure}

The results above demonstrated that an appropriate phase relationship between scales can lead to organisation of coherent structure into ideal hairpin packets associated with the superposition of velocity response modes with matched azimuthal wavenumbers (in the sense that the larger $n$ values are integer multiples of the smallest value and there is no azimuthal staggering of the modes). In this subsection we extend the analysis to investigate structural organisation in the presence of spanwise decorrelation, by removing this matching condition, both for the hairpin vortices and the very long wavelength motions.

\begin{figure}
	\centering
    \begin{minipage}[c]{.5\textwidth}
        \includegraphics[width=\textwidth]{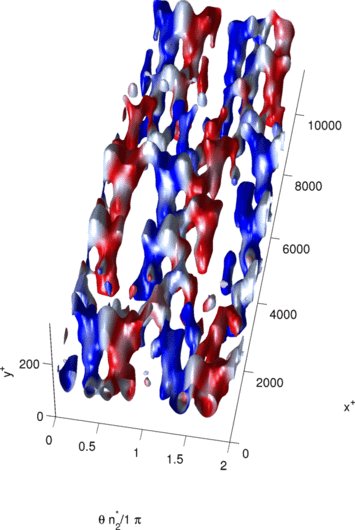}
    \end{minipage}
    \begin{minipage}[c]{\textwidth}
	    \includegraphics[width=\textwidth]{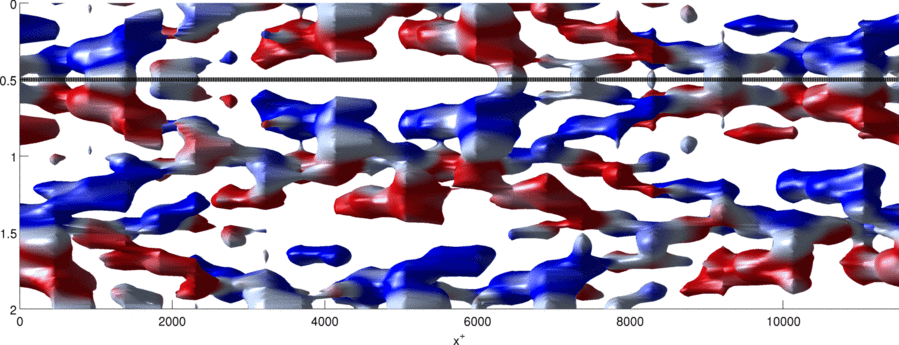}
    	\includegraphics[width=\textwidth]{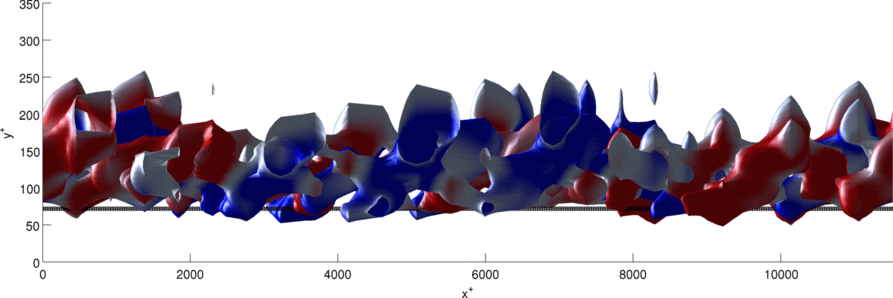}
    \end{minipage}
    \caption{Isosurfaces of constant swirling strength ($25\%$ of maximum over the volume) for the decorrelated modulating packet $K_D$, coloured by wall-normal vorticity. Note the pair of retrograde vortices at $x^+\simeq 4000$ and $x^+\simeq 6000$, just before the main packet begins at $x^+\simeq 7000$.
   Note also that these narrower vortices are also shorter in the wall-normal direction.
	 The lower views show where the cuts for the previous two figures are taken: the cut at constant $\theta$ for figure \ref{fig:uv_outerx} and at constant $y^+$ for figure \ref{fig:uw_outer2x}.    
        }
		\label{fig:swirl_outer2decorr}
\end{figure}

The mode combination $K_D$ was chosen to explore the variation of $n$ on the modulating packet by considering a triadically-consistent mode combination in which the azimuthal phase variation is not matched as in $K_C$. This decorrelation of the resultant velocity field impacts the appearance of coherent vortical structure in the following way, as shown most clearly in the two-dimensional plots of figures~\ref{fig:uv_outerx} and \ref{fig:uw_outer2x}, and the isocontours of swirl shown in figure~\ref{fig:swirl_outer2decorr}. Firstly, the local wall-normal gradients of streamwise velocity associated with $\k_2$ still augment the swirl associated with the heads of prograde hairpin vortices and suppress retrograde ones, in much the same manner as in the ideal, $K_C$ case. Thus the simple thresholding of swirl for the velocity response modes continues to identify vortical structure as predominantly spanning low-momentum regions. In addition, the decorrelation compromises the coherence of hairpin packets; the lack of azimuthal matching of wavenumbers and phases leads to the asymmetric footprints of wall-normal vorticity shown in figure~\ref{fig:uw_outer2x}, indicative of the appearance of more complex structures such as the canes and arches, etc., identified from analysis of DNS by \cite{Robinson91}, but still clustered around the large-scale low momentum regions.
Thus, the spatial variation due to the mix of wavenumbers present in a real flow, a subset of which are investigated here, efficiently explains a number of structural phenomena.

\begin{figure}
    \centering
    \includegraphics[width=14cm]{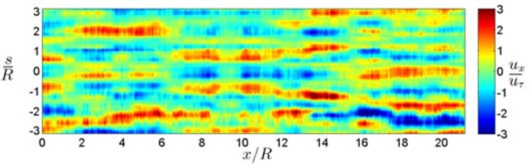}\\
	{\includegraphics[width=12.2cm]{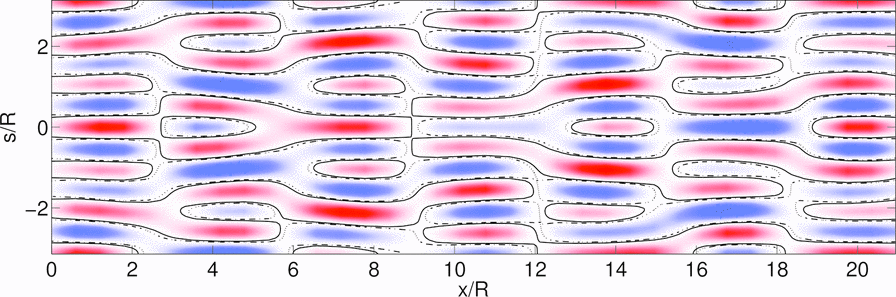}\hspace{1cm}}~\\
	\ablabels{6}{2}
	\caption{(a) Filtered POD data (first 4 modes, \citet{Hellstroem11}) from a $Re\simeq 10^4$ turbulent pipe showing contours of spanwise velocity fluctuations, clearly picking out the VLSM mode as dominant in the flow at $y=0.2R$; (b) contour plot of the spanwise velocity fluctuations for the mode combination $K_E$ predicted at $Re=10^4$ and $y=0.2R$. Note that the difference in $n$ of the modes is responsible for the perceived joining-up of large-scale modes and the spanwise decorrelation effect.}
    \label{fig:Leo}
\end{figure}

The very large scale motions (VLSM), with representative wavenumbers $\k_{\mathrm{VLSM}}=(1,\pm10,2/3)$, were examined at length in \cite{McKeon2010}. That paper made predictions of the structural form of the motions and of the variation with Reynolds number of the wall-normal location of the peak streamwise velocity. The VLSMs were considered to be of special interest because they are (according to the definition in that work) the largest structures that are attached to the wall. As such, they scale with both inner and outer characteristics.

The temporal hot-wire measurements of \cite{Monty07} projected into the spatial domain identified even longer regions of apparent coherence in the wall parallel plane, with meandering in the spanwise sense. Subsequently, \cite{Hellstroem11} performed a snapshot proper orthogonal decomposition (POD) study on temporal velocity field information obtained in a pipe flow. The flow was at Reynolds number $Re=12500$ and the data were obtained using cross-stream PIV, projected into the spatial domain using Taylor's hypothesis. The prevailing structure regenerated from summation of the first four POD modes at $y/R=0.2$ (reproduced here in figure~\ref{fig:Leo}(a)) led to favourable comparisons with the predictions of the VLSM mode shapes in \cite{McKeon2010}, with the caveat that the experimentally-measured observations seemed to be dominated by longer scales than the $k=1$ mode used by \cite{McKeon2010}, even once the meandering associated with the superposition of modes was accounted for.
We would add that at this wall-normal location, the VLSM and the other modes in $K_E$ will be convecting at a different speed ($c=2/3$) than any locally critical modes which presumably would be observed. Thus, care should be taken when interpreting measurements using Taylor's hypothesis for structure spanning a wide range of scales.

The meandering effect described by \citet{Monty07} has been explained in various ways \citep{SharmaMcKeonTSFP, Dennisvort11, Hellstroem11} to arise from a decorrelation of the energetically-dominant VLSMs due to motions at other scales. Figure~\ref{fig:Leo}(b) shows a wall-parallel cut at $y/R=0.2$ of the streamwise velocity field arising from the superposition of the velocity modes comprising packet $K_E$ in table~\ref{tab:modes}. The three modes therein were selected by eye to match the key large scale features of Hellstr\"{o}m \etal's results shown in figure~\ref{fig:Leo}(a). The decorrelation associated with the range of azimuthal wavenumbers, $n$, present in $K_E$ naturally gives rise to the apparent meandering effect and also creates a spatial localisation of coherence in the azimuthal direction because of the azimuthal analogue of the beat effect described in section \ref{sec:beats}.

\subsection{Shear stress variation and comparison with models}
We now consider briefly the contributions to the shear stresses arising from individual response modes and the correspondence between the location of the maxima and observations of structure discussed above. Consideration of the mode shapes in figure~\ref{fig:1stmodeshapes}(a-c) emphasises that the shear stress arising from a particular mode will be strongly dependent on the relative shapes of the streamwise and wall-normal responses. The typical shapes for the modes considered here involve isocontours of streamwise velocity that are heavily inclined in the downstream direction, i.e. have decreasing phase, $\phase(y)$, with increasing $y$, but have little or no phase variation for the wall-normal velocity.

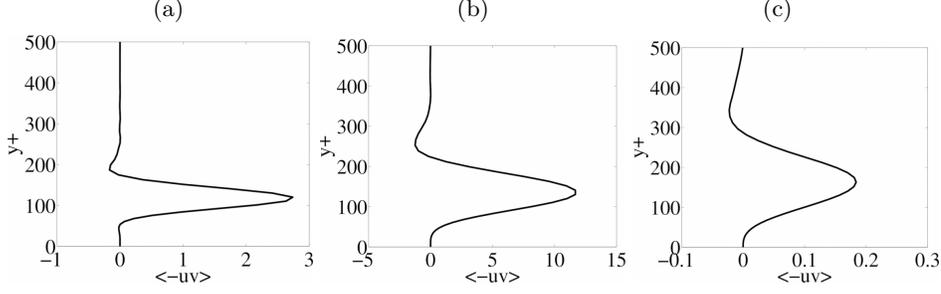
\begin{figure}
\centering
\begin{tikzpicture}[overlay]
	\node [] (a) at (-0.3\textwidth,-0.1) {(a)};
	\node [] (b) at (0\textwidth,-0.1) {(b)};
	\node [] (c) at (0.3\textwidth,-0.1) {(c)};
\end{tikzpicture}\\ \vspace{0.4cm}
\includegraphics[width=0.3\columnwidth]{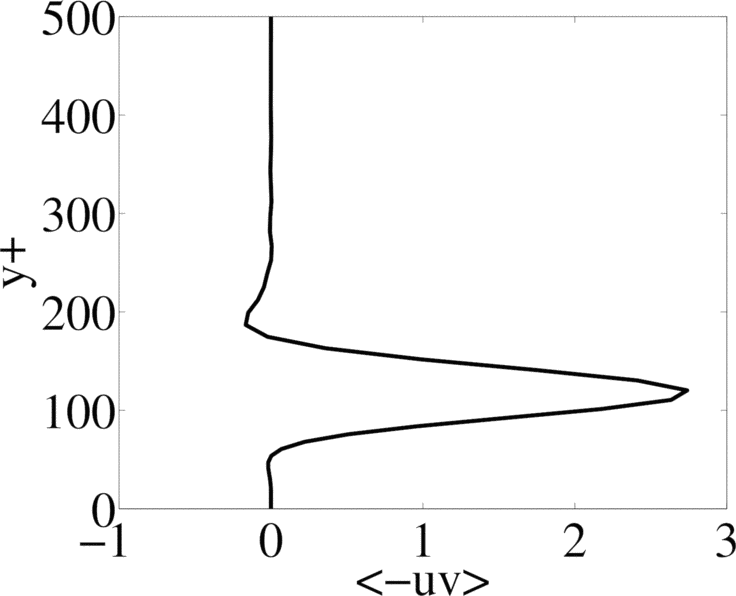}
\includegraphics[width=0.3\columnwidth]{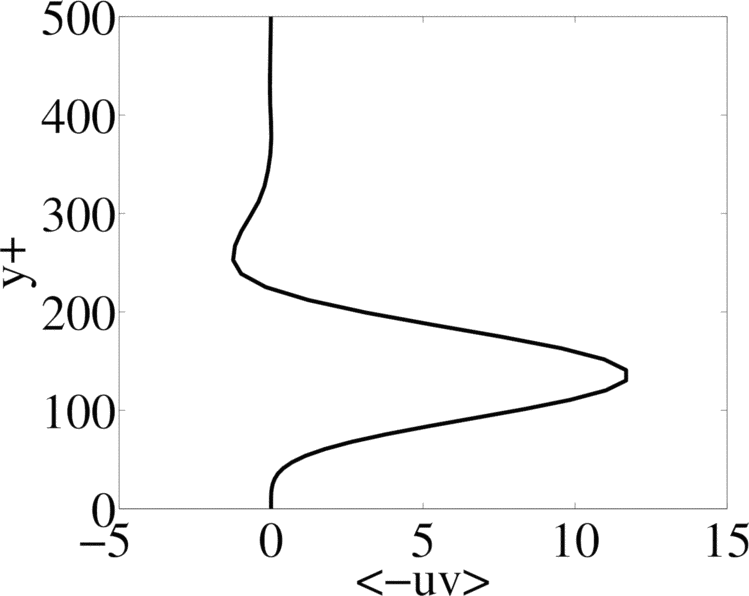}
\includegraphics[width=0.3\columnwidth]{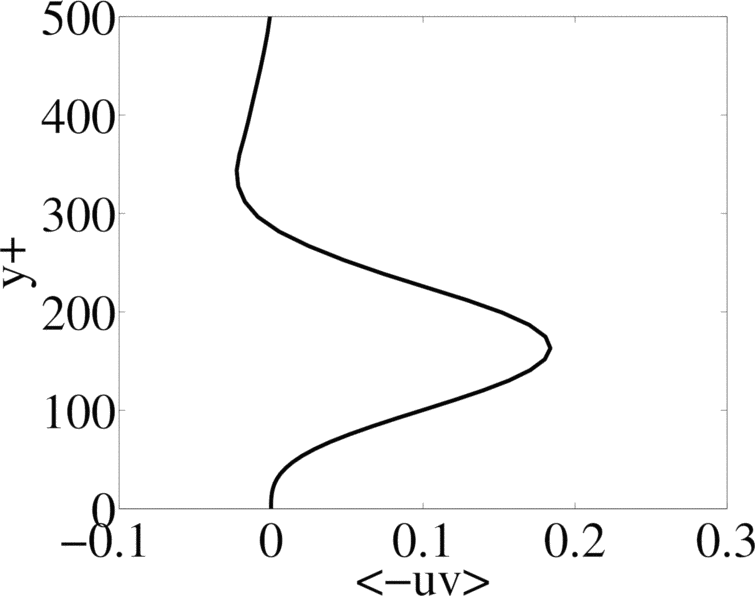}
\caption{Wall-normal profiles of Reynolds stress of the following modes identified in table~\ref{tab:modes}: (a) $\k_1=(6,6,2/3)$, (b) $\k_2=(1,6,2/3)$ and (c) $\k_4=(0.3,3,2/3)$.}
\label{fig:restressprofiles}
\end{figure}

Consider the shear stress from an individual response mode with streamwise and wall-normal velocity fields with magnitudes ${u_\k}(y)$ and ${v_\k}(y)$ and phases $\phase_u(u)$ and $\phase_v(y)$,
\begin{eqnarray}
u(y) = {u}_\k(y) \left(\e^{i[\k\cdot\x+\phase_u(y)]} + \e^{-i[\k\cdot\x+\phase_u(y)]}\right)\\
v(y) = {v}_\k(y) \left(\e^{i[\k\cdot\x+\phase_v(y)]} + \e^{-i[\k\cdot\x+\phase_v(y)]}\right).
\end{eqnarray}
The contribution to the shear stress can be decomposed into mean and fluctuating parts,
\begin{eqnarray*}
uv(y) &=& {u}_\k(y){v}_\k(y) \left (\e^{i[\k\cdot\x+\phase_u(y)]} + \e^{-i[\k\cdot\x+\phase_u(y)]}\right)
\left(\e^{i[\k\cdot\x+\phase_v(y)]} + \e^{-i[\k\cdot\x+\phase_v(y)]}\right)\\
&=&{u}_\k(y){v}_\k(y) \left(\e^{i[\phase_u(y)-\phase_v(y)]}+\e^{-i[\phase_u(y)-\phase_v(y)]}\right)\\
&&+{u}_\k(y){v}_\k(y) \left(\e^{i[2\k\cdot\x+\phase_u(y)+\phase_v(y)]} + \e^{-i[2\k\cdot\x+\phase_u(y)+\phase_v(y)] }\right)
\end{eqnarray*}
so
\begin{equation}
uv(y) = 2{u}_\k(y){v}_\k(y)\left[\cos\left(\phase_u(y)-\phase_v(y)\right)+\cos\left(2\k\cdot\x+\phase_u(y)+\phase_v(y)\right)\right],
\end{equation}
or a contribution to the mean shear stress whose magnitude depends on the relative phases of $u$ and $v$ plus a fluctuating component at $2\k$. Recall that response modes give the velocity \textit{relative to the local mean}. The contribution to the instantaneous shear stress arising from a particular mode will be localised around the wall-normal location(s) at which $u$ and $v$ have a relative phase that is close to an integer multiple of $\pi$ and non-negligible amplitudes. For the modes of this study, these two conditions tend to be met in the interior of the flow, in a relatively narrow region away from the wall. This causes the contribution to the mean shear stress to have the type of profile shown in figure~\ref{fig:restressprofiles}, which shows the wall normal distributions of shear stress associated with the modes comprising packet $K_C$. Note the large magnitude of shear stress associated with the VLSM mode, in agreement with the observations of \cite{Guala06} and \cite{WuBaltzerAdrian12}.

Further inspection reveals that for the majority of response modes, the shear stress peak occurs in the region of the critical layer, where $u$ and $v$ are $\pi$ out of phase such that the peak contribution to the mean shear stress is always negative, i.e. $\ave{-uv}_{peak}>0$, and at the streamwise locations corresponding to the (co-located) minima in the streamwise gradient of $v$ and wall-normal gradient of $u$. 
The gradient of the Reynolds stress is therefore high and positive just below the critical layer and high and negative just above. We note that laminar, inviscid theory predicts the Reynolds stress gradient to be infinite at the critical layer, and zero elsewhere (locally) around it \citep{DrazinReid}.

We conclude the presentation of the results by noting that the connection between the present discussion on Reynolds stress and the structural, statistical and spectral results discussed above and information in real flows is greatly facilitated by the ability to linearly superpose response modes to obtain trends applicable to the full field.  Unfortunately this technique is not amenable to the quadrant and VITA analyses performed by many previous workers, since those approaches are conditioned on events in the physical domain in which all modes participate and periodicity is returned only in the averaged sense.

\section{Discussion}
\label{section:discussion}

The extensive foregoing analysis, together with the results of \cite{McKeon2010}, has demonstrated that simple linear combinations of velocity response modes reproduce key statistical and structural features of wall turbulence. Further, they can give insight into the underlying mechanisms for extraction of energy from the mean flow and for structural dynamics.  We now briefly discuss and connect the key results, outline the mechanisms responsible for this success and make connections to observations and phenomenological models in the literature.

\subsection{Connections to the attached eddy hypothesis, edge states and experimental observations of structure}

The Reynolds shear stress distributions associated with linear superpositions of velocity response modes and the associated observations of structure are entirely consistent with expectations for the behavior of the Reynolds stress in fully turbulent flow. They are also highly reminiscent of the form predicted in the literature by the attached eddy theory of \cite{Townsend1956, Townsend1976} and \cite{Perrychong82} and subsequent authors, as well as the scale hierarchies of the mean momentum balance approach of \cite{Klewicki07}. We explore the similarities in more detail below.

The attached eddy and mean momentum balance analyses predict self-similar families, or hierarchies, of coherent structures that have a shear stress signature that is localised away from the wall. We have shown that the first velocity response modes for $\k$ values that are attached to the wall (in the sense of Townsend, with a footprint that reaches down to the wall) constitute such structure. With respect to the mean momentum balance approach, this is perhaps no surprise since both analyses deal with the Navier-Stokes equations themselves.  The velocity response modes likely identify the velocity and vortical structure giving rise to self-similar local distributions of Reynolds stress associated with the hierarchies.

Similarly, the velocity response modes appear to provide the full velocity field associated with the distribution of self-similar attached eddy structures of the attached eddy hypothesis, or the $\psi_{ii}$ function of \cite{Perryhenchong}.  Whereas the attached eddy hypothesis makes a phenomenological argument for the distribution of attached eddies that explains the statistical behavior of the velocity fluctuations near the wall, our approach identifies structure from the most amplified response modes associated with the Navier-Stokes equations (incorporating appropriate boundary conditions), but lacks the appropriate distribution of amplitudes. This is the opposite interpretation to the vortex induction arguments given by \cite{Adrian07}, the apparent disparity arising because of the assumption of the mean velocity profile as opposed to the attempt to determine the mean velocity profile in the latter case. As such, the approaches appear to be complementary.

The VLSM mode, $\k_2$, clearly belongs to a class of modes that, for high enough Reynolds number, should be considered both ``attached'' to the wall and ``inactive'' in the sense of \cite{Townsend1956}: the response has a velocity footprint that reaches down to the wall but a mean shear stress that is localised sufficiently far from the wall that it has no contribution close to the wall, but can contribute significantly to the overall shear stress. This last effect is expected to increase with Reynolds number, in agreement with the observations of \cite{Adrian07}, who summarises the large contributions to the mean shear stress arising from the very large scales in pipe and channel flows (from the original work of \cite{Guala06} and \cite{BalakumarPTRSA}, respectively), and of \cite{MarusicIntJ10} concerning the increasing significance of turbulence production in log region with increasing Reynolds number.

With respect to the origin of the VLSMs, \cite{Kim99} proposed that they consist of long packets of coherently aligned hairpins. 
  The picture developed here, however, suggests that the packet nature is likely due to a ``beat" effect between modes with the specific beat frequency linked to the dominant energetic VLSM. Furthermore, although the VLSM acts to organise the hairpins, the two structures may be predicted separately. It is however our suspicion that the VLSM and hairpin response mode combinations interact in a mutually supportive manner. The recent papers on hairpin-like exact solutions mentioned in the introduction, in particular that by \cite{Cherubini11} in which a self-sustaining edge state looking like a hairpin packet was found, is suggestive that this is the case.
This indicates that the roll-streak interaction mechanism is not uniquely responsible for driving turbulence and that an analogous hairpin-streak interaction mechanism may exist alongside the near-wall cycle.

\begin{figure}
\centering
\includegraphics[width=0.9\columnwidth]{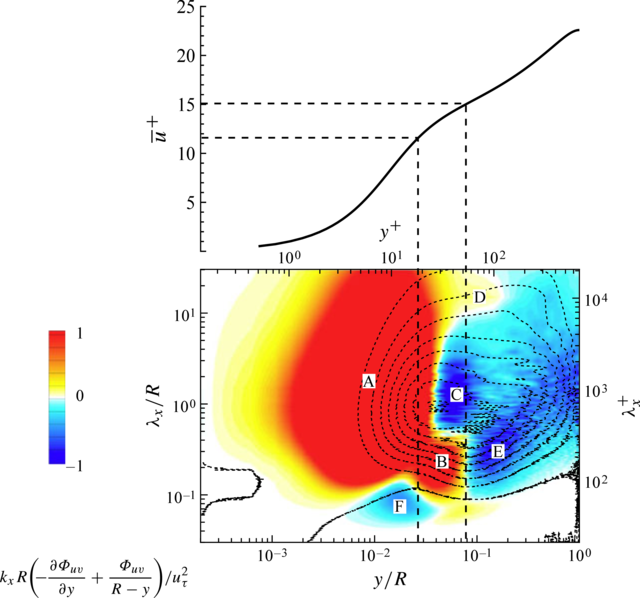}
\caption{Adaptation of figures 2(a) and 17(b) of \cite{WuBaltzerAdrian12}. The lower plot shows the spectral variation of net force in the space of wall-normal locations and streamwise wavelengths for pipe flow at $R^+=685$. The solid line on the upper plot shows the mean velocity profile and the dropped dotted lines show $\uz/U_{CL}=1/2$ and $\uz/U_{CL}=2/3$.}
\label{fig:WuBaltzerAdrian}
\end{figure}

The Reynolds stress distributions associated with individual velocity response modes, which can be superposed, also illuminate the underlying structure behind recent results concerning the wavenumber decomposition of the gradient of the Reynolds stress \citep{Guala06,WuBaltzerAdrian12}. Figure~\ref{fig:WuBaltzerAdrian} includes a reproduction of the net force spectra of figure 17 in \cite{WuBaltzerAdrian12}, consisting of the spectral decomposition of the sum of the gradient of Reynolds stress and a contribution from the Reynolds stress itself arising from the cylindrical coordinate system,
\begin{equation}
    \mathrm{net~force}(y) = -\D{\ave{uv}(y)}{y} + \frac{\ave{uv}(y)}{R-y}.
\end{equation}
Note that in the framework of \cite{WuBaltzerAdrian12}, the full four-dimensional velocity field has been averaged over $n$ and $\omega$. In figure \ref{fig:WuBaltzerAdrian}, the net force associated with streamwise wavenumbers representative of the VLSM modes, $\lambda_x \approx 10$, has a wall-normal extent of positive sign (marked D in figure~\ref{fig:WuBaltzerAdrian}) that reaches much further from the wall than observed for all higher wavenumbers. The change in sign for these wavenumbers occurs at or close to the location where $\uz/U_{CL} = 2/3$, which is the location of the critical layer for the VLSM. These results indicate that such observations are a consequence of the VLSMs being critical modes with critical layers that exist further from the wall than for the shorter modes, as shown in figure~\ref{fig:restressprofiles}(b), such that the associated region of positive gradient extends well into the overlap region of the mean velocity and the sign change resides at or near the critical layer. This effect dominates over the other nearby wavenumber combinations with different wall-normal extents simply because of the energetic dominance of the VLSMs.

Some other observations regarding figure~\ref{fig:WuBaltzerAdrian} are worthy of further comment. Firstly, the locations of the other obvious changes in the sign of the Reynolds stress gradient, corresponding to the transitions from regions B-E and A-C, are coincident with the wall-normal locations of mean velocities equal to $\uz/U_{CL}=2/3$ and $\uz/U_{CL}\simeq 1/2$, respectively. These transitions correspond to the lowest convection velocity of structures observed for the near-wall region in the literature (a summary of the relevant literature is given in \cite{LeHew11}), suggesting that this region is dominated by critical modes traveling at this range of minimum velocities. The B-E transition appears to also occur where $\uz/U_{CL} = 2/3$, identified by \cite{McKeon2010} as an important phase velocity associated with the VLSMs but this is probably an effect of the low Reynolds number associated with the DNS. Remarkably, certain modes close to the wall (not shown) can give rise to positive Reynolds stress, or positive then negative wall-normal gradient of $\ave{uv}$, in the same wavelength range identified by \cite{WuBaltzerAdrian12} denoted F in figure~\ref{fig:WuBaltzerAdrian}. This is another topic worthy of further study.

\subsection{Structure arising from the Navier-Stokes equations and links to the classical laminar, inviscid theory}

We believe this work to constitute the first demonstration of the development of hairpin vortex packets based on analysis of the Navier-Stokes equations and mean profile alone.
While many researchers have identified the non-normality of the linearised Navier-Stokes operator as a source of finite-horizon energy amplification, or transient energy growth, the formulation of \cite{McKeon2010} permits further identification of the loss of symmetry of the resolvent associated with the presence of the wall as the root cause of coherent structure.
The key relevant insight is that the large response near the critical layer can generate localised Reynolds stresses that can in turn excite the linear critical layer amplification. It seems likely, therefore, that underlying all of the nonlinear solutions reported in the literature must be some combination of two linear amplification mechanisms: the critical layer amplification effect manifested as a near-singular response of the linearised system to forcing, and the interaction with the mean shear manifested as operator non-normality.

Regarding non-normality, an operator $X$ is normal when $XX^* = X^* X$ and non-normal otherwise. Since the adjoint $X^*$ is defined only with respect to some inner product (such that $\left<a,Xb\right> = \left<X^*a,b\right>$), non-normality is also defined with respect to the same inner product. Self-adjoint operators are therefore normal.
As is well known, the pipe flow equations are invariant under the translation in $t$, $x$ and $\theta$. Consequently, the resolvent is self-adjoint under the integrals over these coordinates. This is readily proved by considering the translation operator in each coordinate; it commutes with the resolvent and its eigenvectors are the associated Fourier modes. This results in an orthogonality condition and the natural use of the Fourier decomposition. The nonlinear interaction between Fourier modes then obeys the rules of triadic interaction. In the symmetric directions, therefore, the forcing and response modes are equal and are the Fourier modes. The presence of the wall changes this state of affairs: since the symmetry is lost in the direction normal to the wall, the resolvent is not self-adjoint with respect to the inner product associated with integration over that direction. In this case, we see that $0 \leq \inprod{\psi_a}{\phi_b}_{r}\leq 1$, we lose the orthogonality property in the wall-normal direction and associated rules of triadic interaction, the forcing and response modes are no longer equal and the fluctuations may now gain energy from the interaction with the mean flow. The potential for momentum production due to this interaction is quantified by this loss of orthogonality.

\begin{figure}
\centering
{\includegraphics[width = .75\textwidth]{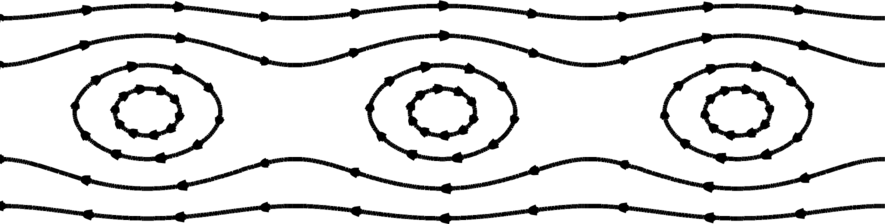}} 
\caption{Kelvin's ``cats' eyes'' streamlines associated with inviscid potential flow in the region of a critical layer \citep{DrazinReid}.}
\label{fig:catseyes}
\end{figure}

It appears from figure \ref{fig:peaklocs1} that the hairpin head is located near the critical layer. This is strongly suggestive that the hairpin heads arise from a mechanism essentially similar to that described by the famous cats' eyes diagram of Lord Kelvin for the laminar, inviscid case -- although the laminar, inviscid theory predicts the Reynolds stress gradient to be infinite at the critical layer, and zero elsewhere \citep{DrazinReid}. The visual resemblance is demonstrated in figure~\ref{fig:catseyes} by a sketch of the corresponding local streamline pattern in the presence of the mean flow.
In the present case, the response modes are three-dimensional, rather than the two-dimensional form implied by Squires' theorem. A hairpin head can therefore be interpreted as a pair of cat's eye vortices meeting obliquely, with regularization at the critical layer due to viscosity.

In a further interesting link to the inviscid laminar theory, for $k=1$ and at high $Re$, the point $c=2/3$ corresponds (on the complex plane) to region where the S-family of eigenvalues of the \emph{laminar} $\Lk$ impinges on the $i\omega$ axis. This explains its association with a high leading singular value. In addition, this is the region where the S-, A- and P-families of eigenvalues meet, and is associated with high sensitivity to perturbation \citep{Reddy93}. Since at this point neither of the branches associated with either the centre eigenmodes or the wall eigenmodes is dominant, the response of the flow has characteristics of both.

\subsection{Taylor's hypothesis, convective velocity, critical layer localisation and coherence}

The response modes under investigation are propagating in the streamwise and spanwise directions and distributed in the wall-normal direction. Their wave-like nature implies that the streamwise propagation velocity of each mode is given by $\omega/k=c$.
Note that $\u_0\simeq c$ modes will dominate in any region with low viscosity. The result of this is that the energy of the critical-type response modes is localised in the region around the critical layer. This is less true for globally-energetic modes, which are less localised and cannot detach.
Such a localisation is manifested by the narrowness of the ridge in the joint $(k,\omega)$ spectrum, where a narrower ridge implies lower dispersion and higher coherence.
This coherence, when coupled with a localisation of energy in the wall-normal direction near the critical layer, is equivalent to an invariance when travelling in the reference frame moving at $c$. The localisation in the wall-normal direction and the low dispersion of corresponding structure is predicted by the radial dependence of the forcing and response modes. In turn, this is explained by the high response of the resolvent in the region of the critical layer.
This effectively justifies the use of Taylor's hypothesis when searching for coherent structure: the application of Taylor's hypothesis is synonymous with coherent, critical response modes. In this sense, the model provides a predictive explanation of coherent structure observed in wall turbulence.

We have only shown idealised hairpin packets here, but the effects described above will be even more important in the case of modes with different convection velocities, where packets will appear to grow and decay, or simply ``evolve'', as a consequence of the interference between modes.
The apparent evolution of collections, or packets, of hairpin vortices therefore occurs because of relative motion between modes. A packet will grow (decay) in space and time if the phase velocity associated with the smaller modes is larger (smaller) than that of the VLSM. Thus a spread in the distribution of convective velocities in the assembly of velocity response modes can capture the experimental observations of the evolution of vortical structure. This is fully within the scope of our approach but is not explored here.
In real flows, deviations from perfect matching of $c$ are of course possible. While a \emph{slight} dispersion negates the problem of choosing an appropriate phase relationship between the modes, it is possible that the nonlinear feedback may lock the relative phase between modes, as described below.

\subsection{Self-organisation, phase and nonlinear feedback}

Turning now to coherent structure, we have shown that the velocity gradients associated with response mode combinations result in very complex swirl fields and that interference between modes results in the apparent disappearance of periodicity.

The presence of an energetic very large scale mode such as $\k_2$ generates local wall-normal gradients of the streamwise velocity. This locally increases or decreases shear, effectively increasing or decreasing the local Reynolds number. Because the swirl diagnostic is nonlinear, this also appears to align hairpins (and their remnants such as canes and arches) to the low-momentum regions, in packets with the period of the VLSM.
The large scale separation of the two types of structure (their separation in wavenumber-frequency space) justifies similar quasi-static explanations. 
Further, the existence of retrograde vortices alongside the prograde ones is entirely consistent with observations in the literature, including the preferred orientations of prograde vortices with respect to retrograde ones discussed by \cite{Natrajan07}.

The analysis also suggests the nonlinear travelling wave solutions reported in the literature for transitional flow continue to be important in high Reynolds number turbulence. We have supported this idea by applying a simple method which examines the linear amplification mechanism that selects such structure.
We expect that if the forcing resulting from a mode combination produces, via the resolvent, a component of the original mode combination, then the mode must be self-sustaining above a certain amplitude. Such a combination would have to be triadically compatible and in the radial direction, the generated forcing must be at the same locality as the mode combination. In actuality, the phase relationship between forcing and response is also important.

We may go on to speculate that another role of the VLSM in a real flow is to cause an effective phase-locking of a wide range of the smaller scale modes. This effect would become increasingly dominant with increasing Reynolds number as the VLSM becomes more and more energetic relative to other individual modes in the flow. Indeed, it seems plausible that the quadratic nonlinear feedback acts to provide phase-locked loop behaviour (analogous to how the nonlinear term acts to keep a soliton localised in the direction of travel), because the nonlinearity is multiplicative between frequencies and so has the right form to act as a phase detector. In this sense the nonlinear feedback can be considered as another mechanism by which the flow can self-organise in a relative phase sense.
Further investigation of this phenomenon is outside the scope of this manuscript, but forms the basis for future work.

\subsection{A new turbulence kernel}

The preceding sections have demonstrated that the linear combination of velocity response modes constituting $K_C$ captures both coherent vortical structure and the essence of higher order statistical results.  In particular, the structure shown in figure~\ref{fig:swirl_outer2} is highly reminiscent of the dominant organization to be expected from the hairpin packet paradigm that has dominated the literature in recent years, while equations~\ref{eqn:skewkc} and \ref{eqn:corrcoeffkc} reveal a distinct similarity between the skewness of the streamwise velocity fluctuations and the apparent amplitude modulation of the small scales by the VLSMs associated with $K_C$ and observations in experiment.

The statistical picture that emerges from $K_C$, which was synthesised using insight from experimental observations at high Reynolds number, confirms that this simple kernel captures key aspects of the turbulence.  In particular, it appears that the statistics of the log-law region are essentially controlled by the VLSM. The ability of the three-mode model to identify the similarity between the amplitude modulation correlation coefficient and the skewness is a consequence of the interpretation of both these measures (and the odd order moments) as providing information on the relative phase between modes, as alluded to by \cite{Bandy84} and \cite{Chung09}. This provides a missing piece of the previous work of \cite{McKeon2010}. The importance of the VLSM to the even moments and the amplitude modulation has been well-documented by \cite{Mathis09} for example. In the case of an energetically dominant VLSM, we expect triadic interactions of the sort described by $K_C$ to dominate the skewness and the amplitude modulation, thus explaining the puzzle of the connection between them that has been discussed by \cite{SchlatterOrlu10} and \cite{Mathis11}. These effects arise from a linear combination of modes with different phases, rather than a nonlinear effect.

That the type of triadic interaction capable of capturing the amplitude modulation also generates recognizable coherent vortical structure supports our hypothesis of its importance to wall turbulence. We suggest, then, that a subset of three modes that includes the VLSM and two smaller modes constitutes a new kernel of turbulence in the sense of \cite{SmithetalPTRSA91}. Such kernels form studiable sub-units that contain important dynamics associated with the flow, a well-known example being the autonomous near-wall cycle~\citep{Waleffe97}, and which can be summed together to regenerate the full picture of wall turbulence (or at least create a synthetic proxy for the real flow). It is also likely, but is not proven here, that such a kernel self-sustains, so that the forcing arising from any two components of triadic interactions like $K_C$ is sufficient to excite the third component.

\section{Conclusion}

\cite{McKeon2010} developed a formulation of the Navier-Stokes equations where the resolvent acts as a directional amplifier to an unstructured forcing generated by the nonlinear interaction between modes. It was demonstrated that this predicts a velocity response mode associated with the first singular value of the resolvent at any given scale and that these modes capture statistical features associated with pipe flow turbulence. This was essentially a demonstration that resolvent is well approximated by a low-rank operator. This reflects the fact that the underlying dynamics are well approximated by low-dimensional dynamics, which can provide a vast reduction in the dimension of the problem.
In the present work, we have built on this formulation to demonstrate that the response modes efficiently encode information on coherent structure as defined phenomenologically in the introduction. While the response modes can be superposed to assemble more complex flow fields, any measure of swirling structure is inherently nonlinear and therefore the perceived structure associated with an assembly of modes is not the linear sum of the structure associated with individual modes. This, perhaps, lies at the heart of the difficulty of observation and characterization of coherent structure in real flows.

The response modes contain coherent structures as objects mathematically derived from the Navier-Stokes equations under the assumption of a mean turbulent profile. As such, we posit that they form a desirable basis for a low-order model when compared to approaches that replicate coherent structures found in experimental data (such as POD) or describe them phenomenologically. The minimal-order model is derived from the lowest rank (rank-1) approximation of the resolvent; assuming the resolvent is compact, there is finite approximation error associated with this truncation. Though the inclusion of additional response modes will improve the approximation, this has not proved to be necessary for the present purpose.

Whilst the mechanisms underlying turbulence are captured by the resolvent, to actually observe hairpins as structure in an experimental flow, four ingredients are necessary: (1) a nonlinear diagnostic such as swirl, (2) an averaging (or equivalently a frequency domain analysis), (3) coherence, provided by the assumption of Taylor's hypothesis or a critical layer response and (4) a flow boundary, to provide characteristic scales and a shear, which acts in the mean sense to provide net vorticity responsible for the hairpin head and locally in response to the VLSM to organise hairpins around areas of low momentum.

The directional amplification of the resolvent provides a mechanism to understand the interaction of simple external forcing strategies, such as the introduction of a single $\k$ disturbance at the wall using dynamic roughness explored by \cite{Jacobidynamic11}, as well as the good approximations to the mean velocity profile obtained using stochastic forcing and a two-dimensional, three-component flow model \citep{Gayme2D3C,Bourguignon11}. Together with the nonlinear feedback term, the directional amplification is understood to underlie the robustness of the flow features observed in real flows. There are obvious related implications for flow estimation and control.

Finally, we have identified a new turbulence kernel, composed of these response modes, that sheds some light on a current controversy in the literature and, importantly, links statistical and structural observations by consideration of the relative phase between modes.

Our study here has been focused on $R^+=1800$, at the upper end of the range of Reynolds numbers reachable using current DNS, but we stress that the only constraint on studying higher Reynolds numbers lies in the numerical precision required to deal with the large response associated with the most amplified forcing.

\vspace{0.1in}
The support of the Air Force Office of Scientific Research Aerothermodynamics and Turbulence portfolio, under grant  $\#$FA9550-08-1-0049 (Program Manager John Schmisseur) is gratefully acknowledged by BJM. A substantial portion of this work was completed while AS was with
the Department of Automatic Control and Systems Engineering at the University of Sheffield, UK.

\bibliographystyle{jfm}

\begin{thebibliography}{82}
\expandafter\ifx\csname natexlab\endcsname\relax\def\natexlab#1{#1}\fi

\bibitem[Adrian(2007)]{Adrian07}
{\sc Adrian, R.~J.} 2007 Vortex organization in wall turbulence. {\em Phys.
  Fluids\/} {\bf 19}~(041301).

\bibitem[Adrian {\em et~al.\/}(2000{\natexlab{{\em a\/}}})Adrian, Christensen
  \& Liu]{Adrianinstfields00}
{\sc Adrian, R.~J., Christensen, K.~T. \& Liu, Z.-C.} 2000{\natexlab{{\em
  a\/}}} Analysis and interpretation of instantaneous turbulent velocity
  fields. {\em Expts. in Fluids\/} {\bf 29}, 275--290.

\bibitem[Adrian {\em et~al.\/}(2000{\natexlab{{\em b\/}}})Adrian, Meinhart \&
  Tomkins]{Adrian00}
{\sc Adrian, R.~J., Meinhart, C.~D. \& Tomkins, C.~D.} 2000{\natexlab{{\em
  b\/}}} Vortex organization in the outer region of the turbulent boundary
  layer. {\em J. Fluid Mech.\/} {\bf 422}, 1--54.

\bibitem[del \'Alamo \& Jim\'enez(2006)]{delAlamo06}
{\sc del \'Alamo, J.~C. \& Jim\'enez, J.} 2006 Linear energy amplification in
  turbulent channels. {\em J. Fluid Mech.\/} {\bf 559}, 205--213.

\bibitem[Bailey \& Smits(2009)]{BaileyPOD09}
{\sc Bailey, S. C.~C. \& Smits, A.~J.} 2009 The structure of large- and very
  large-scale motions in turbulent pipe flow. {\em AIAA Paper\/} {\bf AIAA
  2009-3684}.

\bibitem[Balakumar \& Adrian(2007)]{BalakumarPTRSA}
{\sc Balakumar, B.~J. \& Adrian, R.~J.} 2007 Large- and very-large-scale
  motions in channel and boundary layer flows. {\em Phil. Trans. Royal Soc.
  A\/} {\bf 365}, 665--681.

\bibitem[Bandyopadhyay \& Hussain(1984)]{Bandy84}
{\sc Bandyopadhyay, P.~R. \& Hussain, A. K. M.~F.} 1984 The coupling between
  scales in shear flows. {\em Phys. Fluids\/} {\bf 27}~(9), 2221--2228.

\bibitem[Benney \& Bergeron(1969)]{Benney69}
{\sc Benney, D.~J. \& Bergeron, R.~F.} 1969 {A new class of nonlinear waves in
  parallel flows}. {\em Stud. Appl. Math.\/} {\bf 48}, 181--204.

\bibitem[Bourguignon \& McKeon(2011)]{Bourguignon11}
{\sc Bourguignon, J.-L. \& McKeon, B.~J.} 2011 A streamwise-constant model of
  turbulent pipe flow. {\em Phys. Fluids\/} {\bf 23}~(095111).

\bibitem[Carlier \& Stanislas(2005)]{Carlier05}
{\sc Carlier, J. \& Stanislas, M.} 2005 Experimental study of eddy structures
  in a turbulent boundary layer using particle image velocimetry. {\em J. Fluid
  Mech.\/} {\bf 535}, 143--188.

\bibitem[Chakraborty {\em et~al.\/}(2005)Chakraborty, Balachandar \&
  Adrian]{Chakraborty05}
{\sc Chakraborty, P., Balachandar, S. \& Adrian, R.~J.} 2005 On the
  relationships between local vortex identification schemes. {\em J. Fluid
  Mech.\/} {\bf 535}, 189--214.

\bibitem[Chernyshenko {\em et~al.\/}(2006)Chernyshenko, Cicca, Iollo, Smirnov,
  Sandham \& Hu]{Chernyshenko06}
{\sc Chernyshenko, S.I., Cicca, G.M., Iollo, A., Smirnov, A.V., Sandham, N.D.
  \& Hu, Z.W.} 2006 Analysis of data on the relation between eddies and streaky
  structures in turbulent flows using the placebo method. {\em Fluid
  Dynamics\/} {\bf 41}~(5), 772--783.

\bibitem[Cherubini {\em et~al.\/}(2011)Cherubini, de~Palma, Robinet \&
  Bottaro]{Cherubini11}
{\sc Cherubini, S., de~Palma, P., Robinet, J.-Ch. \& Bottaro, A.} 2011 Edge
  states in a boundary layer. {\em Phys. Fluids\/} {\bf 23}~(5), 051705.

\bibitem[Chung \& McKeon(2010)]{Chung09}
{\sc Chung, D. \& McKeon, B.~J.} 2010 Large-eddy simulation investigation of
  large-scale structures in a long channel flow. {\em J. Fluid Mech.\/} {\bf
  661}, 341--364.

\bibitem[Cossu {\em et~al.\/}(2009)Cossu, Pujals \& Depardon]{Cossu09}
{\sc Cossu, C., Pujals, G. \& Depardon, S.} 2009 Optimal transient growth and
  very large scale structures in turbulent boundary layers. {\em J. Fluid
  Mech.\/} {\bf 619}, 79--94.

\bibitem[Deguchi \& Nagata(2010)]{Deguchi10}
{\sc Deguchi, K \& Nagata, M} 2010 􏰃 traveling hairpin-shaped fluid vortices
  in plane couette flow. {\em Phys. Rev. E\/} {\bf 82}, 056325.

\bibitem[Dennis \& Nickels(2011{\natexlab{{\em a\/}}})]{Dennisvort11}
{\sc Dennis, D. \& Nickels, T.} 2011{\natexlab{{\em a\/}}} Experimental
  measurement of large-scale three-dimensional structures in a turbulent
  boundary layer. {P}art 1. {V}ortex packets. {\em J. Fluid Mech.\/} {\bf 673},
  180--217.

\bibitem[Dennis \& Nickels(2011{\natexlab{{\em b\/}}})]{DennisVLSM11}
{\sc Dennis, D. \& Nickels, T.} 2011{\natexlab{{\em b\/}}} Experimental
  measurement of large-scale three-dimensional structures in a turbulent
  boundary layer. {P}art 2. {L}ong structures. {\em J. Fluid Mech.\/} {\bf
  673}, 218--244.

\bibitem[Drazin \& Reid(2004)]{DrazinReid}
{\sc Drazin, P.~G. \& Reid, W.~H.} 2004 {\em Hydrodynamic {S}tability\/}, 2nd
  edn. {\em Cambridge Mathematical Libraries\/} . Cambridge University Press.

\bibitem[Duguet {\em et~al.\/}(2008)Duguet, Willis \& Kerswell]{Duguet08}
{\sc Duguet, Y., Willis, A.~P. \& Kerswell, R.~R.} 2008 Transition in pipe
  flow: the saddle structure on the boundary of turbulence. {\em J. Fluid
  Mech.\/} {\bf 613}, 255--274.

\bibitem[Duguet {\em et~al.\/}(2010)Duguet, Willis \& Kerswell]{Duguet10}
{\sc Duguet, Y., Willis, A.~P. \& Kerswell, R.~R.} 2010 Slug genesis in
  cylindrical pipe flow. {\em J. Fluid Mech.\/} {\bf 663}, 180--208.

\bibitem[Eckhardt {\em et~al.\/}(2007)Eckhardt, Schneider, Hof \&
  Westerweel]{Eckhardt07}
{\sc Eckhardt, B., Schneider, T.~M., Hof, B. \& Westerweel, J.} 2007 Turbulence
  transition in pipe flow. {\em Annu. Rev. Fluid Mech.\/} {\bf 39}~(1),
  447--468.

\bibitem[Falco(1977)]{Falco77}
{\sc Falco, R.~E.} 1977 Coherent motions in the outer region of turbulent
  boundary layers. {\em Phys. Fluids\/} {\bf 20}, S124--S132.

\bibitem[Falco(1991)]{Falco91}
{\sc Falco, R.~E.} 1991 A coherent structure model of the turbulent boundary
  layer and its ability to predict reynolds number dependence. {\em
  Philosophical Transactions of the Royal Society of London. Series A: Physical
  and Engineering Sciences\/} {\bf 336}~(1641), 103--129.

\bibitem[Ganapathisubramani {\em et~al.\/}(2003)Ganapathisubramani, Longmire \&
  Marusic]{Bharat03}
{\sc Ganapathisubramani, B., Longmire, E.~K. \& Marusic, I.} 2003
  Characteristics of vortex packets in turbulent boundary layers. {\em J. Fluid
  Mech.\/} {\bf 478}, 35--46.

\bibitem[Gao {\em et~al.\/}(2011)Gao, Ortiz-Due{\~{n}}as \& Longmire]{Gao11}
{\sc Gao, Q., Ortiz-Due{\~{n}}as, C. \& Longmire, E.~K.} 2011 Analysis of
  vortex populations in turbulent wall-bounded flows. {\em J. Fluid Mech.\/}
  {\bf 678}, 87--123.

\bibitem[Gayme {\em et~al.\/}(2010)Gayme, McKeon, Papachristodolou, Bamieh \&
  Doyle]{Gayme2D3C}
{\sc Gayme, D.~F., McKeon, B.~J., Papachristodolou, A., Bamieh, B. \& Doyle,
  J.~C.} 2010 Streamwise constant model of turbulence in plane {C}ouette flow.
  {\em J. Fluid Mech.\/} {\bf 665}, 99--119.

\bibitem[Generalis \& Itano(2010)]{Generalis10}
{\sc Generalis, Sotos~C. \& Itano, Tomoaki} 2010 Characterization of the
  hairpin vortex solution in plane couette flow. {\em Phys. Rev. E\/} {\bf 82},
  066308.

\bibitem[Gibson {\em et~al.\/}(2009)Gibson, Halcrow \& Cvitanovi\'c]{Gibson09}
{\sc Gibson, J.~F., Halcrow, J. \& Cvitanovi\'c, P.} 2009 Equilibrium and
  travelling-wave solutions of plane couette flow. {\em Journal of Fluid
  Mechanics\/} {\bf 638}, 243--266.

\bibitem[Guala {\em et~al.\/}(2006)Guala, Hommema \& Adrian]{Guala06}
{\sc Guala, M., Hommema, S.~E. \& Adrian, R.~J.} 2006 Large-scale and
  very-large-scale motions in turbulent pipe flow. {\em J. Fluid Mech.\/} {\bf
  554}, 521--542.

\bibitem[Hall \& Sherwin(2010)]{Hall10}
{\sc Hall, P. \& Sherwin, S.~J.} 2010 Streamwise vortices in shear flows:
  harbingers of transition and the skeleton of coherent structures. {\em J.
  Fluid Mech.\/} {\bf 661}, 178--205.

\bibitem[Head \& Bandyopadhyay(1981)]{HB81}
{\sc Head, M.~R. \& Bandyopadhyay, P.~R.} 1981 New aspects of turbulent
  boundary-layer structure. {\em J. Fluid Mech.\/} {\bf 107}, 297--337.

\bibitem[Hellstr\"{o}m {\em et~al.\/}(2011)Hellstr\"{o}m, Sinha \&
  Smits]{Hellstroem11}
{\sc Hellstr\"{o}m, L. H.~O., Sinha, A. \& Smits, A.~J.} 2011 Visualizing the
  very-large-scale motions in turbulent pipe flow. {\em Phys. Fluids\/} {\bf
  011703}.

\bibitem[Hutchins \& Marusic(2007{\natexlab{{\em a\/}}})]{Hutchins07}
{\sc Hutchins, N. \& Marusic, I.} 2007{\natexlab{{\em a\/}}} Evidence of very
  long meandering features in the logarithmic region of turbulent boundary
  layers. {\em J. Fluid Mech.\/} {\bf 579}, 1--28.

\bibitem[Hutchins \& Marusic(2007{\natexlab{{\em b\/}}})]{HutchinsPTRSA07}
{\sc Hutchins, N. \& Marusic, I.} 2007{\natexlab{{\em b\/}}} Large-scale
  influences in near-wall turbulence. {\em Phil. Trans. Royal Soc. A\/} {\bf
  365}, 647--664.

\bibitem[Hutchins {\em et~al.\/}(2011)Hutchins, Monty, Ganapathisubramani, Ng
  \& Marusic]{Hutchins11}
{\sc Hutchins, N., Monty, J.~P., Ganapathisubramani, B., Ng, H. C.~H. \&
  Marusic, I.} 2011 Three-dimensional conditional structure of a
  high-{R}eynolds number turbulent boundary layer. {\em J. Fluid Mech.\/} {\bf
  673}, 255--285.

\bibitem[Itano \& Generalis(2009)]{Itano09}
{\sc Itano, Tomoaki \& Generalis, Sotos~C.} 2009 Hairpin vortex solution in
  planar couette flow: A tapestry of knotted vortices. {\em Phys. Rev. Lett.\/}
  {\bf 102}, 114501.

\bibitem[Jacobi \& McKeon(2011)]{Jacobidynamic11}
{\sc Jacobi, I. \& McKeon, B.~J.} 2011 Dynamic roughness-perturbation of a
  turbulent boundary layer. {\em J. Fluid Mech.\/} {\bf 688}, 258--296.

\bibitem[Jacobi \& McKeon(2012)]{Jacobi12}
{\sc Jacobi, I. \& McKeon, B.~J.} 2012 Scale interactions in the turbulent
  boundary layer. {\em (In preparation)\/} .

\bibitem[Jeong \& Hussain(1995)]{Jeong95}
{\sc Jeong, J. \& Hussain, F.} 1995 On the identification of a vortex. {\em J.
  Fluid Mech.\/} {\bf 285}, 69--94.

\bibitem[Kerswell(2005)]{Kerswell05}
{\sc Kerswell, R~R} 2005 Recent progress in understanding the transition to
  turbulence in a pipe. {\em Nonlinearity\/} {\bf 18}~(6), R17.

\bibitem[Kim \& Adrian(1999)]{Kim99}
{\sc Kim, K.~C. \& Adrian, R.~J.} 1999 Very large-scale motion in the outer
  layer. {\em Phys. Fluids\/} {\bf 11}, 417--422.

\bibitem[Klewicki {\em et~al.\/}(2007)Klewicki, Fife, Wei \&
  McMurtry]{Klewicki07}
{\sc Klewicki, J.~C., Fife, P., Wei, T. \& McMurtry, P.} 2007 A physical model
  of the turbulent boundary layer consonant with the mean momentum balance
  structure. {\em Phil. Trans. Royal Soc. A\/} {\bf 365}, 823--839.

\bibitem[Lee \& Sung(2011)]{Sung11}
{\sc Lee, J.~H. \& Sung, H.~J.} 2011 Very-large-scale motions in a turbulent
  boundary layer. {\em J. Fluid Mech.\/} {\bf 673}, 80--120.

\bibitem[LeHew {\em et~al.\/}(2011)LeHew, Guala \& McKeon]{LeHew11}
{\sc LeHew, J., Guala, M. \& McKeon, B.~J.} 2011 A study of the
  three-dimensional spectral energy distribution in a zero pressure gradient
  turbulent boundary layer. {\em Expts. in Fluids\/} {\bf 51}~(4), 997--1012.

\bibitem[Marusic {\em et~al.\/}(2010)Marusic, Mathis \&
  Hutchins]{MarusicIntJ10}
{\sc Marusic, I., Mathis, R. \& Hutchins, N.} 2010 High {R}eynolds number
  effects in wall turbulence. {\em Int. J. Heat Fluid Flow\/} {\bf 31},
  418--428.

\bibitem[Maslowe(1986)]{Maslowe86}
{\sc Maslowe, S~A} 1986 Critical layers in shear flows. {\em Annu. Rev. Fluid
  Mech.\/} {\bf 18}~(1), 405--432.

\bibitem[Mathis {\em et~al.\/}(2009)Mathis, Hutchins \& Marusic]{Mathis09}
{\sc Mathis, R., Hutchins, N. \& Marusic, I.} 2009 Large-scale amplitude
  modulation of the small-scale structures of turbulent boundary layers. {\em
  J. Fluid Mech.\/} {\bf 628}, 311--337.

\bibitem[Mathis {\em et~al.\/}(2011)Mathis, Marusic, Hutchins \&
  Sreenivasan]{Mathis11}
{\sc Mathis, R., Marusic, I., Hutchins, N. \& Sreenivasan, K.~R.} 2011 The
  relationship between the velocity skewness and the amplitude modulation of
  the small scale by the large scale in turbulent boundary layers. {\em Phys.
  Fluids\/} {\bf 23}~(121702).

\bibitem[McKeon {\em et~al.\/}(2004)McKeon, Li, Jiang, Morrison \&
  Smits]{mckeonmean04}
{\sc McKeon, B.~J., Li, J., Jiang, W., Morrison, J.~F. \& Smits, A.~J.} 2004
  Further observations on the mean velocity distribution in fully developed
  pipe flow. {\em J. Fluid Mech.\/} {\bf 501}, 135--147.

\bibitem[McKeon \& Sharma(2010)]{McKeon2010}
{\sc McKeon, B.~J. \& Sharma, A.~S.} 2010 A critical layer model for turbulent
  pipe flow. {\em J. Fluid Mech.\/} {\bf 658}, 336--382.

\bibitem[McKeon {\em et~al.\/}(2010)McKeon, Sharma \& Jacobi]{ArXiv10}
{\sc McKeon, B.~J., Sharma, A.~S. \& Jacobi, I.} 2010 Predicting structural and
  statistical features of wall turbulence. {\em ArXiV\/} {\bf 1012-0426}.

\bibitem[McKeon {\em et~al.\/}(2013)McKeon, Sharma \& Jacobi]{McKeonPoF13}
{\sc McKeon, B.~J., Sharma, A.~S. \& Jacobi, I.} 2013 Experimental manipulation
  of wall turbulence: a systems approach. {\em Phys. Fluids\/} {\bf (to
  appear)}.

\bibitem[Meinhart \& Adrian(1995)]{Meinhart95}
{\sc Meinhart, C.~D. \& Adrian, R.~J.} 1995 On the existence of uniform
  momentum zones in a turbulent boundary layer. {\em Phys. Fluids\/} {\bf
  7}~(4), 694--696.

\bibitem[Meseguer \& Trefethen(2003)]{Meseguer03}
{\sc Meseguer, A. \& Trefethen, L.~N.} 2003 Linearized pipe flow to {R}eynolds
  number $10^7$. {\em J. Comp. Phys.\/} {\bf 186}, 178--197.

\bibitem[Monty {\em et~al.\/}(2009)Monty, Hutchins, Ng, Marusic \&
  Chong]{Monty09}
{\sc Monty, J.~P., Hutchins, N., Ng, H. C.~H., Marusic, I. \& Chong, M.~S.}
  2009 A comparison of turbulent pipe, channel and boundary layer flows. {\em
  J. Fluid Mech.\/} {\bf 632}, 431--442.

\bibitem[Monty {\em et~al.\/}(2007)Monty, Stewart, Williams \& Chong]{Monty07}
{\sc Monty, J.~P., Stewart, J.~A., Williams, R.~C. \& Chong, M.~S.} 2007
  Large-scale features in turbulent pipe and channel flows. {\em J. Fluid
  Mech.\/} {\bf 589}, 147--156.

\bibitem[Morris {\em et~al.\/}(2007)Morris, Stolpa, Slaboch \&
  Klewicki]{Morris07}
{\sc Morris, S.~C., Stolpa, S.~R., Slaboch, P.~E. \& Klewicki, J.} 2007
  Near-surface particle image velocimetry measurements in a transitionally
  rough-wall atmospheric boundary layer. {\em J. Fluid Mech.\/} {\bf 580},
  319--338.

\bibitem[Mullin(2011)]{Mullin11}
{\sc Mullin, T.} 2011 Experimental studies of transition to turbulence in a
  pipe. {\em Annu. Rev. Fluid Mech.\/} {\bf 43}~(1), 1--24.

\bibitem[Natrajan {\em et~al.\/}(2007)Natrajan, Wu \& Christensen]{Natrajan07}
{\sc Natrajan, V.~K., Wu, Y. \& Christensen, K.~T.} 2007 Spatial signatures of
  retrograde spanwise vortices in wall turbulence. {\em J. Fluid Mech.\/} {\bf
  574}, 155--167.

\bibitem[Perry \& Chong(1982)]{Perrychong82}
{\sc Perry, A.~E. \& Chong, M.~S.} 1982 On the mechanism of wall turbulence.
  {\em J. Fluid Mech\/} {\bf 119}, 173--217.

\bibitem[Perry {\em et~al.\/}(1986)Perry, Henbest \& Chong]{Perryhenchong}
{\sc Perry, A.~E., Henbest, S. \& Chong, M.~S.} 1986 A theoretical and
  experimental study of wall turbulence. {\em J. Fluid Mech.\/} {\bf 195},
  163--199.

\bibitem[Perry \& Marusic(1995)]{PerryMarusic95}
{\sc Perry, A.~E. \& Marusic, I.} 1995 A wall-wake model for the turbulence
  structure of boundary layers. {P}art 1. {E}xtension of the attached eddy
  hypothesis. {\em J. Fluid Mech.\/} {\bf 298}, 361--388.

\bibitem[Pringle {\em et~al.\/}(2009)Pringle, Duguet \& Kerswell]{Pringle09}
{\sc Pringle, C. C.~T., Duguet, Y. \& Kerswell, R.~R.} 2009 Highly symmetric
  travelling waves in pipe flow. {\em Phil. Trans. Royal Soc. A\/} {\bf 367},
  457--472.

\bibitem[Reddy {\em et~al.\/}(1993)Reddy, Schmid \& Henningson]{Reddy93}
{\sc Reddy, S.~C., Schmid, P.~J. \& Henningson, D.~S.} 1993 Pseudospectra of
  the {O}rr-{S}ommerfeld equation. {\em SIAM J. Appl. Math\/} {\bf 53}~(1),
  15--47.

\bibitem[Robinson(1991)]{Robinson91}
{\sc Robinson, S.~K.} 1991 Coherent motions in the turbulent boundary layer.
  {\em Annu. Rev. Fluid Mech.\/} {\bf 23}, 601--639.

\bibitem[Schlatter \& {\"{O}}rl{\"{u}}(2010)]{SchlatterOrlu10}
{\sc Schlatter, P. \& {\"{O}}rl{\"{u}}, R.} 2010 Quantifying the interaction
  between large and small scales in wall-bounded turbulent flows: {A} note of
  caution. {\em Phys. Fluids\/} {\bf 22}~(051704).

\bibitem[Schlatter {\em et~al.\/}(2009)Schlatter, {\"{O}}rlu, Li, Brethouwer,
  Fransson, Johansson, Alfredsson \& Henningson]{Schlatter09long}
{\sc Schlatter, P., {\"{O}}rlu, R., Li, Q., Brethouwer, G., Fransson, J. H.~M.,
  Johansson, A.~V., Alfredsson, P.~H. \& Henningson, D.~S.} 2009 Simulations of
  spatially evolving turbulent boundary layers up to {${R}e_\theta=4300$}. {\em
  Int J. Heat Fluid Flow\/} {\bf 31}, 251--261.

\bibitem[Sharma \& McKeon(2011)]{SharmaMcKeonTSFP}
{\sc Sharma, A.~S. \& McKeon, B.~J.} 2011 Very large scale motions in pipe
  turbulence derived from a simple critical-layer model. In {\em Turbulence and
  Shear Flow Phenomena, TSFP-7\/}.

\bibitem[Smith {\em et~al.\/}(1991)Smith, Walker, Haidari \&
  Sobrun]{SmithetalPTRSA91}
{\sc Smith, C.~R., Walker, J. D.~A., Haidari, A.~H. \& Sobrun, U.} 1991 On the
  dynamics of near-wall turbulence. {\em Phil. Trans. Royal Soc.\/} {\bf 336},
  131--175.

\bibitem[Smith \& Bodonyi(1982)]{Smith82}
{\sc Smith, F.~T. \& Bodonyi, R.~J.} 1982 Amplitude-dependent neutral modes in
  the {H}agen-{P}oiseille flow though a circular pipe. {\em Proc. Royal Soc.
  A\/} {\bf 384}, 463.

\bibitem[Smits {\em et~al.\/}(2011)Smits, McKeon \& Marusic]{SmitsARFM11}
{\sc Smits, A.~J., McKeon, B.~J. \& Marusic, I.} 2011 High {R}eynolds number
  wall turbulence. {\em Annu. Rev. Fluid Mech.\/} {\bf 43}, 353--375.

\bibitem[Theodorsen(1952)]{Theodorsen52}
{\sc Theodorsen, T.} 1952 Mechanism of turbulence. In {\em Proc. 2nd Midwestern
  Conf. on Fluid Mech.\/}, pp. 1--19. Ohio State University, Columbus, Ohio.

\bibitem[Tomkins \& Adrian(2005)]{Tomkins05}
{\sc Tomkins, C.~D. \& Adrian, R.~J.} 2005 Energetic spanwise modes in the
  logarithmic layer of a turbulent boundary layer. {\em J. Fluid Mech.\/} {\bf
  545}, 141--162.

\bibitem[den Toonder \& Nieuwstadt(1997)]{toonder}
{\sc den Toonder, J. M.~J. \& Nieuwstadt, F. T.~M.} 1997 Reynolds number
  effects in a turbulent pipe flow for low to moderate {R}e. {\em Phys.
  Fluids\/} {\bf 9}, 3398--3409.

\bibitem[Townsend(1956)]{Townsend1956}
{\sc Townsend, A.~A.} 1956 {\em The Structure of Turbulent Shear Flow\/}.
  Cambridge, UK: Cambridge University Press.

\bibitem[Townsend(1976)]{Townsend1976}
{\sc Townsend, A.~A.} 1976 {\em The Structure of Turbulent Shear Flow\/}.
  Cambridge, UK: Cambridge University Press.

\bibitem[Waleffe(1997)]{Waleffe97}
{\sc Waleffe, F.} 1997 On a self-sustaining process in shear flows. {\em Phys.
  Fluids\/} {\bf 9}~(4), 883--900.

\bibitem[Waleffe(2003)]{Waleffe03}
{\sc Waleffe, F.} 2003 Homotopy of exact coherent structures in plane shear
  flows. {\em Phys. Fluids\/} {\bf 15}~(6).

\bibitem[Wedin \& Kerswell(2004)]{Wedin04}
{\sc Wedin, H \& Kerswell, R~R} 2004 Exact coherent structures in pipe flow:
  Travelling wave solutions. {\em J. Fluid Mech.\/} {\bf 508}, 333--371.

\bibitem[Wu {\em et~al.\/}(2012)Wu, Baltzer \& Adrian]{WuBaltzerAdrian12}
{\sc Wu, X., Baltzer, J.~R. \& Adrian, R.~J.} 2012 Direct numerical simulation
  of a {$30R$} long turbulent pipe flow at {$R^+=685$}: large- and very
  large-scale motions. {\em J. Fluid Mech.\/} {\bf 698}, 235--281.

\bibitem[Wu \& Christensen(2006)]{Wu06}
{\sc Wu, Y. \& Christensen, K.~T.} 2006 Population trends of spanwise vortices
  in wall turbulence. {\em J. Fluid Mech.\/} {\bf 568}, 55--76.

\end{thebibliography}

\end{document}